\documentclass[aip,preprint,nofootinbib]{revtex4-1}

\usepackage{algpseudocode}  
\newcounter{algorithm}
\newenvironment{algorithm}{%
  \refstepcounter{algorithm}%
  \par\medskip\noindent\begin{minipage}{\linewidth}%
  \renewcommand{\caption}[1]{\textbf{Algorithm \thealgorithm: ##1}\par\smallskip}%
}{%
  \end{minipage}\par\medskip
}
\usepackage{graphicx}  
\usepackage{subcaption} 
\usepackage{amsmath}
\usepackage{amssymb}
\usepackage{physics}
\usepackage{bm} 
\usepackage[english]{babel}
\usepackage{braket} 
\usepackage{pifont}
\usepackage{xcolor}
\usepackage{xspace}

\usepackage{hyperref}

\newcommand{\mvec}[1]{\bm{#1}}

\newcommand{\octopus}{Octopus\xspace}
\newcommand{\phonopy}{phonopy\xspace}

\begin{document}

\title{Extending the reach of real-space real-time TDDFT: new developments in the \octopus code}

\author{Nicolas Tancogne-Dejean$^*$}
\affiliation{Research Center Future Energy Materials and Systems of the University Alliance Ruhr and Interdisciplinary Centre for Advanced Materials Simulation, Faculty of Physics and Astronomy, Ruhr University Bochum, Universitätsstraße 150, D-44801 Bochum, Germany}
\affiliation{Max Planck Institute for the Structure and Dynamics of Matter, Luruper Chaussee 149, D-22761 Hamburg, Germany}

\author{Martin L\"uders$^*$}
\affiliation{Max Planck Institute for the Structure and Dynamics of Matter, Luruper Chaussee 149, D-22761 Hamburg, Germany}

\author{Esra Ilke Albar}
\affiliation{Max Planck Institute for the Structure and Dynamics of Matter, Luruper Chaussee 149, D-22761 Hamburg, Germany}
\affiliation{Fritz-Haber-Institute of the Max-Planck-Society, 14195 Berlin, Germany}

\author{Heiko Appel}
\affiliation{Max Planck Institute for the Structure and Dynamics of Matter, Luruper Chaussee 149, D-22761 Hamburg, Germany}

\author{Franco P Bonafé}
\affiliation{Max Planck Institute for the Structure and Dynamics of Matter, Luruper Chaussee 149, D-22761 Hamburg, Germany}

\author{Alexander Buccheri}
\affiliation{Max Planck Institute for the Structure and Dynamics of Matter, Luruper Chaussee 149, D-22761 Hamburg, Germany}

\author{Alessandro Casalino}
\affiliation{Max Planck Computing and Data Facility, Gießenbachstraße 2, D-85748 Garching, Germany}

\author{Benshu Fan}
\affiliation{Max Planck Institute for the Structure and Dynamics of Matter, Luruper Chaussee 149, D-22761 Hamburg, Germany}

\author{Hans Fangohr}
\affiliation{Max Planck Institute for the Structure and Dynamics of Matter, Luruper Chaussee 149, D-22761 Hamburg, Germany}
\affiliation{University of Southampton, Southampton, SO17 1BJ, United Kingdom}

\author{Umberto De Giovannini}
\affiliation{Dipartimento di Fisica e Chimica-Emilio Segr\`e, Universit\`a degli Studi di Palermo, Via Archirafi 36, 90123 Palermo, Italy}
\affiliation{Max Planck Institute for the Structure and Dynamics of Matter, Luruper Chaussee 149, D-22761 Hamburg, Germany}

\author{Henning Glawe}
\affiliation{Max Planck Institute for the Structure and Dynamics of Matter, Luruper Chaussee 149, D-22761 Hamburg, Germany}

\author{Hannes H\"ubener}
\affiliation{Max Planck Institute for the Structure and Dynamics of Matter, Luruper Chaussee 149, D-22761 Hamburg, Germany}

\author{Simon V. B. Jensen}
\affiliation{Max Planck Institute for the Structure and Dynamics of Matter, Luruper Chaussee 149, D-22761 Hamburg, Germany}
\affiliation{Center for Free-Electron Laser Science, Luruper Chaussee 149, D-22761 Hamburg, Germany}

\author{Christian J\"ons}
\affiliation{Max Planck Institute for the Structure and Dynamics of Matter, Luruper Chaussee 149, D-22761 Hamburg, Germany}
\affiliation{Center for Free-Electron Laser Science, Luruper Chaussee 149, D-22761 Hamburg, Germany}

\author{Lukas Konecny}
\affiliation{Institute of Inorganic Chemistry, Slovak Academy of Sciences, SK-84536 Bratislava, Slovakia}
\affiliation{Department of Inorganic Chemistry, Faculty of Natural Sciences, Comenius University,
SK-84215 Bratislava, Slovakia}
\affiliation{Max Planck Institute for the Structure and Dynamics of Matter, Luruper Chaussee 149, D-22761 Hamburg, Germany}

\author{Martin Lang}
\affiliation{Max Planck Institute for the Structure and Dynamics of Matter, Luruper Chaussee 149, D-22761 Hamburg, Germany}

\author{Ruipeng Li}
\affiliation{Max Planck Institute for the Structure and Dynamics of Matter, Luruper Chaussee 149, D-22761 Hamburg, Germany}

\author{I-Te Lu}
\affiliation{Max Planck Institute for the Structure and Dynamics of Matter, Luruper Chaussee 149, D-22761 Hamburg, Germany}
\affiliation{National Yang Ming Chiao Tung University, Hsinchu 300093, Taiwan}

\author{Jehferson Mello}
\affiliation{Max Planck Institute for the Structure and Dynamics of Matter, Luruper Chaussee 149, D-22761 Hamburg, Germany}

\author{Henri Menke}
\affiliation{Max Planck Computing and Data Facility, Gießenbachstraße 2, D-85748 Garching, Germany}

\author{Sumit Naskar}
\affiliation{Institute for Physical Chemistry, Lessingstr. 4, Friedrich Schiller University Jena, 07743, Jena, Germany}

\author{Zhengwei Nie}
\affiliation{Dipartimento di Fisica e Chimica-Emilio Segr\`e, Universit\`a degli Studi di Palermo, Via Archirafi 36, 90123 Palermo, Italy}

\author{Sebastian T. Ohlmann}
\affiliation{Max Planck Computing and Data Facility, Gießenbachstraße 2, D-85748 Garching, Germany}

\author{Micael J. T. Oliveira}
\affiliation{Australia's Climate Simulator (ACCESS-NRI), Australian National University, 5 Liversidge St., 2601 Canberra, Australia}

\author{Subhojit Pal}
\affiliation{Dipartimento di Fisica e Chimica-Emilio Segr\`e, Universit\`a degli Studi di Palermo, Via Archirafi 36, 90123 Palermo, Italy}
\affiliation{Centre of Excellence ENSEMBLE3 Sp. z o. o., Wolczynska Str. 133, 01-919, Warsaw, Poland}

\author{Jonas Reimann}
\affiliation{Max Planck Institute for the Structure and Dynamics of Matter, Luruper Chaussee 149, D-22761 Hamburg, Germany}
\affiliation{Center for Free-Electron Laser Science, Luruper Chaussee 149, D-22761 Hamburg, Germany}

\author{Rajarshi Sinha-Roy}
\affiliation{Université Lyon 1, CNRS, Institut Lumière Matière, UMR5306, F-69100, Villeurbanne, France}

\author{Jiuyu Sun}
\affiliation{Max Planck Institute for the Structure and Dynamics of Matter, Luruper Chaussee 149, D-22761 Hamburg, Germany}
\affiliation{MIIT Key Laboratory of Semiconductor Microstructure and Quantum Sensing, Nanjing University of Science and Technology, Nanjing 210094, China}


\author{Francesco Troisi}
\affiliation{Max Planck Institute for the Structure and Dynamics of Matter, Luruper Chaussee 149, D-22761 Hamburg, Germany}
\affiliation{Parity Quantum Computing Germany Gmbh, Schauenburgerstraße 6, 20095 Hamburg, Germany}

\author{Angel Rubio}
\email{angel.rubio@mpsd.mpg.de}
\affiliation{Max Planck Institute for the Structure and Dynamics of Matter, Luruper Chaussee 149, D-22761 Hamburg, Germany}
\affiliation{Nano-Bio Spectroscopy Group and ETSF, Universidad del Pa\'is Vasco, 20018 San Sebasti\'an, Spain}
\affiliation{Center for Computational Quantum Physics, Flatiron Institute, 162 5th Avenue, New York, NY 10010.}

\def\thefootnote{*}\footnotetext{These authors contributed equally to this work}\def\thefootnote{\arabic{footnote}}

\begin{abstract}
\octopus is an open-source package implementing time-dependent density functional theory in a real-space, real-time formulation. Here we report on the developments since our previous contribution to the Special Topic Issue on Electronic Structure Software [J. Chem. Phys. {\bf 152}, 124119 (2020)]. New exchange–correlation functionals extend the range of systems that can be treated, while advances in relativistic and all-electron methods open the door to core-electron spectroscopy. New capabilities for lattice dynamics — stress-tensor calculation, geometry optimization, support for machine-learned force fields, and phonon calculations — enable a fully ab initio description of light-induced lattice dynamics in real time. We also present developments that move beyond the dipole approximation, capturing light–matter interaction more faithfully. Improved methods for exciton dynamics complement a set of algorithmic advances — new propagation schemes, imaginary-time evolution, and a refactored grid handling more than a billion points — that together substantially improve performance and scalability. Combined with technical improvements such as AMD GPU support and a modernized build system, these developments consolidate \octopus as a versatile, high-performance platform for exploring electronic excitations, light–matter interaction, and nonequilibrium dynamics in molecules, nanostructures, and extended systems.
\end{abstract}


\maketitle 

\section{Introduction}

The study of electronic, optical, and magnetic properties of matter is a fast evolving field, at the interface between different fields of physics, chemistry, and biology. The last years have seen the swift development of novel research directions, which have triggered extensive experimental and theoretical developments. To name a few, one can refer to the development of intense laser physics in solids or the realization of cavity engineering of chemical reactions and materials properties. Simulation tools and numerical methods are also evolving fast, and modern electronic structure codes do not rely on the same algorithms as 20 years ago, reflecting the fast pace of the advances in hardware and computing architectures. Furthermore, the last few years have also seen the emergence of machine-learned universal force fields, which are now revolutionizing material science. The \octopus code has continuously been adapting to these emerging research directions, with novel theoretical and numerical developments. \\
This paper thus presents the recent developments of the \octopus project ~\cite{castro2004first, andrade2010linear,  MARQUES200360, castro_octopus:_2006, andrade_real-space_2015, tancogne2020octopus}, especially discussing the recently added features and changes that were not discussed in the last special Topic Issue on Electronic Structure Software in 2020 ~\cite{tancogne2020octopus}.
These features encompass many aspects of density functional theory (DFT) and time-dependent DFT (TDDFT) codes: novel functionals, new capabilities for novel applications (including core-level spectroscopy, lattice dynamics, light-matter interaction beyond the dipole approximation, exciton dynamics). We focus on code developments which are already available to users, and do not discuss the roadmap of the code.\\
Beyond the new theoretical developments, we present several implementations of algorithms that have drastically improved the robustness of the code, in particular our implementation of the Chebyshev filtering, for better ground-state convergence, and novel time propagators. Other noticeable improvements include a mixing preconditioner or the implementation of the tetrahedron method, to name a few. Further developments allow for instance the code to run on different architectures, or to work on much larger real-space grids beyond one billion grid points. More details are also available in the new tutorials, the continuously improving online manual and variable descriptions, that can be found on the \octopus webpage~\cite{website}.

The paper first present in Secs.~\ref{sec:functionals} - \ref{sec:dissipative} the main theoretical developments of the code, that are organized around novel applications in light-matter interaction and non-equilibrium dynamics in quantum materials.
We then present more algorithmic and technical developments, that have been instrumental to move the \octopus code toward novel architectures, more robust algorithms, and more reliable tests for correctness and performances. Finally, we draw our conclusions in the last section.

\section{Novel Functionals}
\label{sec:functionals}

\subsection{Functionals for non-collinear spin}
Non-collinear magnetism has always represented a challenge for spin DFT. One of the main reasons, beyond the intrinsic complexity due to increased spin degrees of freedom, is the limited range of available exchange-correlation functionals that are working with non-collinear spin. The rotation to the locally collinear magnetization for the local spin density approximation, as proposed by K\"ubler \textit{et al.}~\cite{kubler1988density} was already present in the \octopus code. However, we recently extended the capabilities of the code by including generalized gradient approximation (GGA) functionals for non-collinear spin, following the usual approach of promoting the existing collinear functionals doing a rotation in the local frame defined by the magnetization. Together with Hartree-Fock, this thus allows to properly treat GGA-hybrid functionals in \octopus. Beyond this, we also recently derived and implemented truly novel functionals for non-collinear spins, which include a mGGA exchange functional that follows the logic of the Becke-Roussel exchange mGGA, and a correlation mGGA functional based on the Colle-Salvetti functional~\cite{PhysRevB.107.165111}. We also extended the most common self-interaction correction scheme (SIC) to the case of non-collinear spins, namely the Perdew-Zunger SIC, and the averaged-density SIC (ADSIC)~\cite{10.1063/5.0179087}. Finally, we extended the Slater potential and the optimized effective potential scheme to also work for non-collinear spins~\cite{PhysRevB.107.165111} following the expression derived in Ref.~\cite{PhysRevB.98.035140}. Using these schemes, we can model more non-collinear systems.

\subsection{Force-balance equations}\label{sec:force-balance}

In TDDFT, the equation of motion of the electronic current plays a central role in the formal derivation of the theory~\cite{PhysRevLett.52.997,PhysRevLett.82.3863}. 
This equation of motion is often referred to as the local force-balance equation (FBE), by analogy with the local balance of forces in classical mechanics. The FBE is a particularly interesting as it provides an alternative way to generate functionals $v_{xc}[\rho]$ which are not based on the energy. Conventional energy-based functionals are inherently tailored to ground-state problems. In contrast, the FBE can be naturally applied to both in and out-of-equilibrium states. It thus provides a complementary perspective on the physical system.
\\
Recently, the FBE has been used to derive novel exchange-correlation functionals, which have been developed and tested within the \octopus code.
\par
To get a relation between the force density and the potential, we start with Hamiltonian
\begin{equation}
    \hat{H}_{\rm tot} = \sum_i \hat{H}_i = \hat{T} + \hat{V} + \hat{W} + \dots
.
\end{equation}
The resulting equation of motion for the electronic current density is given by the Heisenberg equation of motion
\begin{equation}
    \partial_t \mathbf{j}
    =
    -
    \sum_i
    \frac{i}{\hbar}
    \left[
    \mathbf{j}
    ,
    \hat{H}_i
    \right]
    +
    \left\langle
    \partial_t
    \hat{H}_i
    \right\rangle
.
\end{equation}
The time derivative of the current density is given by all local (electronic) forces.
Writing the equation of motion for the many-body system (with its many-body wavefunction $\Psi$) and the Kohn-Sham system (with its Slater determinant wavefunction $\Phi$), we arrive at
\begin{equation}\label{equ:FBE_EoM}
    \partial_t
    (
    \mathbf{j} [\Psi]
    -
    \mathbf{j} [\Phi]
    )
    =
    \rho \nabla v_{\rm Hxc}
    +
    \mathbf{F}_T [\Psi]
    -
    \mathbf{F}_T [\Phi]
    +
    \mathbf{F}_W[\Psi]
    +
    \dots
.
\end{equation}
Here, $F_T$ denote the force coming from the kinetic term in the Hamiltonian, and $F_W$ the one from the interaction term. The Hartree-exchange-correlation potential arises from the difference of the forces from the local potential in the many-body and Kohn-Sham systems.
To utilize this equation, we distinguish between the ground state and the time-dependent case.
\par
Ground state systems are described within density functional theory, where we follow the derivation of Tancogne-Dejean et al.~\cite{10.1063/5.0177346}.
The electronic Hamiltonian is given by $\hat{H}=\hat{T}+\hat{V}+\hat{W}$, whereas the auxiliary system has no interaction $\hat{W}$.
For the ground state solution, $\partial_t \mathbf{j} = 0$ as all time-derivatives vanish and the left-hand side of the equation of motion (\ref{equ:FBE_EoM}) vanishes.
Dividing the remaining right-hand side by $\rho$ results in the static force-balance equation that links the exchange-correlation potential to the exchange-correlation force density
\begin{equation}
    \nabla v_{\rm xc}
    =
    - \rho^{-1}
    \mathbf{F}_{\rm xc}
    := -\mathbf{f}_{\rm xc}
,
\end{equation}
with $\rho\nabla v_{\rm H} + \mathbf{F}_{\rm xc} = \mathbf{F}_T [\Psi] - \mathbf{F}_T [\Phi] + \mathbf{F}_W[\Psi]$.
The gradient of the exchange correlation potential balances out the exchange correlation force, showing that $\mathbf{f}_{\rm xc}$  is a conservative force.
The potential $v_{\rm xc}$ can therefore be found solving a Poisson equation
\begin{equation}\label{equ:FBE_GS}
    v_{\rm xc}^{\rm FBE} = \int \frac{\nabla' \cdot \mathbf{f}_{\rm xc} (\mathbf{r}')}{4\pi |\mathbf{r} - \mathbf{r}'|} \mathrm{d} \mathbf{r}'
.
\end{equation}
By approximation the force $\mathbf{f}_{\rm xc}$, an approximated potential can therefore be obtained.
This method was recently used to define an exchange potential by replacing the $\mathbf{f}_{\rm xc}$ by an exchange-only force, resulting in an exchange potential very similar to the optimized effective potential from exact exchange~\cite{10.1063/5.0177346}.
\par
In the time-dependent case, we cannot neglect the time variation of the total current.
Instead, one can employ the continuity equation to obtain the time variation of the longitudinal part of the current between the Kohn-Sham and the many-body system, or equally the second time-derivative of the density~\cite{PhysRevLett.82.3863}.
This gives $\partial_t^2 (\rho[\Phi] - \rho[\Psi]) = 0 = - \nabla \cdot \partial_t (\mathbf{j}[\Phi] - \mathbf{j}[\Psi])$,
because the densities of the physical $\Psi \mapsto \rho$ and the auxiliary system $\Phi \mapsto \rho$ are equal.
Rearranging the right-hand side, we get the defining equation for the potential
\begin{equation}\label{equ:FBE_TD}
    \nabla \cdot (\rho \nabla v_{\rm xc}^{FBE_{SL}} )
    =
    -
    \nabla \cdot \mathbf{F}_{\rm xc}
,
\end{equation}
referred to as the Sturm-Liouville equation in TDDFT~\cite{10.1063/1.5123608,marques2012fundamentals,maitra2016perspective}.
For a given density and force density approximation, the Sturm-Liouville equation can be solved.
As for the foundation of TDDFT, we assume here the uniqueness and existence of the solution.
\par
In the \octopus code, derivatives are approximated by high-order finite differences.
This allows to rewrite the Sturm-Liouville problem as a discrete matrix problem that is solved iteratively as a linear problem
\begin{equation}
    \underbrace{\nabla \cdot \rho \nabla}_{:=A}
    \underbrace{v}_{:=x}
    =
    \underbrace{-\nabla \cdot \mathbf{F}}_{:=B}
.
\end{equation}
For a given force density $\mathbf{F}$, the vector expression $B$ is simply evaluated.
The Sturm-Liouville operator $A$, on the other hand, is a density weighted differential operator on the real space grid.
It is straightforward to show that the Sturm-Liouville operator is a self-adjoint operator. However, the above formulation does not guarantee that property with a succession of skew-adjoint finite-difference operators. We found that the numerical stability of the Sturm-Liouville solver is strongly improved by rewriting the linear solver as a sum of self-adjoint operators as
\begin{equation}
    \nabla \cdot 
    (\rho \nabla  x)
    =
    \frac{1}{2}
    \left(
      \nabla^2 (\rho x)
      -
      x \nabla^2 \rho
      +
      \rho \nabla^2 x
    \right)
.
\end{equation}
We note in passing that we employed the same strategy for making the nonlocal term of mGGAs more stable too.\\
The initial guess of the iterative procedure is the Poisson solution for the force density $\mathbf{f}_{\rm xc}$, and we employed the preconditioned quasi-minimal residual solver~\cite{CHEN2006198} for solving the linear problem.
We found that a simple Jacobi preconditioner (using $\nabla\cdot[ \rho \nabla x] \approx \rho \nabla^2 x$ and replacing the finite-difference Laplacian by the central weight only) already improves the convergence without increasing the numerical cost of a step of the solver.
We expect only short-range deviation from the Poisson solution, i.e. contributions with a decay faster than $-1/r$ in the potential.
For conservative forces, like in spherical symmetric systems, the Poisson and the Sturm-Liouville solution give analytically and numerically the same result.
\par
To define an approximated potential by solving either force balance equation, equation (\ref{equ:FBE_GS}), or the Sturm-Liouville equation, equation (\ref{equ:FBE_TD}), approximated forces are needed.
To describe an exchange contribution, the force density induced by the exchange of the interacting Kohn-Sham orbitals is implemented in the \octopus code~\cite{10.1063/5.0177346}.
In the same spirit, a correlation functional was derived, following the Colle-Salvetti approach~\cite{tancogne2025local}, allowing to get an approximation of correlation contribution to the interaction force.

\subsection{QEDFT Functional Developments}
\label{sec:qedft}

TDDFT treats the electromagnetic field as a classical external perturbation and, even when coupled self-consistently to Maxwell's equations, cannot describe electromagnetic vacuum fluctuations.
Quantized photon modes retain zero-point fluctuations in the field-free limit, enabling cavity-induced modifications of ground-state properties without external driving~\cite{Ruggenthaler2018,Ruggenthaler2023,Lu2025AOP,GarciaVidal2021,Bloch2022,Schlawin2022,Basov2025,Bretscher2026}. 
QEDFT incorporates these effects by treating photons and electrons on equal footing~\cite{Tokatly2013,Ruggenthaler2014}.
Within the long-wavelength approximation and velocity gauge, the Pauli--Fierz Hamiltonian is
\begin{align}
\hat{H}_{\rm{PF}} = \sum_{i=1}^{N_{e}}\frac{1}{2}\left(-i\bm{\nabla}_{i}
+ \frac{1}{c}\hat{\mvec{A}}\right)^{2}
+ \hat{V}_{\mathrm{ext}} + \hat{W}
+ \sum_{\alpha=1}^{M_{p}}\omega_{\alpha}
\left(\hat{a}^{\dagger}_{\alpha}\hat{a}_{\alpha}+\tfrac{1}{2}\right),
\label{eq:pauli-fierz}
\end{align}
with
\begin{align}
\hat{\mvec{A}}
=c\sum_{\alpha=1}^{M_{p}}
\frac{\lambda_{\alpha}}{\sqrt{2\omega_{\alpha}}}
\mvec{\varepsilon}_{\alpha}
(\hat{a}_{\alpha}+\hat{a}^{\dagger}_{\alpha})
\equiv c\sum_{\alpha=1}^{M_{p}}
\lambda_{\alpha}\mvec{\varepsilon}_{\alpha}\hat{q}_{\alpha}.
\label{eq:mode-expansion}
\end{align}
Here $\lambda_{\alpha}\propto V_{\mathrm{eff},\alpha}^{-1/2}$ is determined by the effective mode volume $V_{\mathrm{eff},\alpha}$. 
The diamagnetic $\hat{\mvec{A}}^2$ term can be absorbed into dressed photon modes with renormalized frequency $\tilde{\omega}_{\alpha}$ and coupling $\tilde{\lambda}_{\alpha}$~\cite{Rokaj2022,PhysRevA.109.052823}; for a single polarization,
$\tilde{\omega}_{\alpha}^{2}=\omega_{\alpha}^{2}+N_e\lambda_{\alpha}^{2}$.
Thus, the modes entering practical QEDFT calculations are the dressed modes.

The QEDFT mapping replaces the interacting electron-photon problem by an auxiliary Maxwell--Kohn--Sham system,
\begin{align}
  i\partial_{t}\varphi_{j}(\mvec{r},t)
    &= \left[\frac{1}{2}\left(-i\bm{\nabla}+\frac{1}{c}\mvec{A}(t)\right)^{2}
       + v_{\mathrm{ext}}(\mvec{r},t)
       + v_{\mathrm{Hxc}}(\mvec{r},t) + v_{\mathrm{pxc}}(\mvec{r},t)\right]
       \varphi_{j}(\mvec{r},t),
       \label{eq:mks-electrons}\\[4pt]
  \left(\partial_{t}^{2}+\tilde{\omega}_{\alpha}^{2}\right) q_{\alpha}(t)
    &= -\,\lambda_{\alpha}\,\mvec{\varepsilon}_{\alpha}\cdot\mvec{J}_{\mathrm{p}}(t)
       + j^{\mathrm{ext}}_{\alpha}(t),
       \label{eq:mks-photons}
\end{align}
where $\mvec{J}_{\mathrm{p}}$ is the paramagnetic current and
$\mvec{A}(t)=c\sum_{\alpha}\lambda_{\alpha}\mvec{\varepsilon}_{\alpha}q_{\alpha}(t)$.
The electron-photon exchange-correlation potential $v_{\mathrm{pxc}}$ is the essential ingredient absent from ordinary TDDFT.
\octopus has supported QEDFT since the electron-photon optimized effective potential (OEP) formulation~\cite{Pellegrini2015,Flick2018OEP}, including real-time propagation, ground-state photon OEP, and linear-response QEDFT in both Casida-like~\cite{Flick2019} and Sternheimer~\cite{Welakuh2022} formulations. 
The photon modes are specified through their frequencies, polarizations, and effective volumes; the latter must be obtained from a separate electromagnetic calculation or experiment and are therefore inputs rather than predictions of the QEDFT calculation.

\begin{figure}[t]
  \centering
  \includegraphics[width=\linewidth]{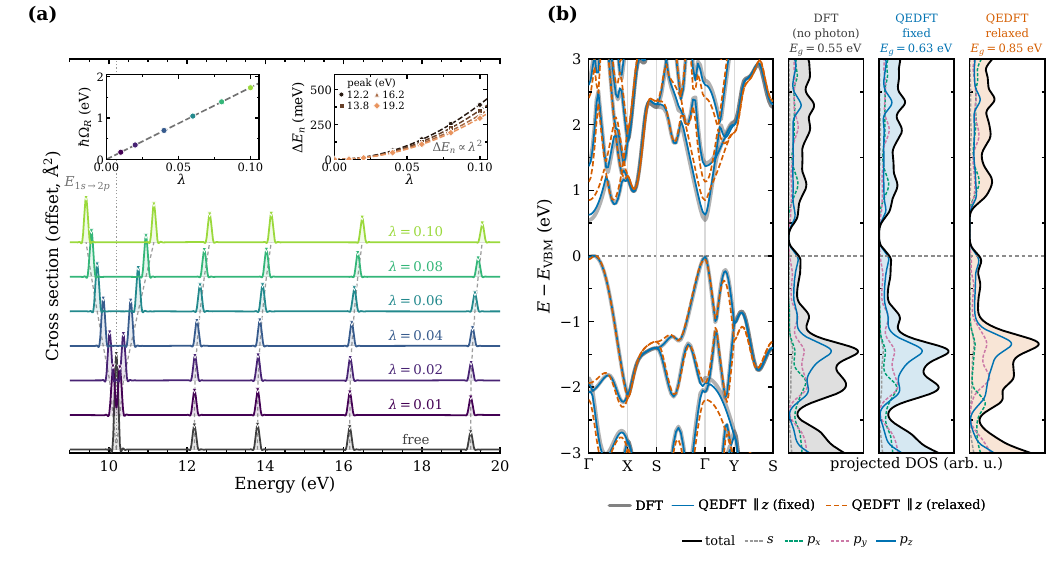}
  \caption{(a) Absorption cross section of a hydrogen atom outside the cavity (grey) and inside a
  cavity mode with the polarization along the $x$ direction and resonant with the $1s\rightarrow 2p_{x}$ transition at $10.19$~eV, obtained using the pxLDA functional, see Eq.~\eqref{eq:px-poisson}. The coupling strengths is $\lambda=0.01$--$0.10$; curves are offset vertically and markers track each peak, with dashed
  guides connecting them across $\lambda$. 
  Left inset: the splitting is linear in the coupling, $\hbar\Omega_{R}\propto\lambda$. 
  Right inset: the
  off-resonant transitions at $12.18$, $13.80$, $16.16$ and $19.25$~eV acquire no partner and
  instead blue-shift quadratically, $\Delta E_{n}\propto\lambda^{2}$, by $0.30$--$0.39$~eV at
  $\lambda=0.10$.
  (b) Cavity-mediated electronic structure of monolayer black phosphorus obtained using DFT (grey), QEDFT
  with a single out-of-plane mode with the cavity frequency $\omega=0.1$~eV, $\lambda/\omega=0.5$, and the polarization $\mvec{\varepsilon}$ along the $z$ direction at the DFT geometry (blue), and with the atoms relaxed inside the
  cavity (orange, dashed). The right panels show the total and orbital-resolved density of states. 
  %
  }
  \label{fig:qedft}
\end{figure}

Recent developments instead construct the electron-photon functional from the force-balance equation~\cite{10.1063/1.5123608,PhysRevA.109.052823}. 
The photon force contains a correlator between fluctuations of the vector potential and paramagnetic current. 
Since the photon equation is linear, the field fluctuations can be adiabatically eliminated in favor of the current fluctuations~\cite{Schafer2021},
\begin{align}
\delta\hat{\mvec{A}}
\longrightarrow
-c\sum_{\alpha=1}^{M_p}
\frac{\lambda_\alpha^2}{\tilde{\omega}_\alpha^2}
\mvec{\varepsilon}_\alpha
\left(\mvec{\varepsilon}_\alpha\cdot
\delta\hat{\mvec{J}}_{\mathrm{p}}\right).
\label{eq:slaving}
\end{align}
The resulting photon-free formulation retains the quantum fluctuations through a current-current correlator weighted by $\lambda_\alpha^2/\tilde{\omega}_\alpha^2$, while eliminating the photon Fock space. 
It therefore has the computational cost of a conventional DFT calculation and, being formulated in the velocity gauge, is applicable to periodic and extended systems.

Keeping the leading exchange contribution gives the electron-photon exchange potential $v_{\mathrm{px}}$. 
In the present \octopus implementation, it is obtained from the following force-balance Poisson equation.
Evaluating the current-current correlator in the homogeneous limit yields the local-density approximation
\begin{equation}
\nabla^{2} v^{\mathrm{LDA}}_{\mathrm{px}}(\mvec{r})
= -\sum_{\alpha=1}^{M_{p}}
\frac{2\pi^{2}\tilde{\lambda}_{\alpha}^{2}}
{\tilde{\omega}_{\alpha}^{2}}
\left(\mvec{\varepsilon}_{\alpha}\cdot\bm{\nabla}\right)^{2}
\left[\frac{3n(\mvec{r})}{8\pi}\right]^{2/3},
\label{eq:px-poisson}
\end{equation}
defining the electron-photon exchange local-density approximation (pxLDA)~\cite{PhysRevA.109.052823}.
The corresponding exchange energy can be evaluated separately, although its consistency with the potential is presently guaranteed only for isotropic photonic environments~\cite{PhysRevA.109.052823}. 
An energy functional consistent with the general Sturm--Liouville force-balance potential is under development.

Figure~\ref{fig:qedft} illustrates the range of systems accessible with pxLDA. Panel (a) shows a hydrogen atom resonantly coupled to a cavity mode. 
The $1s\rightarrow2p_x$ transition splits into two polariton branches with a Rabi splitting $\hbar\Omega_R\propto\lambda$, while off-resonant transitions undergo quadratic blue shifts. 
These effects are obtained from an ordinary real-time propagation without an explicit photon Fock space.
Panel (b) shows monolayer black phosphorus coupled to a low-frequency out-of-plane mode. 
Here the cavity is off-resonant with electronic transitions and modifies the ground-state density and band structure rather than producing a polariton. 
The band gap increases from $0.55$ to $0.63$~eV at fixed geometry and to $0.85$~eV after structural relaxation, demonstrating that cavity-induced structural changes can dominate the electronic contribution. 
The projected density of states further identifies the $p_z$ states as the orbitals most directly affected by the out-of-plane mode.

Further developments include a consistent energy functional for the Sturm--Liouville formulation, extension from discrete effective modes to a continuum for Lamb shifts, lossy cavities, and spontaneous emission~\cite{Kudlis2022}, relativistic linear-response QEDFT for heavy-element systems~\cite{Konecny2025}, and coupling to nuclear motion~\cite{FlickNarang2018}.
Together with the semiclassical Maxwell--TDDFT framework of Sec.~\ref{sec:lightmatter-leaving-dipole}, these developments extend \octopus from quantized cavity modes to fully structured electromagnetic environments.

\section{Core Electron Spectroscopy}\label{sec:core-level}

\subsection{Relativistic electronic structure}

While the description of electronic structure based on the non-relativistic Schr\"{o}dinger equation is sufficient for light
elements and processes involving the valence electrons, for accurate description of heavy elements and core electrons,
relativistic treatment of electronic structure based on the Dirac equation becomes necessary.~\cite{Pyykko1979,Pyykko1988,Pyykko2012}
The most accurate relativistic description is provided by the four-component Dirac--Coulomb(--Breit) Hamiltonian.~\cite{Dirac1928I, Dirac1928II, Breit1929, Gaunt1929, Dyall2007}
However, practical calculations using such a Hamiltonian may become computationally expensive for realistic molecular
or solid state systems. Therefore, approximate relativistic methods based on two-component Hamiltonians are commonly
developed and implemented.~\cite{Saue2011}
Two-component methods aim to decouple the four-component Dirac Hamiltonian either at the level of operators or their matrix representations.
Popular and historically important two-component Hamiltonians include
the second-order Douglas--Kroll--Hess Hamiltonian,~\cite{Douglas1974,Hess1985,Hess1996,Wolf2002}
the zeroth-order regular approximation (ZORA) Hamiltonian,~\cite{Chang1986,Lenthe1993}
and the recently popular exact two-component Hamiltonians~\cite{Kutzelnigg2005, Ilias2007, Knecht2022}.
For the implementation in the \octopus program, we chose the ZORA method
due to its implementational simplicity in the real-space formalism
(not requiring the construction and storage of any
four-component objects or large matrices with the dimension of the real-space grid, nor performing any non-local transformations)
combined with satisfactory performance for many molecular and material properties particularly in terms of predicting trends and providing qualitative interpretation.~\cite{Lenthe1996, Lenthe1999, Fronzoni2005, Mori2014, Ludowieg2022}

The ZORA method is based on eliminating the small component from the four-component one-electron Dirac equation (here written with the rest energy shifted by $mc^2$)
\begin{equation}
\begin{pmatrix}
V & c \boldsymbol{\sigma}\cdot\boldsymbol{p} \\
c \boldsymbol{\sigma}\cdot\boldsymbol{p} & V-2mc^2
\end{pmatrix}
\begin{pmatrix}
\Phi(\boldsymbol{r}) \\
\chi(\boldsymbol{r})
\end{pmatrix}
=
E
\begin{pmatrix}
\Phi(\boldsymbol{r}) \\
\chi(\boldsymbol{r})
\end{pmatrix},
\end{equation}
where $V$ is the scalar one-electron potential, $m$ electron mass, $c$ the speed of light,
$\boldsymbol{\sigma}$ is the vector of Pauli matrices, $\boldsymbol{p}$ is the momentum operator, and $E$ is the electronic energy.
The electron's wave function is composed of the large component $\Phi(\boldsymbol{r})$ and
the small component $\chi(\boldsymbol{r})$, both of which are two-component spinors.
Exact elimination of the small component gives the two-component equation with energy-dependent Hamiltonian
\begin{equation}
\left[
V+
\boldsymbol{\sigma}\cdot\boldsymbol{p}
\frac{c^2}{2mc^2+E-V}
\boldsymbol{\sigma}\cdot\boldsymbol{p}
\right]\Phi
=
E\Phi .
\end{equation}
In the ZORA, the denominator is expanded using the regular expansion as
\begin{equation}
\frac{1}{2mc^2+E-V}
=
\frac{1}{2mc^2-V}
\left(
1+\frac{E}{2mc^2-V}
\right)^{-1},
\end{equation}
and only the zeroth-order term in $E/(2mc^2-V)$ is retained. 
This yields the energy-independent ZORA equation
\begin{equation}
\label{eq:ZORA}
\left[
V+
\boldsymbol{\sigma}\cdot\boldsymbol{p}
\frac{c^2}{2mc^2-V}
\boldsymbol{\sigma}\cdot\boldsymbol{p}
\right]\Phi^\mathrm{ZORA}
=
E^\mathrm{ZORA}\Phi^\mathrm{ZORA},
\end{equation}
that is the final equation to be solved for molecular orbitals.
In Kohn-Sham DFT, the ZORA Kohn-Sham Hamiltonian reads as
\begin{equation}
\label{eq:ZORAeq-2terms}
H^\mathrm{KS-ZORA}
=
V
+
\boldsymbol{p}
\frac{c^2}{2mc^2-V}
\boldsymbol{p}
+
\frac{c^2}{(2mc^2-V)^2}
(\boldsymbol{\sigma} \times \nabla V ) \cdot \boldsymbol{p},
\end{equation}
where one can identify scalar-relativistic and spin-orbit terms.
The solution of this equation forms the basis for two-component relativistic electronic
structure calculations in \octopus. Moreover, a scalar relativistic version of ZORA obtained
by neglecting the last term in Eq.~\eqref{eq:ZORAeq-2terms} is available as well for cases
when only the scalar relativistic effects come into play.\\
One advantage of the ZORA implementation in real-space is that the reference energy is fixed by the vacuum level, making this approach free of gauge invariance problem.


In order to deal with all-electron calculations, we also implemented the recently developed analytic, norm-conserving, regularized Coulomb (ANC) potential of Gigy~\cite{gygi2023all}. This allows to run all-electron calculations and complement the previously implemented methods of $\delta$-charge, which requires the atom to be on a grid point, or Gaussian charge~\cite{modine1997adaptive}. Together with the improvements of the calculation of the weight of the Laplacian, see Sec.~\ref{sec:improved_stencil}, it implies that \octopus can perform all-electron calculation on curvilinear meshes, including scalar and fully-relativistic effects.\\

\section{Towards full description of the lattice dynamics}

The excitation of a solid by an ultrashort laser pulse can transiently or permanently drive it
into a different phase, changing its electronic, structural, or magnetic order on femtosecond to
picosecond timescales. Such photoinduced phase transitions are a central topic of
ultrafast materials science, both because they give access to transformation pathways that are
not accessible thermally, and because the possibility of switching a material between phases
with light underlies proposed ultrafast electronic and photonic devices. 
Being able to describe such light-induced phase transition requires not only to evolve the electronic and ionic degrees of freedom, which can be done in the simplest level by TDDFT+Ehrenfest dynamics~\cite{andrade2009modified}, but also the evolution of the lattice itself. 
While the quantum nature of the nuclei is neglected by this standard TDDFT+Ehrenfest dynamics, there are various suggestions to recover them by multi-trajectory Ehrenfest approaches and other approaches \cite{ten2022real,albareda2021conditional,lively2021simulating,PhysRevLett.130.166401}. In the simplest approach, we run independent trajectories, which are initialized by sampling the Wigner distribution~\cite{LivelyEtAl2021, LivelyEtAl2024}, which requires knowledge of the phonon eigenmodes.

Over the last few years, we have implemented important pieces required to reach this goal. This first includes accurate computation of stress tensor, implementation of geometry optimization algorithms capable of relaxing the cell, as well as several other methods that we describe in this section.

\subsection{Stress tensor and geometry optimization}


We consider the Parrinello-Rahman Lagrangian describing electrons in their ground state under the constraint of an external pressure $p$, with moving nuclei, and a moving lattice
\begin{equation}
 \mathcal{L} = E_{\rm KS} + \sum_{I}\frac{1}{2}M_I \dot{\mathbf{R}}^2_I - \sum_{mn} \sum_{\mathbf{k}}^{BZ} \lambda_{mn}(\mathbf{k}) (\langle m\mathbf{k}|n\mathbf{k}\rangle - \delta_{mn}) + \frac{1}{2}W\mathrm{Tr}\{\mathbf{\dot{h}}^T\mathbf{\dot{h}}\} -pV\,,
\end{equation}
where $\mathbf{h}=\{\mathbf{a},\mathbf{b},\mathbf{c}\}$, i.e. the matrix made of the lattice vectors, $V=\mathrm{det}(\mathbf{h})$ is the cell volume, $W$ is the fictitious lattice mass, and $E_{\rm KS}$ is the Kohn-Sham energy given by
\begin{equation}
 E_{\rm KS} = E_{\rm II} + E_{\rm ext} + F[\rho]\,,
\end{equation}
where $E_{\rm II}$ is the direct Coulomb interaction energy between the ions, $E_{\rm ext}$ is the external potential energy, and $F[\rho]$ is the universal functional $F[\rho] = \sum_i \langle \psi_i | \frac{\mathbf{p}_i^2}{2m_i}|\psi_i\rangle + E_{\rm H} +  E_{\rm xc}[\rho]$, containing the kinetic energy of the Kohn-Sham system, the Hartree, and the exchange-correlation functional.\\
This Lagrangian describes the ground-state of atoms, molecules and solids, as well as Born-Oppenheimer molecular dynamics, which are both implemented in \octopus. It this Lagrangian formulation, one can extend the approach to the real-time simulation of light-driven materials, and combined with the Maxwell equation, which is why we formulate the stress tensor in terms of the Lagrangian of the light-matter system.\\
The stress tensor can be computed from the derivative of the total energy with respect to an infinitesimal deformation $\epsilon$ of the cell such that $\mathbf{x}\to \mathbf{\tilde{x}}=\mathbf{\epsilon}\mathbf{x}$, $\mathbf{R}\to\mathbf{\tilde{R}}=\mathbf{\epsilon}\mathbf{R}$, $\mathbf{k}\to\mathbf{\tilde{k}}=\mathbf{\epsilon}^{-T}\mathbf{k}$, leading to a Lagrangian $\tilde{\mathcal{L}}$ after deformation. \\
The stress tensor is finally given by 
\begin{equation}
 \sigma_{\alpha\beta} = \frac{1}{V}\frac{\partial \tilde{\mathcal{L}}}{\partial \epsilon_{\alpha\beta}}\Big|_{\mathbf{\epsilon=I}}\,.
\end{equation}

From this expression, we recover the usual expressions of the literature. Importantly, to maintain the scalability of the real-space approach, we do not use expressions in momentum space, but instead derived expressions that are computed directly in real space, similar to the approach of Ref.~\cite{sharma2018calculation}. This applies in particular to the kinetic stress tensor and the Hartree stress tensor. The code was validated against plane-wave implementations in Abinit and Quantum ESPRESSO, and convergence can be checked by the fulfillment of the kinetic and Hartree pressure sum rules: $-3 P_{\rm kin} V = 2 E_{\rm kin}$ and $-3P_{\rm Hartree} V = E_H$. \octopus supports LDA, GGA, mGGA, and hybrid functional stress tensors.
Given the implementation of \octopus for the vector potential, the kinetic stress tensor for the electrons is directly gauge invariant, meaning that it includes the vector potential properly.
Finally, we also implemented the ion kinetic stress tensor.

Based on the stress tensor, we extended the geometry optimization capabilities of \octopus to support volume optimization and shape optimization. In both cases, we regenerate the real-space grid by keeping the same number of grid points, only adjusting their Cartesian coordinates, while keeping the reduced coordinates fixed. This approach works well close to equilibrium, but if one strongly quench or extend one of several directions during the optimization, the results can be biased. In order to get numerically-stable results, we implemented a more robust computation of the weights of the finite-difference operators, see below.\\

Using the Euler-Lagrange equation, one obtains from the Lagrangian following the equations of motion, including the one for the lattice
\begin{equation}
 \ddot{h}_{\alpha\beta} = \frac{1}{W}(\sigma_{\alpha\gamma} - p\delta_{\alpha\gamma})V h^{-T}_{\gamma \beta}\,.
\end{equation}
To improve numerical stability, we found it convenient to employ the right polar decomposition $h = RU$, where $U=\sqrt{h^Th}$ and R is a SO(3) rotation matrix. $U=U^T$ is positive definite. With this, we can eliminate global rotations of the cell analytically, and get the modified equation of motion
\begin{equation}
 \ddot{U} = \frac{1}{W}(R^T\sigma R - pI)V U^{-1}\,.
\end{equation}
For simulations in which k-point symmetries are used, we always symmetrize the stress tensor using both symmorphic and non-symmorphic symmetries, to prevent numerical loss of symmetries.

\subsection{Machine-learned universal force field}

Machine-learned interatomic potential is a very growing field in material science~\cite{batatia2022mace,yu2024systematic,wines2025chips}, and the recent generations of universal force fields now reach close-to-DFT-codes accuracies, at order-of-magnitude reduced numerical cost. It is therefore quite interesting to implement support of these  new numerical tools. In \octopus, we decided to implement these universal force fields for two applications: one is to serve as a pre-optimization tool. If the feature is activated, we first run the universal force field together with ASE~\cite{hjorth2017atomic}, to obtain a relaxed geometry from the force field, and then run DFT on top of this. This approach allows to obtain very quickly a good guess of the geometry, only leaving few steps for the DFT-based geometry optimizer to run. The second approach, that we found useful, consists of using the universal force field as a calculator that replaces DFT, providing the energy, force, and stress. This can thus serve as a debug tool where the geometry optimization algorithm of \octopus are run using the input from the model. \\
We demonstrated this approach using Mattersim~\cite{yang2024mattersim}, which we interfaced with \octopus using ASE, but the implementation is designed to be capable of working with any other models in the future, allowing in particular to retrain some models and use them for learning excited-state forces obtained from real-time TDDFT.

\subsection{Robust finite-difference operators}
\label{sec:improved_stencil}

To enable lattice dynamics, the finite-difference (FD) representation of the differential operators, mainly the Laplacian, needs to be robust across different lattice types because space-group and symmetry may change in the course of a simulation.

We employ the following generic procedure to obtain coefficients for FD
stencils for gradients and the Laplacian, although the method can be
easily extended to other differential operators. We require the action of the
differential operator to be exact on a set of monomials evaluated on the points
of the stencil which corresponds to a polynomial fit of the corresponding
order.

For a generic coordinate transformation $\xi^i = \phi({x^i})$,
the Jacobian is defined as
\begin{equation}
  \frac{\partial \xi^i}{\partial x^j} = \left( J^{-1} \right)^i_j \,,\nonumber
\end{equation}
and the metric tensor as
\begin{equation}
  g^{ij} = \left( J^{-1}\left( J^{-1} \right)^T \right)^{ij}\,. \nonumber
\end{equation}
The gradient can be written as
\begin{equation}
\left( \nabla f \right)_j = \frac{\partial f}{\partial x^j} = \frac{\partial \xi^i}{\partial x^j} \frac{\partial f}{\partial \xi^i}\,,\nonumber
\end{equation}
and the Laplacian as
\begin{equation}
  \Delta f = \frac{1}{\sqrt{\det(g)}} \frac{\partial}{\partial \xi^i} \left( \sqrt{\det(g)} g^{ij} \frac{\partial f}{\partial \xi^j} \right)
  = g^{ij} \frac{\partial^2 f}{\partial \xi_i \partial \xi_j} + \frac{\partial^2 \xi^k}{\partial x^i \partial x^i} \frac{\partial f}{\partial \xi_k}\,,\nonumber
\end{equation}
where the coefficient of the second term corresponds to the trace of the Hessian.\\
For affine transformations, the metric is constant and the second term is zero. For Cartesian coordinates, we also have $g^{ij}=\delta^{ij}$.\\
The finite-difference representation of an operator $D$ in $d$ dimensions can be written as
\begin{equation}
  D f (\mathbf{\xi}) = \sum_{i=1}^n c_i f(\mathbf{\xi}_i)\,,\nonumber
\end{equation}
where $n$ is the stencil size and $c_i$ is the coefficient corresponding to the function value at the grid
point $\mathbf{\xi}_i$. Now this is evaluated on a set of monomials
\begin{equation}
  f_j(\mathbf{\xi}) = \prod_{k=1}^d (\xi_k - \xi_{p,k})^{\alpha_{j,k}}\,,\nonumber
\end{equation}
where $\mathbf{\xi}_p$ is the central point of the stencil (0 for translationally invariant grids),
and $\alpha_{j,k}$ is the exponent.
Now we require the evaluation of the operator on these monomials to be exact,
\begin{equation}
  D f_j(\mathbf{\xi}) = d_j, \ j = 1, \dots, m\,,\nonumber
\end{equation}
where the right-hand side can be evaluated analytically. This corresponds to the matrix
equation
\begin{equation}
  f_j(\mathbf{\xi}_i) c_i = d_j\,,\nonumber
\end{equation}
where the summation over $i$ is implied. We choose the number of monomials $m$ to be
equal to the number of stencil points $n$ and solve the corresponding linear system. \\
For the gradient, the right-hand side is the action of the gradient on
the monomials, evaluated at 0, which yields
\begin{equation}
  d_j = 
  \begin{cases}
    1, & \alpha_{j,i} = 1, \alpha_{j,k} = 0\ \text{for}\ k\ne i \\
    0, & \text{else}
  \end{cases}
\end{equation}
and we always compute the gradient in reduced or curvilinear coordinates
with a stencil along the axis and transform it with the Jacobian to the
Cartesian frame.\\
For the Laplacian, the right-hand side is the action of the Laplacian
on the monomials, evaluated at 0, which yields
\begin{equation}
  d_j =
  \begin{cases}
    2 g^{mn}, & \sum_{k=1}^d \alpha_{j,k} = 2,\ m,n\ \text{non-zero indices} \\
    \frac{\partial^2 \xi^m}{\partial x^n \partial x^n}, & \sum_{i=1}^d \alpha_{j,i} = 1,\ m\ \text{non-zero index}  \\
    0, & \text{else}
  \end{cases},
\end{equation}
where the summation over $n$ is implied in the second line.
The first line corresponds to monomials for which the sum of the exponents is 2,
e.g., $x^2$ or $xy$, and the second line corresponds to monomials for which
only one term has a non-zero exponent, e.g. $x$. The 
term in the second is only non-zero if the metric varies in space, i.e., for generic
curvilinear coordinates. For the special case of Cartesian coordinates,
the second line vanishes and the first is simply $2\delta^{mn}$.

For orthogonal grids, we choose a star-like stencil by default with four points
in each direction from the center, i.e., 27 points in 3d. For non-orthogonal
grids, we adapt the method of \citet{natan2008} by adding
additional axes to the stencil as needed for non-orthogonal directions. For
computing the weights, we then use the same methods as outlined above.
When choosing the additional axes, they must not lie in the same plane, otherwise
the coefficient matrix is degenerate, which can happen for certain triclinic
cells when selecting the closest axes naively. We make sure to avoid this
condition such that the coefficient matrix is always well-behaved. For generic curvilinear
grids, we enlarge the star stencil to make sure cross-derivatives are well
represented.

By default, we compute the coefficient matrix and the right-hand side in
reduced coordinates, except for curvilinear coordinates, for which we use
Cartesian coordinates because that usually results in lower errors. The
method can be selected in the input file with the boolean variable
\texttt{StencilPrimitiveCoordinates}. We have validated the implementation by
ensuring that the error of the FD computation of the
Laplacian and the gradient of some analytically known functions are small enough
on a wide range of grids, including all 14 Bravais lattices.
The scheme as presented here allows us to compute FD coefficients for arbitrary stencil sizes, lattice types and grid geometries in a generic way with a single algorithm.

\subsection{Multi-trajectory Ehrenfest dynamics}

Multi-trajectory Ehrenfest (MTEF) dynamics is an approximate approach to the fully quantum mechanical dynamics of the coupled electron-nuclei system. It is based on mixed quantum-classical Liouville equation, describing the approximate evolution of the density matrix. To this end, one uses the 
partial Wigner transform of an operator $\hat{O}$:
\begin{equation}
    \hat{O}_{\rm W}({\bf R},{\bf P}) = {(2 \pi)}^{-d N} \int d{\bf Q} e^{-i {\bf P Q}} 
    \langle {\bf R} + {\bf Q}/2 | \hat{O} | {\bf R} - {\bf Q}/2 \rangle\,,
\end{equation}
where $\mathbf{R}$ and $\mathbf{P}$ are the positions and momenta of the nuclei.

In the mean-field limit, the density operator can be factorized as
\begin{equation}
  \hat{\rho}_{\rm W} = \rho_{\rm n, W}({\bf R}_{i}, {\bf P}_i, t) \hat{\rho}_e(t)\,,
\end{equation}
and the density operator of the nuclei can be expressed in terms of an average over trajectories:
\begin{equation}
  \rho_{\rm n, W}({\bf R}, {\bf P}, t)  = \sum_i w_i \delta({\bf R} - {\bf R}_i(t)) \delta({\bf P} - {\bf P}_i(t))\,.
\end{equation}
The electronic density operator, and the phase-space coordinates of each trajectory follow the Ehrenfest dynamics:
\begin{eqnarray}
    \partial_t \hat{\rho}_e(t) &=& -i [ \hat{H}_{\rm W}({\bf R}_i, {\bf P}_i), \hat{\rho}_e(t) ] \,,\\
    \dot{\bf P}_i &=& - {\rm Tr}[ \hat{\rho}_e(t) \grad_{\bf R} {\hat{H}_{\rm W}({\bf R}, {\bf P})}|_{({\bf R}_i(t), {\bf P})_i(t))} ] \,,\\
    \dot{\bf R}_i &=& {\bf P}_i/M_i\,.
\end{eqnarray}

Observables can then be constructed from an ensemble average over trajectories
\begin{equation}
    \langle \hat{O}(t) \rangle = 
    \sum_i {\rm Tr} [ \hat{O}_{\rm W}({\bf R}_i(t), {\bf P}_i(t)) \hat{\rho_{\rm n, W}}({\bf R}_i(0), {\bf P}_i(0))) \hat{\rho_{e}(t)} ]\,.
\end{equation}

When transforming to phonon reduced coordinates (for details see, e.g., Ref.~\cite{LivelyEtAl2024}) the probabilities for a given ionic displacements and initial velocities are given as functions of the phonon frequencies and the phonon eigenvectors.
 

\octopus has a multi-system framework \cite{tancogne2020octopus}, which allows the time propagation of interacting physical systems. This is used, for instance, for coupled Maxwell-matter systems. This framework also builds the foundation of ensemble calculations, where each replica is one such system. Dedicated routines exist to create a number of identical systems with different initial conditions, and in particular, initial conditions can be drawn by sampling the Wigner distribution. This requires the phonon eigenmodes of the system to be known. This information can be generated by connecting \octopus, using Phonopy \cite{phonopy-phono3py-JPCM, phonopy-phono3py-JPSJ} for solids, or directly by \octopus for molecules. The Phonopy interface is outlined in the next section.


Currently, the implementation allows for non-interacting replicas, i.e. the mean-field MTEF, outlined above. The next steps will be the implementation of interacting replicas, including interactions of each replica with the ensemble average (centroid). This will then open the doors towards methods such as ring-polymer dynamics or methods such as suggested by Zhao et al. \cite{PhysRevLett.130.166401}.

\subsection{Phonon Calculations with \octopus and \phonopy}

For many applications, such as e.g. multi-trajectory calculations, as described in the next section, one needs information about the phonon modes of the system. While \octopus has a built-in capability to calculate vibrational modes of molecules, for periodic solids we decided to resort to the well established \phonopy package \cite{phonopy-phono3py-JPCM, phonopy-phono3py-JPSJ}  which can use various electronic structure codes as force calculators. We now also added the interface to use \octopus to the official \phonopy package.

In order to calculate the phonon modes, from the dynamical matrix, \phonopy requires forces, evaluated for a number of supercells with certain atoms being displaced from their equilibrium position. The size of the supercell determines the q-vectors for which the phonons are being evaluated.

In general, the workflow looks as follows:

\begin{enumerate}
\item{\bf Relax the equilibrium structure with \octopus:}\\
This is a standard \octopus geometry optimization run for the primitive cell. In order to generate the geometry as a {\tt POSCAR} file, which \phonopy requires as input, one has to specify
\begin{verbatim}
%Output 
   geometry | poscar
%
\end{verbatim}

\item{\bf Generate the supercells with distortion}\\
Based on the structure information in the {\tt POSCAR} file, we can use \phonopy to generate a number of
geometry files, corresponding to the distorted supercells. The following example would create a 2x2x2 supercell.
\begin{verbatim}
    phonopy-init --octopus -d --dim 2 2 2 --pa auto
\end{verbatim}
This creates two {\tt geometry-XXX} files, which contain the geometries in the \octopus format.

\item{\bf Calculate the forces}\\
For each geometry, we now can perform the ground state calculation with \octopus, including the respective geometry file with {\tt include geometry-XXX}. It is best practice to run each calculation in a separate folder, say {\tt disp-XXX}.

\item{\bf Calculate the dynamical matrix:}\\
With the converged results at hand, we can use \phonopy to generate the dynamical matrix, which then forms the foundation of all further \phonopy calculations, such as phonon band structures.
\begin{verbatim}
    phonopy-init --fz disp-{000..00N}/static/info
    phonopy-init --fz disp-{000..00N}/static/info
    phonopy-init --fz disp-{000..00N}/static/info
\end{verbatim}

In case of a polar material, we also need to take into account the Born effective charges. 
\begin{equation}
  Z^*_{\alpha, ij} = \frac{\partial F_{\alpha, j}}{\partial {\cal E}_i} = \frac{\partial P_i}{\partial R_{\alpha, j}}
\end{equation}
where $F_{\alpha, j}$ is $j$-component of the force on atom $\alpha$, and ${\cal E}_i$ the $i$ component of the electric field, $P_i$ is the $i$ component of the polarization and $R_\alpha, j$ the $j$ component of the position of atom $\alpha$.

They can be obtained by a linear response calculation with \octopus. In case of periodic systems, one first has to perform a $k \cdot p$ calculation and then perform a Sternheimer calculation. The details of this procedure are explained in the manual and the tutorials on the \octopus web pages\cite{website}. 

Once the dielectric tensor and the Born effective charges have been calculated by \octopus, we can use a helper script {\tt phonopy-octopus-born}
to generate the {\tt BORN} input file, used by the \phonopy calculator.
\end{enumerate}

As a demonstration of the implementation of the previous framework for phonons, we calculate the phonon band structure of BaTiO$_3$ in the cubic phase, which is of high technological interest.

For simplicity, and to compare to the results of the study by Ghosez {\it et al.} \cite{GhosexEtAl1999}, we used the cubic structure at the experimental lattice constant and performed the calculations within the local density approximation. For the calculation of the dynamical matrix, we employed a 3x3x3 supercell. \phonopy generated 4 such cells with displaced atoms for which \octopus ground state calculations were performed to generate the forces on all atoms. For these supercells, we performed $\Gamma-$ only calculations, corresponding to 3x3x3 k-points in the primitive cell. The grid spacing was chosen to be 0.2 a.u.. 

The calculated values for the dielectric tensor and the Born effective charges compare well with those of Ghosez, as can be seen in Tab.~\ref{tab:BaTiO3-Born}.

\begin{table}[]
    \centering
    \begin{tabular}{c|c|c}
         &  this work & Ghosez \\
         \hline
         $\epsilon_{\infty}$       &  6.74 &  6.75 \\
         $Z^*_{\rm Ba}$            &  2.73 &  2.74 \\
         $Z^*_{\rm Ti}$            &  7.37 &  7.32 \\
         $Z^*_{\rm O_\perp}$       & -2.14 & -2.14 \\
         $Z^*_{\rm O_\parallel}$   & -5.82 & -5.78
    \end{tabular}
    \caption{Dieletric tensor and Born effective masses of cubic BaTiO$_3$, calculated by \octopus and taken from Ghosez {\it et al.}\cite{GhosexEtAl1999}}
    \label{tab:BaTiO3-Born}
\end{table}

\begin{figure}
    \centering
    \includegraphics[width=0.6\linewidth]{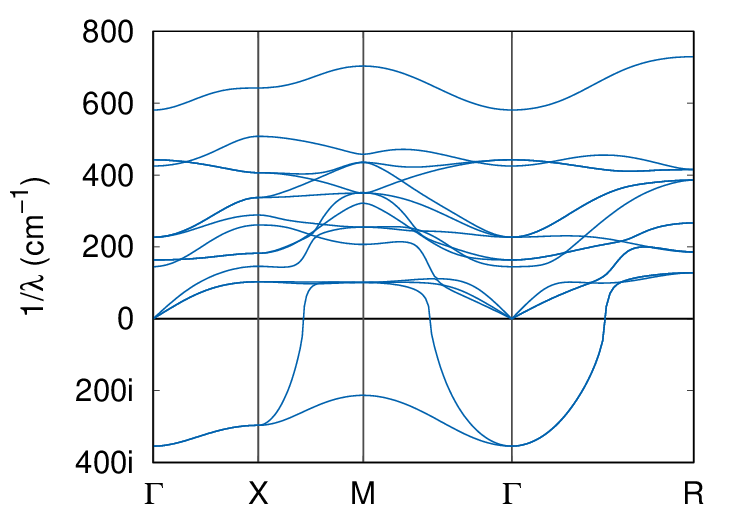}
    \caption{Phonon band structure of cubic BaTiO$_3$ including the non-analytic corrections, based on the Born charges.}
    \label{fig:enter-label}
\end{figure}

\section{Light-Matter Interaction: Leaving the Dipole Oasis}\label{sec:lightmatter-leaving-dipole}

\subsection{Full minimal coupling Maxwell-TDDFT}

The recent advances in cavity quantum materials and polaritonic chemistry have stimulated intense theoretical efforts aimed at elucidating the microscopic mechanisms responsible for the novel effects and at establishing predictive frameworks capable of guiding future experimental design. While the most rigorous theoretical descriptions of light–matter interaction are based on fully quantum approaches like QEDFT, they are typically restricted to a few photon modes and idealized cavity geometries. Also, most implementations rely on the electric dipole approximation and assume spatially uniform fields of a single cavity mode. These constraints limit their applicability to realistic nanophotonic environments, spatially structured fields, and collective effects beyond simple dipolar coupling.

To overcome these limitations, we have developed alternative formulations that retain the fundamental structure of the Pauli–Fierz Hamiltonian while relaxing the requirement of a fully quantized electromagnetic field. In the semiclassical limit, the photonic degrees of freedom are described by Maxwell's equations, whereas the matter subsystem is treated at the quantum mechanical level. Among these formulations is the Maxwell-TDDFT method in the dipole approximation \cite{Jestdt2019}, already presented in previous publications on \octopus features \cite{tancogne2020octopus}. This approach enables a self-consistent propagation of the electronic density together with the electromagnetic fields in space and time, thereby capturing radiation reaction and field propagation effects beyond simplified mode descriptions. 

However, the need to move beyond idealized cavities and, therefore, beyond the dipole approximation, has become increasingly evident as experiments increasingly explore plasmonic nanocavities, waveguides, and hybrid photonic architectures. In these settings, electromagnetic excitations are open, lossy, and strongly structured in space. In several types of spectroscopy, for example, the dipole selection rules are broken, and this provides access to molecular properties never seen before. Namely, beyond-dipole effects are present in extreme ultraviolet and X-ray spectroscopy, strong-field phenomena, magneto-optical spectra, plasmonic-enhanced spectroscopy, and optical vortices, to name a few. 

In previous works, a truncated multipolar expansion of the full light-matter interaction Hamiltonian was used. However, such approach is origin-dependent, and it converges slowly with the expansion order \cite{Lestrange2015, Rivera2016}. For this reason, the full minimal coupling (FMC) approach within Maxwell-TDDFT was developed \cite{Bonaf2025}, which takes into account the full spatially-dependent vector potential $A(r,t)$ in the electronic Hamiltonian. It is described by a Kohn-Sham time-dependent Hamiltonian in the velocity gauge, as shown in the following expression (SI units are used in this section unless stated otherwise):
\begin{align}
\mathcal{H}^{(\mathrm{FMC})}=\frac{1}{2m}\left(-i\hbar \nabla + \frac{|e|}{c}\mathbf{A}(\mathbf{r},t)\right)^{2}+ \frac{|e|}{m} \mathbf{B}(\mathbf{r},t) \cdot \hat{\mathbf{s}} + V_{\mathrm{H}}(\mathbf{r}, t)+ V_{\mathrm{xc}}(\mathbf{r}, t)+V_{\text{nuc}}(\mathbf{r}, t).
\label{eq:fullminham}
\end{align}
Here $\hbar$ denotes the reduced Planck constant, $m$ and $e$ are the electron mass and charge, respectively, $c$ is the speed of light in vacuum, $\mathbf{A}(\mathbf{r},t)$ is the transverse vector potential, and $\mathbf{B}(\mathbf{r},t)$ is the associated magnetic field. Furthermore, $\hat{\mathbf{s}}$ is the spin operator, while $V_{\mathrm{H}}$, $V_{\mathrm{xc}}$, and $V_{\text{nuc}}$ represent the Hartree, exchange–correlation, and nuclear potentials, respectively. In all calculations, we employ the adiabatic local density approximation for the time-dependent exchange–correlation potential.

The total transverse vector potential includes both external and matter-induced contributions. The latter, $\mathbf{A}_{\mathrm{mat}}$, must be obtained by solving Maxwell's equations self-consistently together with the electronic dynamics. Following Ref.~\cite{Jestdt2019}, we adopt the Riemann–Silberstein representation to propagate the electromagnetic fields, in which the Riemann–Silberstein vector $\mathbf{F}(\mathbf{r},t)$ is defined as

\begin{align}
\mathbf{F}^\pm(\mathbf{r},t)= \sqrt{ \epsilon_0/2}\left(\mathbf{E}(\mathbf{r},t) \pm i c \mathbf{B}(\mathbf{r},t)\right),
\label{eq:rsvector}
\end{align}

with $\epsilon_0$ being the vacuum permittivity. Throughout the remainder of this work, we retain only the positive component, $\mathbf{F}^+$. Within this framework, Faraday's law and Ampère's law are combined into a single equation of motion,

\begin{align}
i \partial_t \mathbf{F}(\mathbf{r}, t)= c \boldsymbol{\nabla} \times \mathbf{F}(\mathbf{r}, t)-\frac{i}{\sqrt{2 \epsilon_0}} \mathbf{J}(\mathbf{r}, t),
\label{eq:rsmaxwelleqs}
\end{align}

while the two Gauß laws are similarly combined

\begin{align}
\nabla \cdot \mathbf{F}(\mathbf{r}, t) = \sqrt{\frac{1}{2 \epsilon_0}} \rho(\mathbf{r},t).
\label{eq:rsgausseqs}
\end{align}

Here $\rho(\mathbf{r},t)$ is the electronic density that is given by $\rho(\mathbf{r},t)=-|e| n(\mathbf{r},t)$. The Maxwell-matter coupling is introduced through the source term $\mathbf{J}(\mathbf{r},t)$ in Eq.~\eqref{eq:rsmaxwelleqs}, which corresponds to the total electric current density. This quantity is related to the probability current density $ \mathbf{J}(\mathbf{r}, t)$ via $ \mathbf{J}(\mathbf{r}, t)=-|e| \mathbf{j}(\mathbf{r}, t)$ , where $\mathbf{j}(\mathbf{r}, t)$ is given by \cite{Jestdt2019}
\begin{align}
\nonumber
\mathbf{j}(\mathbf{r},t)&=\frac{\hbar }{2\mathrm{i}m}\sum_{j=1}^{N_\mathrm{occ}}\sum_{\sigma=\pm \frac{1}{2}} \left( \left[\boldsymbol{\nabla}\varphi^{\dagger}_j(\mathbf{r},\sigma,t)\right]\varphi_j(\mathbf{r},\sigma,t) - \varphi_j^{\dagger}(\mathbf{r},\sigma,t)\left[\boldsymbol{\nabla}\varphi_j(\mathbf{r},\sigma,t)\right]\right)
+\frac{|e|}{m c} \mathbf{A}(\mathbf{r},t) n(\mathbf{r},t) \\
&+ \frac{\hbar}{2m} \boldsymbol{\nabla} \times\sum_{j=1}^{N_\mathrm{occ}}\sum_{\sigma=\pm \frac{1}{2}}\varphi_j^{\dagger}(\mathbf{r},\sigma,t)\hat{\mathbf{s}}\varphi_j(\mathbf{r},\sigma,t),
\label{eq:currentfull}
\end{align}
where the first, second, and third terms are the paramagnetic, diamagnetic, and magnetization current densities, respectively. We note that in the presence of an applied vector potential, working with the time-dependent electronic density and inferring the current as $\mathbf{J}_P$ is not formally exact, since different vector potentials can produce the same density response. In such cases, time-dependent current-density functional theory provides a fitting tool, but we leave this for the future. 

Finally, to obtain the matter-induced vector potential, we make use of the time-propagated magnetic field $\mathbf{B}(\mathbf{r},t)$. The Poisson equation for $\nabla \times \mathbf{A}(\mathbf{r},t)=\mathbf{B}(\mathbf{r},t)$ is then solved, which, in the Coulomb gauge, can be expressed as a functional of the instantaneous magnetic field \cite{Stewart2003}:

\begin{align}
\mathbf{A}\left(\mathbf{r},t\right) = \nabla \times \int \frac{\mathbf{B}\left(\mathbf{r}',t\right)}{4 \pi |\mathbf{r} - \mathbf{r}'|} ,\mathrm{d}^3 r^{\prime}.
\label{eq:vecpotmagcoulombs}
\end{align}

The scheme described above has been implemented in \octopus .
The \octopus code implementation supports different grids for the Maxwell and matter system, with arbitrary shapes and spacings, allowing for bridging the different scales. Examples of this implementation are provided below.

\begin{figure}
\centering
\includegraphics[width=1.0\textwidth]{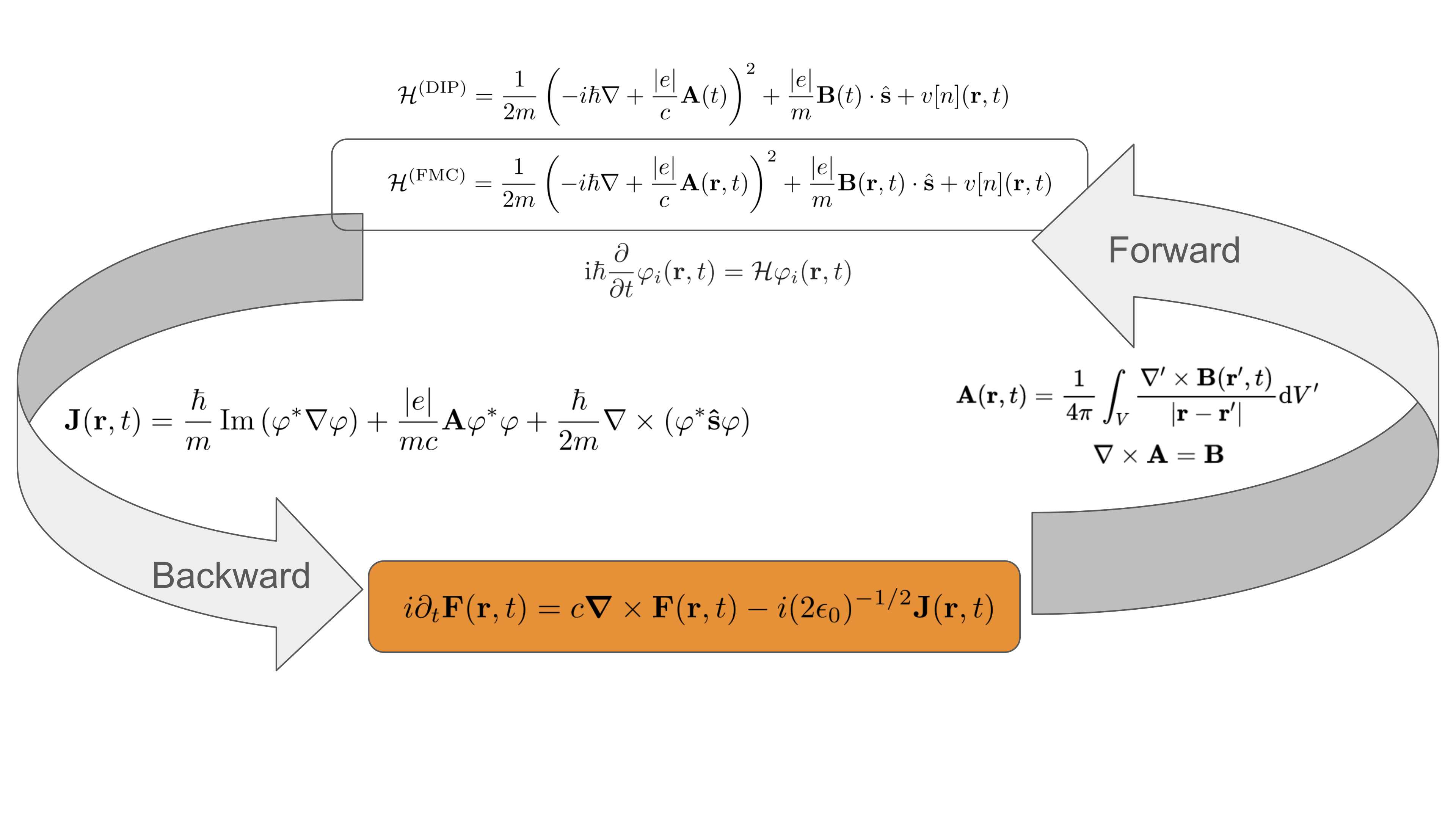}
\caption{The scheme describing the forward-backward coupling between light and matter subsystems beyond the electric dipole approximation. The current contribution $\mathbf{J}(\mathbf{r},t)$ links the matter system to the electromagnetic field. The vector potential $\mathbf{A}(\mathbf{r},t)$ is computed from the instantaneous magnetic field  $\mathbf{B}(\mathbf{r},t)$ and is substituted into the Hamiltonian of choice on the matter side.}
\label{fig:fmc_forward_backward}
\end{figure}

\subsection{Drude medium description for simulations of nanoplasmonic structures}

 
 The Drude model is a well-known effective medium approach for describing the collective electronic response of the material, which is useful for describing the electromagnetic response at larger scales. Similar to the coupling with quantum mechanical systems, the current response can also be used to connect the Drude system to Maxwell's equations, as seen in Fig. \ref{fig:fmc_forward_backward}.

In the case of a Drude linear medium description, the 
frequency-dependent polarization current can be written as $\mathbf{P}(\mathbf{r},\omega) = \epsilon_0 \chi(t) \mathbf{E}(\mathbf{r},\omega)$, where $\chi$ represents the scalar, frequency-dependent, isotropic susceptibility expressed as 
\begin{align}
\chi(\omega) = - \frac{\omega_p^2}{\omega ^2 - i \omega \gamma_p}.
\label{eq:drude_susceptibility}
\end{align}
Here $\omega_p$ is the plasma frequency and $\gamma_p$ is the inverse relaxation time of the material. We combine this formalism with the solutions of Maxwell's equations by solving Eq.~\eqref{eq:rsmaxwelleqs} simultaneously with the auxiliary differential equation (ADE) \cite{taflove} 
\begin{align}
\partial_t \mathbf{J}_P(\mathbf{r},t) + \gamma_P\mathbf{J}_P(\mathbf{r},t) = \epsilon_0\omega_P^2\mathbf{E}(\mathbf{r},t).
\label{eq:ade}
\end{align}
where $\mathbf{J}_P$ is the current response of the nonmagnetic Drude material.
 Valid in the linear regime, the Drude model implementation enables the simulation of Maxwell's equations using nanoplasmonic structures. \octopus allows the user to perform calculations using geometries designed with 
 computer-aided design programs such as OpenSCAD \cite{openscad}. The structure is mapped onto the grid using the CGAL library \cite{CGAL} and the ADE is solved simultaneously with the Maxwell propagation using the 4th-order Runge-Kutta method. This capability of the code has already been used to demonstrate the generation of orbital angular momentum pulses from plane waves with nanoplasmonic spirals \cite{Albar2023}. This framework has further potential, as various structures, including cavity walls and waveguides, can be modeled, and phenomena related to electromagnetic-field confinement and dispersion can be studied.

\subsection{Spatially-structured electromagnetic fields}

To extend beyond the electric dipole approximation, the electromagnetic source should be position-dependent. Such sources can be explicitly propagated with appropriate boundary conditions using the Maxwell solver of the \octopus code, as seen in the previous section. However, solving Maxwell’s equations explicitly can be computationally demanding for some cases, where analytical expressions are known. Therefore, we have implemented spatially dependent laser sources (named \textit{`External Source'} in the code) that evaluate the electric and/or magnetic fields directly on the matter grid and do not require explicit time propagation. The presently available beams are plane waves and Bessel beams.

\begin{figure}[!ht]
	\includegraphics[scale=0.4, width=\linewidth]{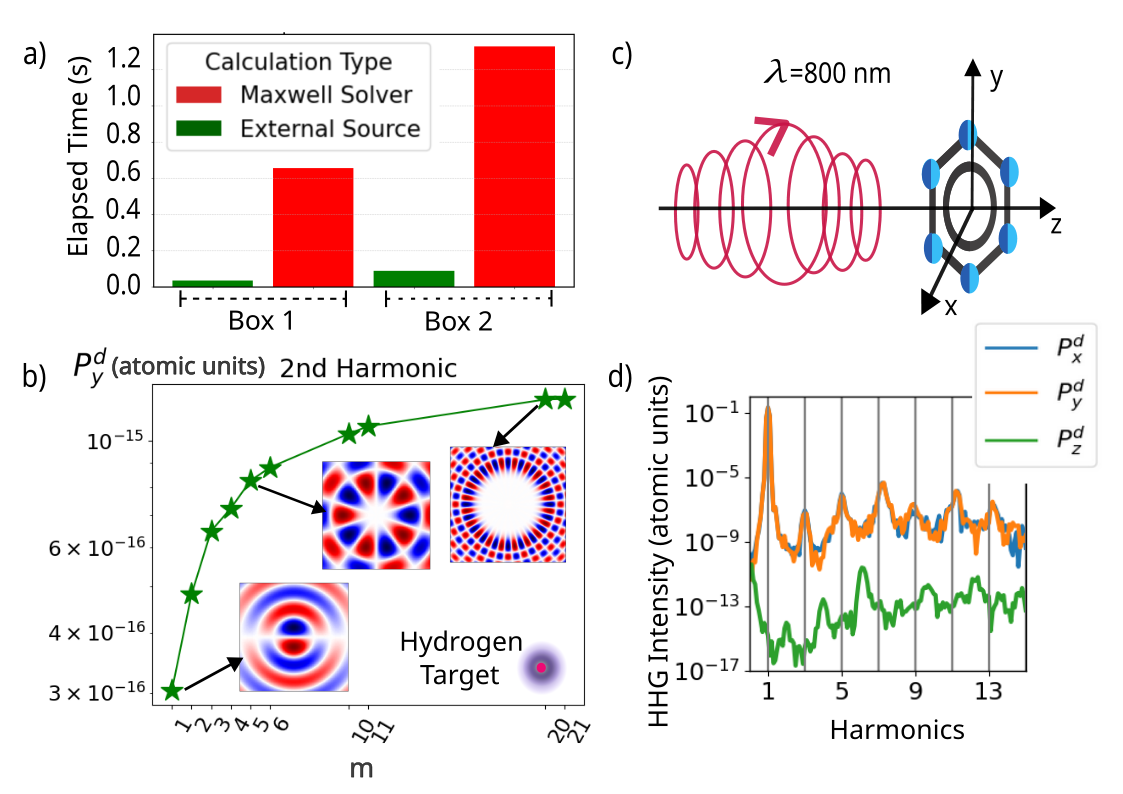}
	\caption{a) Comparison of elapsed calculation time for the \textit{`External Source'} (green bars) and Maxwell solver (red bars). b) Second harmonic generation yield in the y direction with a hydrogen atom excited with Bessel beams of OAM ranging from $m=1$ to $m=21$. Spatial cross sections of beams $m=1,5, 20$ are provided. OAM beam calculations are enabled in \octopus, combining the beyond-dipole full minimal coupling approach with the spatially structured beam in the same simulation. This panel is reproduced from reference \cite{Albar2025} with the authors' permission. c) We target a benzene molecule aligned with the $xy$ plane, using a circularly polarized plane wave pulse with a wavelength of 800 nm. 
    d) Beyond the electric dipole approximation, high-harmonic spectra corresponding to the setup in c); the electric dipole approximation-neglected component in the propagation direction $z$ is captured by the full minimal coupling approach.}
	\label{fig:extsource}
\end{figure}

We show the numerical advantage of this implementation in comparison to the Maxwell solver in panel a) of Fig. \ref{fig:extsource}. Here, plane wave pulses are propagated first in a cubic simulation box with dimensions $3.0 \times 3.0 \times 3.0 $ Å (Box 1), then in a larger box (Box 2) of size $4.0 \times 4.0 \times 4.0 $ Å with equal spacing of $0.2$ Å. The target is a carbon atom, coupled to the electromagnetic field within the electric dipole approximation (EDA). The elapsed time in seconds to complete one timestep ($2.41888$ as) of simulation is reported. The calculation time is drastically reduced for \textit{`External Source'} for both boxes, and the reduction is more prominent as one moves to larger grids. This demonstration was performed on two cores of an AMD Ryzen 7 PRO 7840U CPU.

The spatially dependent laser functionality enables studies of light-matter dynamics using pulses carrying orbital angular momentum (OAM), also referred to as twisted light. 
 One category of such pulses is Bessel beams. They are represented as a function of their frequency $\omega$, helicity $\lambda$, the projection of the linear wave vector $k_z$, and total angular momentum $m$ \cite{AlbarThesis}. The $z$ axis is fixed as the propagation direction in our formulation. The Bessel-wave vector potential can be separated into temporal and spatial parts as $\mathbf{A}_{\varkappa m k_z \lambda}^{(\mathrm{tw})}(\mathbf{r},t) = \mathbf{A}_{\varkappa m k_z \lambda}^{(\mathrm{tw})}(\mathbf{r})e^{-i\omega t}$,  where $\varkappa=\sqrt{\omega^2-k_z^2c^2}$ is the well-defined transversal momentum of the Bessel beam \cite{Kosheleva2020}. 
The corresponding spatial components of
$\mathbf{A}_{\varkappa m k_z \lambda}^{(\mathrm{tw})}(\mathbf{r}) = (A_x(\mathbf{r}), A_y(\mathbf{r}), A_z(\mathbf{r}))$ in Cartesian coordinates are 
\begin{align}
A_x &= \frac{1}{\sqrt{2}} A_0 e^{i k_z  z} e^{i m \varphi_r}(e^{i \varphi_r} d_{-1 \lambda}^{1}\left(\theta_k\right) J_{m+1}\left(\frac{\varkappa r_{\perp}}{ c}\right)+ e^{-i \varphi_r} d_{1 \lambda}^{1}\left(\theta_k\right) J_{m-1}\left(\frac{\varkappa r_{\perp}}{c}\right)),\label{eq:bessel_reformulated}
 \\
\nonumber
A_y &= \left(\frac{ i}{\sqrt{2}}\right) A_0 e^{i k_z z} e^{i m \varphi_r}(-e^{i \varphi_r} d_{-1 \lambda}^{1}\left(\theta_k\right) J_{m+1}\left(\frac{\varkappa r_{\perp}}{ c}\right)
+ e^{-i \varphi_r} d_{1 \lambda}^{1}\left(\theta_k\right) J_{m-1}\left(\frac{\varkappa r_{\perp}}{ c}\right)),
\\
\nonumber
A_z &= -A_0  e^{i k_z  z}  i e^{i m \varphi_r}  d_{0 \lambda}^{1}\left(\theta_k\right)J_{m}\left(\frac{\varkappa r_{\perp}}{c}\right).
\end{align}

Here, $A_0$ is the amplitude of the vector potential, $J_n$ represents the Bessel function of the first kind, $\mathbf{r}=\left(r_{\perp}, \varphi_r, r_{\|} \right)$ is the position vector in cylindrical coordinates. Cylindrical components are connected to Cartesian counterparts such that $r_{\perp} = \sqrt{x^2 + y^2}$, $\varkappa=\sqrt{\omega^2-k_z^2c^2}$, 
$k_z = \frac{\varkappa}{\tan( \theta_k)c}$, and $\tan (\phi_{r}) = \frac{y}{x}$.
$d_{\mu \lambda}^1\left(\theta_k\right)$ is a small Wigner matrix described below in Eq. \eqref{eq:small_wigner}, dependent on both opening angle $
    \theta_k= \arctan \left(\frac{\varkappa}{k_zc} \right)$, and helicity of the beam $\lambda$ as follows:
\begin{equation}
d_{\mu \lambda}^1\left(\theta_k\right) = \begin{cases}\frac{\sin \theta_k}{\sqrt{2}} \lambda, & \text { if } \mu=0 \\ \frac{1+\lambda\cos \theta_k }{2}, & \text { if } \mu=1 \\ \frac{1-\lambda\cos \theta_k }{2}, & \text { if } \mu=-1\end{cases}. \label{eq:small_wigner}
\end{equation}
The real parts of the above formulation are used to compute observables such as the electric and magnetic fields \cite{Greiner1998}. Combined with the full minimal coupling framework, the Bessel beam implementation enables us to report the high harmonic generation spectra of the hydrogen atom under beams of various OAM values tied to the parameter $m$ in Eq. \eqref{eq:bessel_reformulated} \cite{Albar2025}. The beyond-dipole effect of the changing OAM number $m$ of the incoming laser is visible in the even harmonics yield in the $y$ direction, as seen in panel b) of Figure \ref{fig:extsource}. The rich spatial structure of such beams for selected $m$ is also visible in the inset of panel b) of Fig. \ref{fig:extsource}.

%


The spatially dependent laser necessitates the reformulation of the pulse envelope. Within the electric dipole approximation, the conventional time-dependent representation of a laser pulse follows the following expression:
 \begin{align}
    A( t) = A_0 \exp\left[i\left( - \omega t + \phi_p\right)\right] 
    \exp\left( -\frac{\left(  t - t_0 \right)^2}{2 \tau_0^2} \right).
  \label{eq:time_envelope}
\end{align}
Here, $A_0$ denotes the amplitude, $\phi_p$ is the phase of the beam, $\omega$ is the central frequency, and $t$ refers to time. $t_0$ is the point in time where the pulse is initiated, and $\tau_0$ is the width in time, i.e., the duration of the pulse. For a space-dependent pulse, the same pulse is expressed as
\begin{align}
    A(\mathbf{r}, t) = A_0 \exp\left[i\left(\mathbf{k} \cdot \mathbf{r} - \omega t + \phi_p\right)\right] 
    \exp\left( -\frac{\left( \mathbf{k} \cdot (\mathbf{r} - \mathbf{r}_0)/|\mathbf{k}| \right)^2}{2 \sigma^2} \right).
    \label{eq:space_time_envelope}
\end{align}
The dependence of the pulse on the wave vector $|\mathbf{k}|=\omega/c$ is preserved, as well as position-dependent terms in the exponential describing a plane wave: $\exp (i(\mathbf{k}\cdot\mathbf{r} - \omega t + \phi_p))$. $\mathbf{r}_0$ denotes the position where the pulse starts, and the width in space $\sigma_0$ is used. 

Beyond-dipole behavior can still be observed with plane wave sources. Using the above spatially dependent formulation of the envelope, we employ circularly polarized laser beams of 10 cycles, with a central wavelength of 800 nm. With a peak electric field of $3.82\times10^{10}$ V/m, we target a benzene molecule with the setup visible in part c) of Fig. \ref{fig:extsource}. As seen in part d), the spatial variance of the laser requires to switch on the full minimal coupling framework, and based on this we can identify beyond dipole effects in the $z$ component of the spectrum, which has dominant even harmonics \cite{AlbarThesis}. This component is a result of the Lorentz force-caused deviation in the driven electrons' trajectory and is solely detectable beyond the dipole approximation \cite{Ludwig2014}.

\subsection{Non-dipole strong-field approximation} 

In practice, coupling an electronic system to the Maxwell equations imposes certain inherent theoretical restrictions. Issues primarily arise if the symmetry of the electromagnetic wave that arises from the Maxwell equations is incommensurate with the symmetries underlying the description of the electronic system. A principal example is condensed matter systems, which are almost exclusively formulated based on Bloch’s theorem that relies on a conserved lattice translational symmetry. For exceptional cases, one might be able to match the translational symmetries of the electromagnetic field with the translational symmetry of a tailored electronic supercell, making a theoretical description possible albeit computationally costly. However, in general, the incompatible symmetries will impede the conservation of crystal momentum in the electronic system, rendering the simulation of the condensed matter system with spatially-varying electromagnetic field practically unattainable. Serendipitously, a practical solution for such issues for condensed matter systems initially arose from the strong-field atomic physics community targeting ionization and stabilization phenomena. 

A central insight, initially proposed for atomic stabilization~\cite{PhysRevA.48.R4027}, is that for sufficiently strong fields, the paramagnetic $\bm A\cdot \bm p \propto \sqrt{I}\lambda$ term of Eq.~\eqref{eq:fullminham}, responsible for the Lorentz force associated with the cycle-averaged electron motion, becomes negligible compared to the diamagnetic $\bm A\cdot \bm A \propto I\lambda^2$ term responsible for the Lorentz force associated with the electronic quiver motion. This approximation was later similarly corroborated for Compton and Raman scattering~\cite{PhysRevLett.112.073001,Bachau_2017}.  
Although the short-wavelength regime of atomic stabilization could impede the wavelength-scaling of the approximation, the conditions were well fulfilled due to the process' narrow zero-centered electron wavepacket momentum distribution~\cite{PhysRevA.64.013411}. Further developments, centered on also high-frequency ionization, gave rise to elegant gauge transformations granting particularly well-suited Hamiltonians for numerical efficiency~\cite{PhysRevA.90.053411,PhysRevA.92.013405,PhysRevA.93.013423,PhysRevA.93.063425,PhysRevA.97.013415,Brennecke_2018}. The strong-field non-dipole approaches later developed into the long-wavelength regime~\cite{PhysRevA.101.043408}, grounded on semiclassical analysis, in which dominant magnetic field contributions were considered perturbatively along the dipole-induced trajectory, ideas reminiscent of those introduced for enabling Coulomb-corrected strong-field approximated simple man models~\cite{PhysRevA.85.023428,PhysRevLett.93.233002}. The obtained decoupling of the spatial and momentum space operators would become essential for the development of non-dipole Volkov functions \cite{PhysRevA.101.043408} and laser-induced adiabatic states \cite{Madsen2021} for progress in laser-assisted scattering \cite{Jensen_2020}, above-threshold ionization \cite{Lund_2021,PhysRevA.106.043118,PhysRevA.110.023111}, and tunneling ionization~\cite{PhysRevA.105.043107}. Thus, while the historical development of non-dipole approaches was motivated by enabling numerically efficient split-step methods and analytical progress, the key advancements for enabling such would also prove critical for enabling a condensed-matter suitable non-dipole strong-field approximation formalism~\cite{PhysRevLett.134.196902}. The non-dipole strong-field approximated (NDSFA) approach considers a general electric dipole approximated theory, which is corrected by the introduction of an effective field $\bm A^\text{ND}$ that captures leading-order non-dipole effects arising from the diamagnetic term. The minimal coupling Hamiltonian of Eq.~\eqref{eq:fullminham}, thus reduces to~\cite{PhysRevA.101.043408,PhysRevLett.134.196902}
\begin{equation}
	H^\text{NDSFA}  = \frac{\bm p -q \left(\bm A^\text{EDA}(t) + \bm A^\text{ND}(t) \right)}{2 m} + V
\end{equation}
Here, the non-dipole correction $\bm A^\text{ND}$ can be precomputed at negligible computational cost based on knowledge of $\bm A^\text{EDA}$ and its associated propagation direction $\bm k_L$, or be evaluated in real-time through its integral representation if coupled self-consistently to e.g. a macroscopic Maxwell equation solver. For a simple dipole field with well-defined propagation direction, the associated non-dipole correction takes the form of~\cite{PhysRevA.101.043408,PhysRevLett.134.196902}
\begin{equation}
	\bm A^\text{ND}(t) = -  \frac{q}{2mc} \widehat{\bm k}_L \left(\bm A^\text{EDA}(t)\right)^2. \label{eq:ND}
\end{equation}
For more evolved multi-field settings where e.g. interference between multiple laser-pulses occur, each dipole field contribution, $\bm A_i^\text{EDA}$, with its associated electric field $\bm E_i^\text{EDA}$, can be corrected through its respective non-dipole correction, $\bm A_i^\text{ND}$. In cases where individual pulses have no temporal overlap, perpendicular polarization axis or identical propagation axis, $\widehat{\bm k}_{L,i} = \widehat{\bm k}_{L,j}$, Eq.~\eqref{eq:ND} remains valid. However, for more general cases, non-dipole interferences need to be evaluated to grant a total non-dipole correction through the integral representation~\cite{PhysRevLett.134.196902}
\begin{equation}
	\bm A^\text{ND}(t) = \frac{q}{mc} \int dt'  \sum_i \bm A^{\text{EDA}}(t') \cdot \bm E^{\text{EDA}}_i(t')  \widehat{\bm k}_{L,i}.
\end{equation}
The formalism, as implemented in \octopus, can accurately and cost-effectively reproduce magnetic field-induced bulk second-order susceptibilities~\cite{PhysRevLett.134.196902}, while also recently being extended to related semiclassical~\cite{PhysRevA.105.L021101} and semiconductor Bloch equation formalisms~\cite{weeth2025intenselasernondipoleinducedsymmetrybreaking} within the condensed matter setting.

\section{Exciton Dynamics}

The optical response of semiconductor or insulator systems interacting with 
excitation by light is often dominated by features arising due to bound electron-hole pairs, known as excitons. They play a significant role in many processes, such as energy transfer and light absorption in biomolecules, nanostructures, and solids, and have fundamental and technological applications. However, their theoretical description remains to date challenging and is still an active area of research. In the Bethe-Salpeter equation (BSE) approach, excitons appear explicitly as bound states because the electron-hole interaction is directly included via a screened Coulomb potential. In standard TDDFT, based on semi-local functionals, excitons are however heavily suppressed or completely missed because standard xc kernels lack the necessary long-range tail. The existing solutions are often challenging to implement and require significant optimization effort. We have implemented in \octopus several of the cutting-edge methods that allow for dealing with excitons, for instance using hybrid functionals, as well as using long-range corrected potential for TDDFT. 

\subsection{Accelerating Molecular Hybrid Functional Calculations with Interpolative Separable Density Fitting}

Hybrid functionals are a class of approximation to the DFT exchange-correlation functional in which a fraction of non-local exact exchange from Hartree-Fock theory is mixed with a semilocal exchange-correlation functional. This typically improves atomisation energies, bond lengths, and band gaps relative to calculations performed with purely semi-local or mGGA functionals~\cite{schimka2011improved, garza2016predicting, borlido2019large}.

As such, they provide an efficient route to quantitative band gaps and optical properties without the cost of many-body perturbation or wavefunction-based methods, whose conventional implementations typically scale with system size as $O(N^4)$ for GW–BSE~\cite{blase2020bethe} and MP2~\cite{schafer2017quartic}, and $O(N^6)$ for periodic CCSD~\cite{mcclain2017gaussian}. This makes hybrid functionals one of the most viable approaches for otherwise intractable simulations for large systems or long time-dependent trajectories.

Hybrid functional calculations can be made more efficient by exploiting low-rank factorisations of the exchange operator. The adaptively compressed exchange (ACE) formalism replaces the dense, full-rank exchange operator by a low-rank operator constructed on the fly from its action on the manifold of occupied Kohn-Sham states \cite{lin2016adaptively}. By construction, the ACE operator agrees exactly with $V_X$ within the subspace spanned by the states used to build it.

ACE reduces the cost of repeatedly applying exchange once the compressed operator has been constructed, but the construction itself still requires applying the underlying exact-exchange operator. This step remains expensive because it involves $O(N_{e}^2)$ orbital pair products entering the exchange potential. A complementary strategy is therefore to reduce the rank of the pair-product space itself. Interpolative separable density fitting (ISDF), which is closely related to tensor hyper-contraction, is a form of density fitting in which a numerical auxiliary basis is constructed to give a compact representation of orbital pair products~\cite{parrish2012tensor, lu2015compression, rettig2023even}. The exact-exchange operator can then be formulated in terms of these ISDF basis vectors, reducing the number of Poisson solves for molecular systems from $O(N_e^2)$ to $O(N_e)$. When combined with ACE, ISDF reduces the cost of constructing the ACE operator by accelerating the underlying exact-exchange applications required during its construction.

\subsubsection{ACE + ISDF in Octopus}

Octopus implements ISDF as a strategy for accelerating hybrid functional calculations for molecular systems. Our implementation follows the formulation of Hu et al.\cite{hu2017interpolative}, where the action of the exchange operator on the occupied Kohn-Sham states is recast in terms of ISDF interpolation vectors, and the selection of interpolation points is separated from the construction of the interpolation vectors.

Interpolation points are selected using the standard weighted k-means algorithm, which is well suited to the MPI domain decomposition of the real-space grid\cite{dong2018interpolative}. Since weighted k-means is a local optimisation procedure, the final set of points is sensitive to the initial seed. We find that the quality of the interpolation points can be substantially improved by replacing random or weighted-random initialisation with points seeded using the k-means++ algorithm\cite{arthur2007k}.

One downside of selecting interpolation points with a weighted k-means algorithm is that it does not provide an a priori estimate of the resulting error in the exact-exchange energy, which must therefore be assessed a posteriori against a reference calculation. In addition, as the number of interpolation points approaches the numerical rank of the pair-product space, or when the interpolation points are poorly chosen, the rows of $C$ become nearly linearly dependent and the coefficient Gram matrix $CC^T$ becomes ill-conditioned. In practice, we avoid issues with inversion by using an SVD pseudo-inverse. The numerical rank of $CC^T$ can also be used in convergence studies as a proxy for saturation with respect to the number of interpolation points, without requiring an ACE reference calculation.

\subsubsection{Constructing ISDF Vectors}

The ISDF formalism is described in \cite{hu2017interpolative}. Here we summarise the basic ISDF construction in real-space notation. Let $r_p$ denote the real-space grid and $r_\mu$ denote the selected interpolation points, respectively. For two sets of Kohn-Sham orbitals $\{\varphi_i\}_{i=1}^m$ and $\{\psi_j\}_{j=1}^n$, their pair products are approximated by sampling over a small set of interpolation points, $\{r_\mu\}_{\mu=1}^{N_\mu}$, and expanding them in interpolation vectors $\theta = \{\zeta_\mu(\mathbf{r}_p)\}$. For fixed interpolation points, the interpolation vectors are obtained from the least-squares problem, giving:
\begin{equation} \label{eq.least_squares_isdf}
\Theta=ZC^T(CC^T)^{-1}.
\end{equation}
where the pair-product matrix, $Z\in\mathbb{R}^{N_r\times mn}$, and the coefficient matrix, $C\in\mathbb{R}^{N_\mu\times mn}$, are defined as:
\begin{equation}
Z_{p,(ij)}=\varphi_i(\mathbf{r}_p)\psi_j(\mathbf{r}_p),\qquad C_{\mu,(ij)}=\varphi_i(\mathbf{r}_\mu)\psi_j(\mathbf{r}_\mu).
\end{equation}
The separable nature of ISDF means that the two matrices in Eq. \eqref{eq.least_squares_isdf} do not need to be constructed by explicitly summing over all pair products. Instead, contractions over pair indices factor into products of quasi-density matrices:
\begin{equation}
[ZC^T]_{p\mu}=\left[\sum_i\varphi_i(\mathbf{r}_p)\varphi_i(\mathbf{r}_\mu)\right]\left[\sum_j\psi_j(\mathbf{r}_p)\psi_j(\mathbf{r}_\mu)\right],
\end{equation}
and
\begin{equation}
[CC^T]_{\mu\nu}=\left[\sum_i\varphi_i(\mathbf{r}_\mu)\varphi_i(\mathbf{r}_\nu)\right]\left[\sum_j\psi_j(\mathbf{r}_\mu)\psi_j(\mathbf{r}_\nu)\right].
\end{equation}
In the single-orbital-set case relevant to a single-level SCF (self-consistent field) scheme, $\psi=\varphi$. Defining:
\begin{equation}
P_{p\mu}=\sum_i\varphi_i(\mathbf{r}_p)\varphi_i(\mathbf{r}_\mu),\qquad Q_{\mu\nu}=\sum_i\varphi_i(\mathbf{r}_\mu)\varphi_i(\mathbf{r}_\nu),
\end{equation}
with $P\in\mathbb{R}^{N_r\times N_\mu}$ and $Q\in\mathbb{R}^{N_\mu\times N_\mu}$, the interpolation vectors reduce to:
\begin{equation}
\Theta=(P\odot P)(Q\odot Q)^{-1}.
\end{equation}
Following the construction of the ISDF vectors, the action of exact exchange on an orbital $\psi_j$ is then approximated as:
\begin{equation}
\left(V_X[{\varphi_i}]\psi_j\right)(\mathbf{r}_p)\approx
-\sum_{\mu=1}^{N_\mu}
\left[\sum_i f_i\varphi_i(\mathbf{r}_p)\varphi_i(\mathbf{r}_\mu)\right]
V_{\zeta_\mu}(\mathbf{r}_p)\psi_j(\mathbf{r}_\mu).
\end{equation}
where $f_i$ is the occupation number, $V_{\zeta_\mu}(\mathbf{r}_p)$ is the Coulomb potential in the ISDF basis:
\begin{equation} \label{eq.isdf_poisson_eq}
V_{\zeta_\mu}(\mathbf{r})=\int K(\mathbf{r},\mathbf{r'})\zeta_\mu(\mathbf{r'}),d\mathbf{r'},
\end{equation}
and $K(\mathbf{r},\mathbf{r'})$ is the Coulomb or screened Coulomb kernel. This avoids the direct convolution of every pair density $\varphi_i(\mathbf{r})\psi_j(\mathbf{r})$ with the Coulomb kernel. Instead, the convolution is performed once for each ISDF vector, requiring only $N_\mu$ Poisson-like solves to construct the potentials $\{V_{\zeta_\mu}(\mathbf{r})\}$.

\subsubsection{Results and Discussion}

\begin{figure}
\centering
\includegraphics[width=0.8\textwidth]{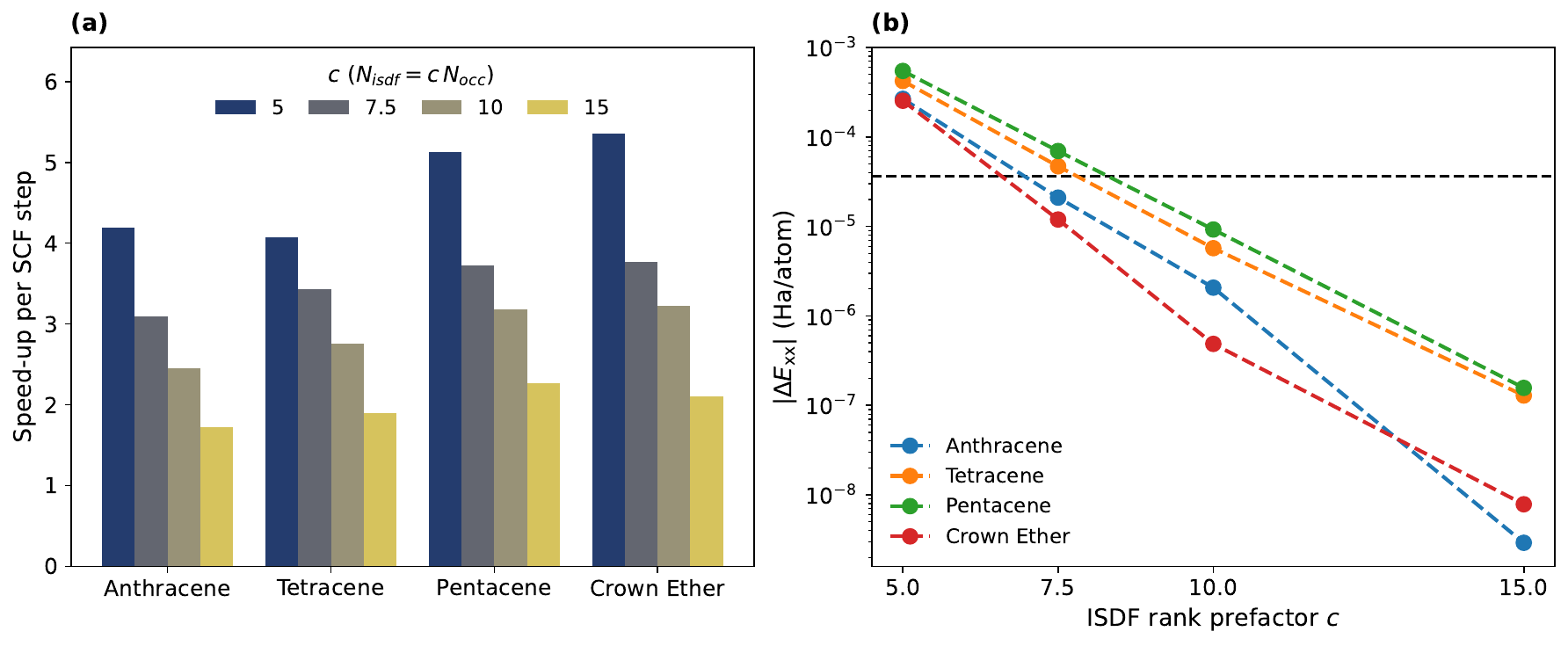}
\caption{ISDF ground-state performance for a set of molecules. In all cases, the interpolation points are chosen once, during the first SCF step. Left panel: speed-up relative to the ACE-only reference calculation for several interpolation point set sizes. Right panel: error in the exact-exchange energy as a function of ISDF interpolation set size.
The dashed horizontal line marks a target precision of 1 meV/atom ($3.67 \times 10^{-5}$ Ha/atom), the order of magnitude typically accepted for density-fitted exchange\cite{rettig2023even}.
}
\label{fig.isdf_performance}
\end{figure}

ISDF is implemented in Octopus with OpenMP threading, and full support for both states and domain distribution over MPI. For molecular systems, we find that the required number of interpolation points scales linearly with system size (number of electrons), and that the error in the exact exchange energy typically drops below 1 meV/atom ($3.67\times10^{-5}$ Ha/atom) when using $N_\mu \sim 7-10 N_e$ interpolation vectors, as shown in Fig. \ref{fig.isdf_performance}.

Solving the Poisson equation, Eq. \eqref{eq.isdf_poisson_eq}, dominates the simulation time and scales linearly as $O(N_\mu)$, for a fixed real-space grid size, since one Poisson-like solve is required for each interpolation vector. The method therefore demonstrates practical speed-ups for systems containing more than 10-12 occupied states. For timing calculations, the interpolation points are fixed after the first SCF step. This has a small effect on the exchange energy, within the same order of magnitude as the error introduced by the ISDF approximation, but at substantially lower computational cost. Extensions to TD, periodic boundary conditions and occupied-virtual pair products for the description of band gaps, are all currently under development.

\subsection{Real-time TDDFT for exciton dynamics}
\subsubsection{LRC Vector Potential}
Real-time TDDFT provides a powerful framework for simulating ultrafast electron dynamics in solids. However, standard semilocal xc functionals (and their adiabatic versions), such as the most used ALDA (adiabatic local density approximation), fails to describe excitonic effects, which are however critical for the optical response of finite-gap systems~\cite{Byun2020}. This problem of missing the long-range electron-hole interaction required to describe bound excitons can be effectively fixed by employing the long-range-corrected (LRC) xc kernel in the real-time domain~\cite{Sun2021}, a strategy originally developed for linear-response TDDFT\cite{reining2002prl}.

In linear response, the LRC kernel has the form~\cite{reining2002prl}
\begin{equation}
f^{\rm LRC}_{\rm xc,GG'}(\mathbf{k})
=
-\frac{\alpha}{|\mathbf{k}+\mathbf{G}|^2}
\delta_{\mathbf{G}\mathbf{G}'},
\end{equation}
where the parameter $\alpha$ controls the strength of the effective long-range electron-hole interaction, $\mathbf{G}$ and $\mathbf{G}'$ represent the reciprocal lattice vectors and $\mathbf{k}$ is a vector in the first Brillouin zone. Since the macroscopic scalar potential associated with this kernel is ill-defined in the optical limit, one could reformulate the LRC contribution as a macroscopic exchange-correlation vector potential. In the velocity gauge, the head-only approximation leads to the equation of motion~\cite{Sun2021}
\begin{equation}\label{eq:LRC}
\frac{d^2 \mathbf{A}^{\rm LRC}_{\rm xc}(t)}{dt^2}
=
\frac{\alpha c}{V} \mathbf{J}(t),
\end{equation}
where $\mathbf{A}_{\rm xc}^{\mathrm{LRC}}(t)$ is the time-dependent xc vector potential and $\mathbf{J}(t)$ is the macroscopic electronic current density, with $V$ and $c$ the unit-cell volume and speed of light, respectively. The total vector potential in the time-dependent Kohn--Sham (TDKS) equation is then $\mathbf{A}(t)=\mathbf{A}_{\mathrm{ext}}(t)+\mathbf{A}_{\rm xc}^{\mathrm{LRC}}(t)$, thus the xc vector potential is determined simultaneously by evolving TDKS equation together with Eq.~\eqref{eq:LRC}.

This formulation incorporates excitonic effects directly into the real-time propagation while introducing negligible additional computational cost compared with ALDA. For materials with weakly bound excitons, such as Si and CsGeCl$_3$, the optical spectra obtained using real-time TDDFT agree well with the corresponding linear-response TDDFT results and represent a substantial improvement over those obtained with semilocal xc approximations~\cite{Sun2021}. Moreover, because the method is formulated in the time domain, it can be applied beyond the linear-response regime, making it suitable for investigating femtosecond exciton dynamics and nonlinear optical phenomena. 

\subsubsection{Modified LRC (MLRC) potential}
The real time LRC approach has achieved success in obtaining consistent optical spectra with the linear response TDDFT. However, because of the violation of zero-force theorem~\cite{Sun2021,Williamsjctc2025}, the real time LRC leads to an instability in the total current especially for materials with strongly bound excitons. One way to solve this problem is to add a mass term~\cite{Dewhurst2025prb}, namely
\begin{equation}\label{eq:proca}
    a_2\frac{d^2}{dt^2}\mathbf{A}^{\rm{MLRC}}_{\mathrm{xc}}(t)+a_0\mathbf{A}^{\rm{MLRC}}_{\mathrm{xc}}(t)=\frac{4\pi c}{V}\mathbf{J}(t),
\end{equation}
where $a_0$ and $a_2$ are adjustable parameters, and $\mathbf{A}_{\rm xc}^{\rm MLRC}(t)$ is the corresponding xc vector potential. The $a_0$ term prevents the unbounded growth of the xc vector potential. When $a_0$ exceeds a system-dependent threshold, the zero-force condition is satisfied on average, thereby stabilizing the real-time propagation. Owing to the formal resemblance of Eq.~\eqref{eq:proca} to the Proca equation for a massive vector field, this approach is also referred to as the Kohn--Sham--Proca scheme~\cite{Dewhurst2025prb}. This scheme has been implemented in the \textsc{Octopus} code.

We use bulk Si as our representative material to show the application of our code. In order to calculate its optical absorption spectrum, we probe the ground state with a delta-kick $\mathbf{E}(t)=\mathbf{E}_0\delta(t)$, and propagate the system with real-time TDDFT. The response of the system is encoded in the total current $\mathbf{J}(t)$, from which we can calculate the optical conductivity by $\mathbf{J}(\omega)=\sigma(\omega)\mathbf{E}(\omega)$. We can thus obtain the dielectric function from $\epsilon(\omega)=1+4\pi i\sigma(\omega)/\omega$. In all calculations, a $32\times32\times32$ k-point grid was employed to ensure convergence, and the lattice constant was set to 5.387~\AA{}. Following previous studies~\cite{Sun2021,Dewhurst2025prb}, we set $\alpha=0.2$ for the LRC calculation and used $a_2=85$ and $a_0=0.2$ for the MLRC calculation.

\begin{figure}[!t]
  \centering
\includegraphics[width=0.7\linewidth]{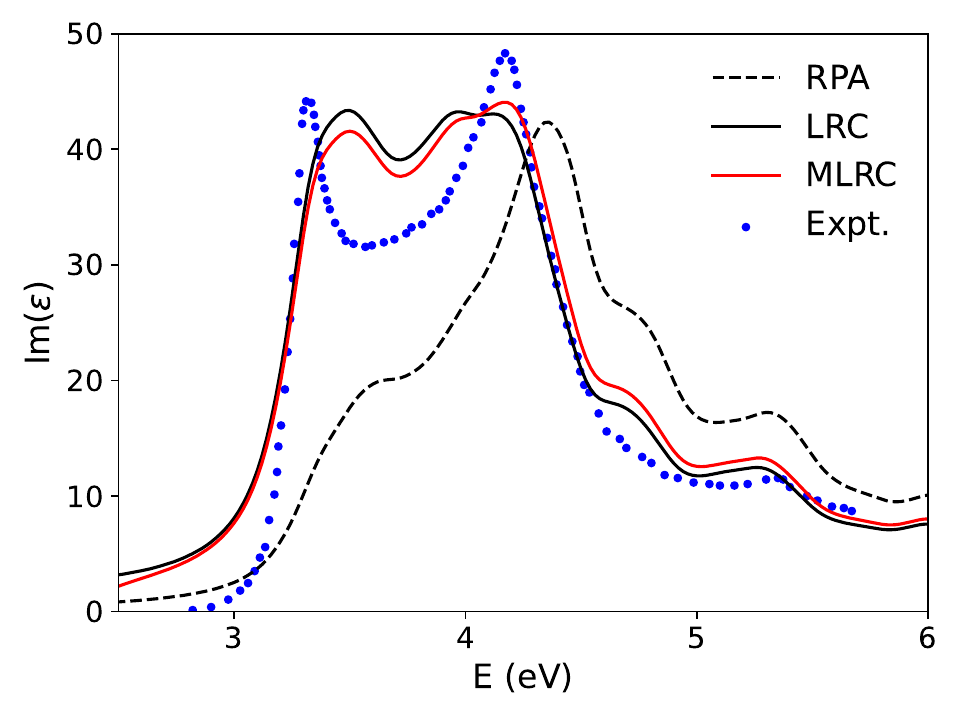}
  \caption{Imaginary part of the the dielectric function $\epsilon$ of bulk Si calculated using RPA, LRC and MLRC approaches. The numerical results are scissor shifted by $0.6~\rm eV$ to compare with the experimental data~\cite{lautenschlager1987temperature}.}
  \label{fig:sidielectric}
\end{figure}

Figure~\ref{fig:sidielectric} compares the optical absorption spectra of bulk Si obtained using real-time TDDFT within the random-phase approximation (RPA)~\cite{bertsch2000prb}, the LRC scheme, and the MLRC scheme. Both the LRC and MLRC approaches substantially enhance the first absorption peak near the optical absorption edge, whereas the RPA considerably underestimates its intensity because it does not include the required long-range electron-hole interaction. The close agreement between the LRC and MLRC results demonstrates that the stabilizing mass term preserves the essential excitonic contribution to the optical response. These results validate our implementation and demonstrate its ability to describe excitonic features in optical spectra directly within real-time simulations at relatively low cost.

\subsection{Time-Resolved Exciton Wave Functions}

A convenient object for analysis of exciton properties and dynamics, such as dissociation rates or charge separation rates both within- and out-of-equilibrium is the time-dependent exciton wave function. The exciton wavefunction can be constructed based on the single-particle transition density matrix (TDM)~\cite{williams2021time}, as these can be shown analogous from Green's function theory~\cite{10.1063/1.4885819}. The TDM thus offers insights into the spatial distribution of a given excitation, provided in terms of single-particle excitations. For TDDFT and time-dependent Hartree--Fock (TDHF) the extracted time-dependent TDM, $\Gamma_s (\bm r,\bm r',t)$, thus encodes the weighted transition amplitudes of exciton annihilation, or hole creation, at position $\bm r$ given electron creation at position $\bm r'$~\cite{LI2011157}. 

The time-dependent Kohn-Sham TDM generally follows~\cite{williams2021time}
\begin{equation}
	\Gamma_s (\bm r, \bm r', t) =\bra{\Psi(t)}\widehat{\rho}(\bm r, \bm r' )\ket{\Psi(t)}- \bra{\Psi_0}\widehat{\rho}(\bm r, \bm r') \ket{\Psi_0},
\end{equation}
as the difference between the time-dependent and the ground state one-body density matrices. 
For real time formulations of TDDFT or TDHF one can equivalently evaluate
\begin{equation} \label{eq:tdexciton}
	\Gamma_s (\bm r, \bm r', t) = \sum_i^{\text{occ}} \left[  \varphi_{i} \left(\bm r ,t \right) \varphi_{i}^{\ast} \left(\bm r' ,t \right)  - \varphi_{i} \left(\bm r\right) \varphi_{i}^{\ast}\left(\bm r' \right) \right],
\end{equation}
with time-dependent single-particle orbitals $\varphi_{\sigma,i} \left(\bm r ,t \right)$, that is propagated from their ground state orbitals $\varphi_{\sigma,i} \left(\bm r ,t=0\right)=\varphi_{\sigma,i} \left(\bm r \right)$. In periodic systems, the usage of Bloch's theorem allows for expressing the TDM in terms of valence band $(v)$ states with crystal momenta $
\bm k$ sampling the first Brillouin zone, as
\begin{equation} \label{eq:tdexciton2}
	\Gamma_s (\bm r, \bm r', t) = \sum_{v,\bm k} \left[  \varphi_{v,\bm k} \left(\bm r ,t \right) \varphi_{v,\bm k}^{\ast} \left(\bm r' ,t \right)  - \varphi_{v,\bm k} \left(\bm r\right) \varphi_{v,\bm k}^{\ast}\left(\bm r' \right) \right] e^{i \bm k \cdot \left(\bm R - \bm R'\right)}.
\end{equation}
We note that the exciton wavefunction is not restricted to the first Brillouin zone, but occupies the full crystal volume, defined by the number of $k$-points times the unit cell volume. 
Related observables are the time-dependent density response, $\delta n(\bm r,t) = n(\bm r,t) - n(\bm r,t=0)$, which arise as the diagonal of the TDM $\Gamma_s (\bm r, \bm r, t) = \delta n(\bm r,t)$, and whose integral remains invariant $\int d\bm r \left[ \Gamma_s (\bm r,\bm r, t) \right] = 0$. A complete analysis of the exciton wavefunction, requires visualizing the full map of both electron and hole coordinates. For many systems, the structure of the TDM is often dominated by the near-diagonal contributions, signifying a close relative electron-hole distance, as expected from bound particles. Thus, the time-dependent TDM can often be conveniently represented as simply a function of $\bm r$ and $t$ with respect to a fixed hole at some atomic center, or in terms of a relative coordinate $\bm R_\text{r}$ compared to a center of mass $\bm R_\text{cm}$ coordinate. To distinguish the nature of the excitation, one can evaluate the extent of the exciton first moment, or the exciton radius $R$\cite{Prete2020,PhysRevA.109.063104}
\begin{equation}
	R = \frac{ \int d\bm r \abs{\bm r} \abs{\Gamma_s (\bm r,0,t)}^2 }{ \int d\bm r \abs{\Gamma_s (\bm r,0,t)}^2 } \,, \label{eq:1}
\end{equation}
which arises normalized to the magnitude of the excitation. For materials with predominantly strongly-bound excitons, $R$ attains values of a few \AA, due to the vanishing relative electron-hole distance, whereas a material with predominantly free carriers have $R$ increasing to the length scale of the crystal. 

\begin{figure}
\centering
\includegraphics[width=0.7\textwidth]{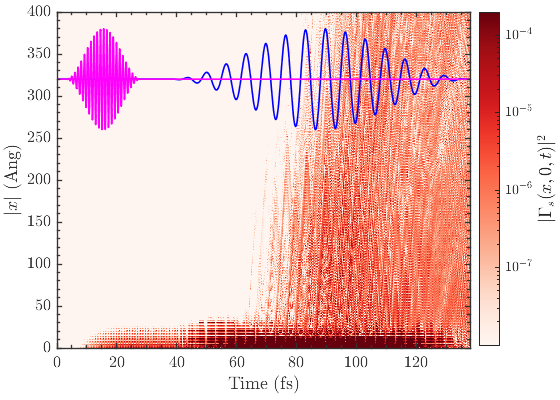}
\caption{Norm-squared exciton wavefunction driven in real-time and real-space for a hydrogen chain model system in TDHF. A $25$ fs exciton-resonant $\omega = 3.86$ eV pump of $5$~MW/cm$^{-2}$ is illustrated with magenta, generating a localized exciton wavefunction during the initial timesteps. Hereafter, a strong $1$~TW/cm$^{-2}$ IR $2000$ nm field drives exciton dissociation to either subsequent recombination or formation of delocalized free carriers during its $100$ fs duration.}\label{fig:exciton}
\end{figure}

To illustrate the exciton wavefunction visualization offered by the time-dependent TDM, in Fig.~\ref{fig:exciton} a strong-field-driven exciton is considered in a one-dimensional hydrogen chain, described using TDHF. Details on the model and simulation parameters are given in the caption and in Ref.~\cite{PhysRevA.109.063104}. Extending upon the visualization of Ref.~\cite{williams2021time}, Fig.~\ref{fig:exciton} showcases a pump-probe simulation for which a pump laser generates an exciton population, and the intense probe pulse drives it on an ultrafast femtosecond timescale. During the initial $25$ fs, the exciton-resonant pump initiates a strongly bound localized exciton, with a first moment $R$ remaining restricted to a few~\AA. After a brief delay, in which the exciton remains localized, a strong probe pulse in the IR regime is added. It generates additional excitons, observed through the increase in the localized time-dependent TDM magnitude, and, at sufficient intensities, drives exciton dissociation, as observed by the bursts of exciton wavefunctions which delocalize across hundreds of~\AA\, originating at each local field maximum. Due to the oscillatory behavior of the strong IR field, some of the dissociated exciton population can return during a following subcycle and recombine to once again form localized electron-hole pairs, while other parts of the time-dependent TDM are driven to dissociate across the full lattice, thus forming an excitation of a free-carrier nature.

\section{Dissipative dynamics in real-time TDDFT}
\label{sec:dissipative}
When a material is exposed to an ultrafast laser pulse, its electronic system is driven far from equilibrium, creating highly excited carrier distributions. While the initial response is determined by the coherent interaction of the electrons with the external electromagnetic field, the subsequent dynamics is governed by scattering mechanisms that redistribute the energy and momentum of the hot carriers. Electron--phonon (e--ph) scattering is one of the primary channels for carrier relaxation, energy dissipation, and the decay of quantum coherence in many semiconductors and quantum materials~\cite{giustino2017electron,bernardi2014ab}. These processes determine how a photoexcited electronic system evolves toward thermal equilibrium and are central to the interpretation of ultrafast optical and photoemission spectroscopy.

The conventional formulation of real-time TDDFT evolves the electronic system unitarily under the TDKS Hamiltonian $\hat H_{\mathrm{KS}}(t)$ and therefore cannot describe irreversible carrier relaxation and decoherence induced by e--ph scattering in a purely electronic propagation. To overcome this limitation, we implemented in \octopus a dissipative extension of real time TDDFT based on an open-system description of the electronic subsystem~\cite{nie2026first}, starting from the total Hamiltonian 
\begin{equation}
  \hat{H}_{\rm tot} = \hat{H}_{\rm PF} + \hat{H}_{\rm ph} + \hat{H}_{e\text{-ph}},
\end{equation}
where $\hat{H}_{\rm PF}$ is the PF Hamiltonian of Eq.~(\ref{eq:pauli-fierz}), %
\begin{equation}
  \hat{H}_{\rm ph} = \sum_{\mathbf{q},\lambda}
    \hbar \omega_{\lambda}(\mathbf{q})\,\hat{b}_{\mathbf{q}\lambda}^\dagger\hat{b}_{\mathbf{q}\lambda}
\end{equation}
is the phonon bath, and
\begin{equation}
  \hat{H}_{e\text{-ph}} = \sum_{\mathbf{k},\mathbf{q}}\sum_{mn,\lambda}
    g_{mn}^\lambda(\mathbf{k},\mathbf{q})\,
    \hat{c}_{m\mathbf{k}+\mathbf{q}}^\dagger\hat{c}_{n\mathbf{k}}
    \!\left(\hat{b}_{\mathbf{q}\lambda}+\hat{b}_{-\mathbf{q}\lambda}^\dagger\right)
\end{equation}
is the e--ph coupling with matrix elements 
$g_{mn}^\lambda(\mathbf{k},\mathbf{q})$. The coherent evolution generated by $\hat H_{\mathrm{KS}}(t)$, where 
the electromagnetic field is treated classically, is supplemented by a collision operator $\mathcal I_{\mathrm{e-ph}}$ that accounts for phonon-mediated scattering. Although we focus here on a first-principles e--ph collision integral, the propagation framework is sufficiently general to accommodate other dissipative collision operators, including phenomenological Lindblad dissipator~\cite{heras2026pulse}. The resulting evolution of the reduced one-particle density matrix of the auxiliary Kohn-Sham system, $\hat\rho^{(1)}(t)$, is
\begin{equation}
\frac{\partial \hat\rho^{(1)}(t)}{\partial t}
= -i\left[ \hat H_{\mathrm{KS}}(t), \hat\rho^{(1)}(t) \right]_{\mathrm{RS}}
+ \mathcal I_{\mathrm{e-ph}} \left[ \hat\rho^{(1)}(t) \right]_{\mathrm{AS}},
\label{eq:open_density_matrix}
\end{equation}
where the first term on the right-hand side describes the conventional coherent real time TDDFT propagation, while $\mathcal I_{\mathrm{e-ph}}$ describes the phonon-mediated scattering processes that remain after the phonon degrees of freedom have been integrated out.

\subsection{Mixed real-space and active-space propagation}
The numerical realization of Eq.~\eqref{eq:open_density_matrix} combines two representations: the coherent evolution generated by $\hat H_{\mathrm{KS}}(t)$ is performed on the real-space (RS) grid using the standard real time TDDFT propagators available in \octopus, whereas the dissipative e--ph collision integral is evaluated in a reduced active space (AS).

\begin{figure}[!htbp]
  \centering
  \includegraphics[width=0.75\linewidth]{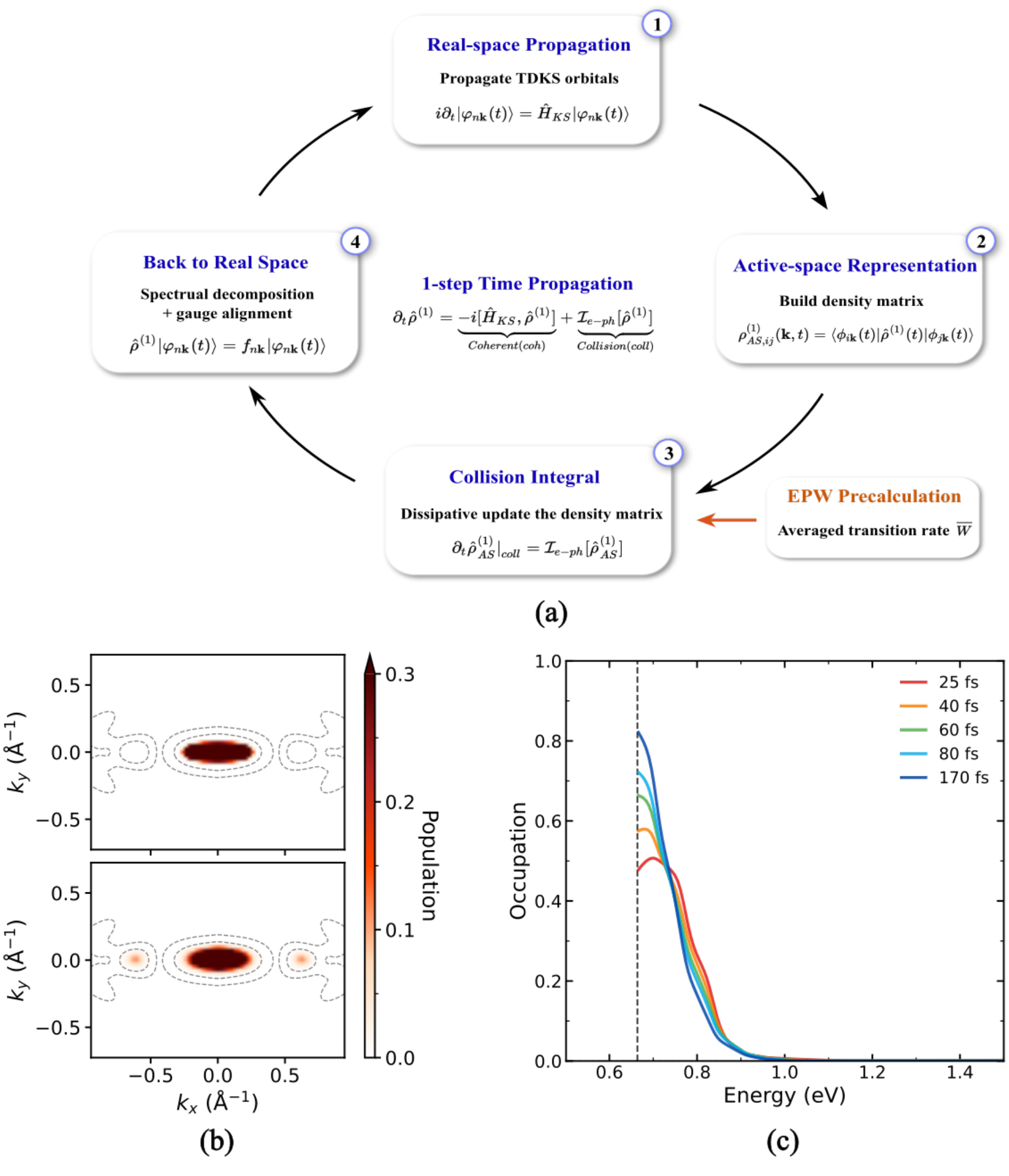}
  \caption{Implementation and application of the collision-integral propagation. (a) Schematic representation of the mixed RS/AS workflow within real time TDDFT. The dissipative contribution to the time evolution of the AS density matrix is evaluated in \octopus through the collision integral in Eq.~(\ref{eq:collision_integral}), using average transition rates in Eq.~(\ref{eq:coarse_rates}) extracted from the modified EPW\cite{epw} calculations. (b) Excited-electron population in the first Brillouin zone at 170~fs. The upper and lower panels show the momentum-space distributions calculated without and with the collision integral, respectively. Dashed lines indicate constant-energy contours. (c) Evolution of the instantaneous energy-resolved electron occupations during thermalization. The vertical dashed line indicates the conduction-band minimum.
  }
  \label{fig:dmphosphorene}
\end{figure}

The overall propagation cycle is illustrated in Fig.~\ref{fig:dmphosphorene}(a). During each time step, the TDKS orbitals are first propagated on the RS grid under the action of $\hat H_{\mathrm{KS}}(t)$. The resulting states are then projected onto the AS, where the collision integral $\mathcal I_{\mathrm{e-ph}}$ is applied. Finally, the updated AS density matrix is transformed back into a set of TDKS orbitals that provides the initial states for the next RS propagation step. In practice, the coherent and dissipative propagators are applied sequentially using a first-order Trotter splitting. The evolution over a single time step $\Delta t$ is
\begin{equation}
    \hat{\rho}^{(1)}(t + \Delta t)
    =
    \exp\!\left(\Delta t\,\mathcal{I}_{\mathrm{e-ph}}\right)
    \,
    \mathcal{T}
    \exp\!\left[
    \int_t^{t+\Delta t}
    d\tau\,
    \mathcal{L}_{\mathrm{vN}}(\tau)
    \right]
    \hat{\rho}^{(1)}(t) +
    \mathcal{O}(\Delta t^2),
\label{eq:collision_propagate}
\end{equation}
where $\mathcal{L}_{\mathrm{vN}}(\tau) = -i[\hat H_{\mathrm{KS}}(\tau),\cdot]_{\mathrm{RS}}$ is the Liouvillian superoperator associated with the coherent real time TDDFT propagation and $\mathcal{T}$ denotes the time-ordering operator. The first-order Trotter splitting introduces an error of order $\Delta t^2$, which can be systematically reduced by decreasing the time step.

The dissipative update is performed in an AS spanned by the instantaneous adiabatic states $|\phi_{n\mathbf{k}}(t)\rangle$, defined by
\begin{equation}
\hat H_{\mathrm{KS}}(t)|\phi_{n\mathbf{k}}(t)\rangle
= \varepsilon_{n\mathbf{k}}(t)|\phi_{n\mathbf{k}}(t)\rangle.
\end{equation}
The states are obtained by diagonalizing the instantaneous Hamiltonian at each time step. This time-dependent basis follows the transient electronic structure during the laser-driven dynamics and provides a natural representation for e--ph scattering. The AS density matrix is
\begin{equation}
\hat{\rho}^{(1)}_{\mathrm{AS}}(\mathbf{k},t)
=
\sum_{mn}
\rho^{(1)}_{mn}(\mathbf{k},t)
|\phi_{m\mathbf{k}}(t)\rangle
\langle\phi_{n\mathbf{k}}(t)|,
\label{eq:active_density_matrix}
\end{equation}
where the matrix elements $\rho^{(1)}_{mn}$ describe the electronic populations and coherence within the AS.

A practical challenge arises from the different numerical requirements of the coherent and dissipative parts of the calculation. The e--ph transition rates entering the collision integral are obtained from first-principles calculations using EPW~\cite{giustino2017electron,ponce2016epw,epw}. Resolving the meV energy scale of phonons while satisfying energy and momentum conservation requires a much finer Brillouin-zone sampling than is computationally feasible for real-time propagation. We therefore map the microscopic transition rates onto the coarser $\mathbf{k}$-point mesh used in the real time TDDFT simulation. The fine mesh is partitioned into momentum clusters associated with each coarse-grid point, and the microscopic transition rates between the corresponding clusters are averaged according to
\begin{equation}
\begin{aligned}
\overline W_{(m,\mathbf k)\leftarrow(n,\mathbf k')}
&=
\frac{1}{N_{\mathbf k}N_{\mathbf k'}}
\sum_{\mathbf k_{\beta}\in\mathcal C(\mathbf k)}
\sum_{\mathbf k_{\alpha}\in\mathcal C(\mathbf k')} \\
&\quad\times\frac{2\pi}{\hbar}\sum_{\lambda}
\left|g^{\lambda}_{mn}(\mathbf k_{\alpha},\mathbf k_{\beta})\right|^2
\Big[n_{\lambda}(\mathbf q)\,
\delta\!\left(\varepsilon_{m\mathbf k_{\beta}}-\varepsilon_{n\mathbf k_{\alpha}}-\hbar\omega_{\lambda}(\mathbf q)\right)
\delta_{\mathbf k_{\alpha} = \mathbf k_{\beta}-\mathbf q} \\
&\qquad + \big(n_{\lambda}(\mathbf q)+1\big)
\delta\!\left(\varepsilon_{m\mathbf k_{\beta}}-\varepsilon_{n\mathbf k_{\alpha}}+\hbar\omega_{\lambda}(\mathbf q)\right)
\delta_{\mathbf k_{\alpha} = \mathbf k_{\beta}+\mathbf q}\Big].
\end{aligned}
\label{eq:coarse_rates}
\end{equation}
Here, $\mathcal C(\mathbf k)$ denotes the set of fine-grid points associated with the coarse-grid point $\mathbf k$. The resulting coarse-grained rates preserve the microscopic scattering channels while allowing the dissipative dynamics to be evaluated efficiently on the propagation mesh.

Using these effective transition rates in Eq.~(\ref{eq:coarse_rates}), the e--ph collision integral is evaluated within the Born--Markov approximation as
\begin{equation}
\begin{aligned}
\mathcal{I}_{\mathrm{e-ph}}[\rho^{(1)}_{mn}(\mathbf{k},t)]
={}& -(1-\delta_{mn})\frac{\gamma_m(\mathbf k,t)+\gamma_n(\mathbf k,t)}{2}\rho^{(1)}_{mn}(\mathbf k,t) \\
&+\delta_{mn}\sum_{i\mathbf k'}\Big[
\overline W_{(n,\mathbf k)\leftarrow(i,\mathbf k')}\rho^{(1)}_{ii}(\mathbf k',t)\big(1-\rho^{(1)}_{nn}(\mathbf k,t)\big) \\
&\qquad\qquad-\overline W_{(i,\mathbf k')\leftarrow(n,\mathbf k)}\rho^{(1)}_{nn}(\mathbf k,t)\big(1-\rho^{(1)}_{ii}(\mathbf k',t)\big)
\Big].
\end{aligned}
\label{eq:collision_integral}
\end{equation}
Here, $m$, $n$, and $i$ label electronic bands, while $\mathbf k$ and $\mathbf k'$ denote crystal momenta. The coarse-grained transition rates $\overline W_{(m,\mathbf k)\leftarrow(n,\mathbf k')}$ encode the phonon-assisted transfer of energy and crystal momentum between electronic states. The state-dependent scattering rate entering the coherence decay is
\begin{equation}
\gamma_n(\mathbf k,t)
= \sum_{i,\mathbf k'} \Big[
\overline W_{(i,\mathbf k')\leftarrow(n,\mathbf k)}\big(1-\rho^{(1)}_{ii}(\mathbf k',t)\big)
+\overline W_{(n,\mathbf k)\leftarrow(i,\mathbf k')}\rho^{(1)}_{ii}(\mathbf k',t)
\Big].
\label{eq:gamma}
\end{equation}

The structure of Eq.~\eqref{eq:collision_integral} has the usual interpretation of a density-matrix collision integral. Its diagonal part has the gain--loss form of a Boltzmann equation and describes the phonon-assisted redistribution of carrier populations across bands and momenta, including Pauli blocking of the final states. The off-diagonal part describes the decay of interband coherence induced by the same microscopic scattering channels. The formulation therefore retains a direct connection to first-principles e--ph scattering theory while extending the population-only Boltzmann picture to include coherence decay.

After the dissipative update, the AS density matrix is diagonalized,
\begin{equation}
\hat{\rho}^{(1)}(t+\Delta t)
|\varphi_{n\mathbf{k}}(t+\Delta t)\rangle
=
f_{n\mathbf{k}}(t+\Delta t)
|\varphi_{n\mathbf{k}}(t+\Delta t)\rangle,
\label{eq:rho_diagonalization}
\end{equation}
yielding the natural orbitals $|\varphi_{n\mathbf{k}}(t+\Delta t)\rangle$ and their nonequilibrium occupations $f_{n\mathbf{k}}(t+\Delta t)$. Because the eigenvectors are defined only up to a unitary transformation, they are gauge aligned with the TDKS orbitals from the previous time step to ensure a smooth temporal evolution. Specifically, $|\psi_i^{\,\mathrm{new}}\rangle=\sum_j U_{ji}|\varphi_j\rangle$, where the unitary matrix $U$ is chosen to maximize the overlap between successive time steps. The reconstructed orbitals $\{|\psi_i^{\,\mathrm{new}}\rangle\}$ are then mapped back to the RS representation and used as the initial TDKS orbitals for the next coherent propagation step, completing the cycle shown in Fig.~\ref{fig:dmphosphorene}(a).

To demonstrate the collision-integral propagation discussed above, we simulate ultrafast carrier relaxation in phosphorene. Figure~\ref{fig:dmphosphorene}(b) compares the excited-electron population in the first Brillouin zone at 170~fs with and without the collision integral. In the absence of the collision term, resonant excitation promotes electrons only near the $\Gamma$ valley, as determined by energy conservation. Including the collision integral enables intervalley scattering and produces a broader momentum distribution.

Figure~\ref{fig:dmphosphorene}(c) follows the thermalization through the instantaneous energy-resolved electron occupations. At 25~fs, the excited electrons form a strongly nonequilibrium distribution slightly above the conduction-band minimum (CBM). The system subsequently cools as the population shifts toward the CBM. By 80~fs, a quasi-equilibrium distribution emerges that is well described by a Fermi--Dirac function. At 170~fs, the fitted carrier temperature is 420~K, approaching the phonon-bath temperature of 300~K. The residual difference of 120~K ($\approx 10.3$~meV) is limited by the energy resolution of the discrete $\mathbf{k}$-point sampling, here a $72 \times 60$ mesh, and decreases for denser Brillouin-zone meshes.

\section{Wannier functions}

Wannier functions (WFs) are a real-space representation of the
electronic structure complementary to the delocalized Bloch states
$\psi_{n\mathbf{k}}$ obtained from DFT. Their localized nature makes them ideally suited as a basis for constructing low-energy tight-binding Hamiltonians. 
These static low-energy lattice models derived from maximally localized Wannier Hamiltonians provide a compact real-space description of the electronic structure with hopping matrix elements and onsite energies, in which a relevant subset of bands is mapped onto localized orbitals. In equilibrium, such models are widely used to analyze band topology~\cite{wang2017first}, transport~\cite{calzolari2004ab}, magnetism~\cite{jiang2025robust}, optical response~\cite{fang2025dataset} and more~\cite{marzariMaximallyLocalizedWannier2012,marrazzo2024wannier}. By serving as input to sophisticated many-body methods, they are an important tool to study correlated-electron physics in real materials from an ab initio starting point~\cite{lechermannDynamicalMeanfieldTheory2006,imadaElectronicStructureCalculation2010}. \newline
Wannier Hamiltonians are also commonly used as the starting point for non-equilibrium simulations, for example by coupling the hopping
amplitudes to an external vector potential through a Peierls substitution~\cite{luttingerEffectMagneticFieldPeriodicPotential1951,peierlsZurTheorieDiamagnetismus1933a,chen2025generalized}. This assumes that the underlying lattice basis remains static, and that the non-equilibrium response is
adequately captured through occupations, phases, and external-field dressing of fixed matrix elements. \newline
Accessing time-dependent WFs opens a route to studying how non-equilibrium dynamics affect Wannier Hamiltonians and the lattice models derived from them. A main goal is to extend the treatment of strongly correlated system with advanced many-body solvers -- which is routinely done in the static case -- to the time-domain.
In particular, it allows changes in onsite energies, hopping amplitudes, and orbital character to be extracted directly from the evolving ab initio
electronic state, rather than through field-dressing of a fixed basis or the addition of phenomenological couplings. Snapshot Wannierization along a TD trajectory therefore provides a means of investigating time-resolved lattice Hamiltonians for driven materials, and
of quantifying when conventional static-model approximations remain valid. Moreover, it generates an intuitive real-space picture of electron dynamics under external fields~\cite{yostPropagationMaximallyLocalized2019,souzaDynamicsBerryphasePolarization2004,leEfficientTimedependentWannier2025}.


A set of $N$ WFs can be constructed via unitary transformation of the Bloch
manifold as
\begin{equation}
    |w_{n\mathbf{R}}\rangle = \frac{V}{(2\pi)^3} \int_{\mathrm{BZ}}
    \left[\sum_{m} U_{mn}^{(\mathbf{k})} |\psi_{m\mathbf{k}}\rangle \right]
    e^{-i\mathbf{k}\cdot\mathbf{R}}\, d\mathbf{k},
\end{equation}
where $\mathbf{R}$ is a real-space lattice vector and $U^{(\mathbf{k})}$ is a
$k$-dependent unitary matrix.
The gauge freedom in choosing $U^{(\mathbf{k})}$ can be exploited to enforce
localization via the Foster-Boys criterion, which minimizes the total
real-space spread
\begin{equation}
    \Omega = \sum_n \left[
        \langle w_n | \mathbf{r}^2 | w_n \rangle
        - \langle w_n | \mathbf{r} | w_n \rangle^2
    \right],
\end{equation}
yielding maximally localized Wannier functions (MLWFs)~\cite{marzariMaximallyLocalizedGeneralized1997}. \newline
The construction of MLWFs is implemented in the \texttt{Wannier90} code, which is widely used in the electronic-structure community, and is supported by a range of DFT codes~\cite{pizziWannier90CommunityCode2020,marrazzo2024wannier}, including \octopus.
For more details on the localization procedure and implementation we refer to the \texttt{Wannier90} documentation~\cite{mostofiUserGuideWannier90,pizziWannier90CommunityCode2020}.

\subsection{\octopus integration with \texttt{Wannier90}}

A key advantage of \texttt{Wannier90} is that it is basis set agnostic. A client DFT code merely needs to supply two matrices as input: $A_{mn}^{(\mathbf{k})}$ and $M_{mn}^{(\mathbf{k,b})}$. Evaluated from the Bloch states, these matrices encode all information needed to construct MLWFs. \newline
The matrix \textbf{M} contains overlaps of periodic Bloch functions at
neighboring $k$-points,
\begin{equation}
    M_{mn}^{(\mathbf{k,b})} = \langle u_{m\mathbf{k}} | u_{n,\mathbf{k+b}}
    \rangle,
\end{equation}
where the $u_{n\mathbf{k}}$ are the periodic parts of the Bloch states and $\mathbf{b}$ is a vector connecting neighboring $\mathbf{k}$-points in the Brillouin zone. \newline
The matrix \textbf{A} contains projections of the Bloch states onto a set
of localized trial orbitals $g_n$,
\begin{equation}
    A_{mn}^{(\mathbf{k})} = \langle \psi_{m\mathbf{k}} | g_n \rangle,
\end{equation}
and serves as the initial guess for the unitary transformation $U^{(\mathbf{k})}$. The choice of trial orbitals is crucial for convergence of the localization algorithm; they are typically chosen to reflect the chemical character of the target WFs and preserve symmetries of the crystal. \newline
An alternative approach for constructing \textbf{A} is the SCDM
algorithm (Selected Columns of the Density Matrix) ~\cite{damleCompressedRepresentationKohn2015,damleDisentanglementEntanglementUnified2018}, which selects an optimal set
of localized columns from the single-particle density matrix 
via pivoted QR
decomposition. 
This method is a localization method that produces Wannier functions, but not maximally localized functions. It requires no chemical intuition and is particularly well-suited for high-throughput calculations where manual selection of trial orbitals is impractical.  \newline
In \octopus, construction of \textbf{A} supports both atom-centered trial orbitals and the SCDM method.
\octopus can additionally evaluate the resulting MLWFs on a real-space grid and export them for further analysis, or use them for defining a localized subspace, for DFT+U calculations~\cite{wcbz-lbr1}. Entangled band manifolds are also supported, and bands far from the energy window of interest can be excluded to reduce computational cost. \newline
All this functionality is bundled in the \texttt{oct-wannier90} utility, which is run as a post-processing step for a completed ground-state or time-dependent calculation. 

\begin{figure}[h]
    \includegraphics[width=0.48\textwidth]{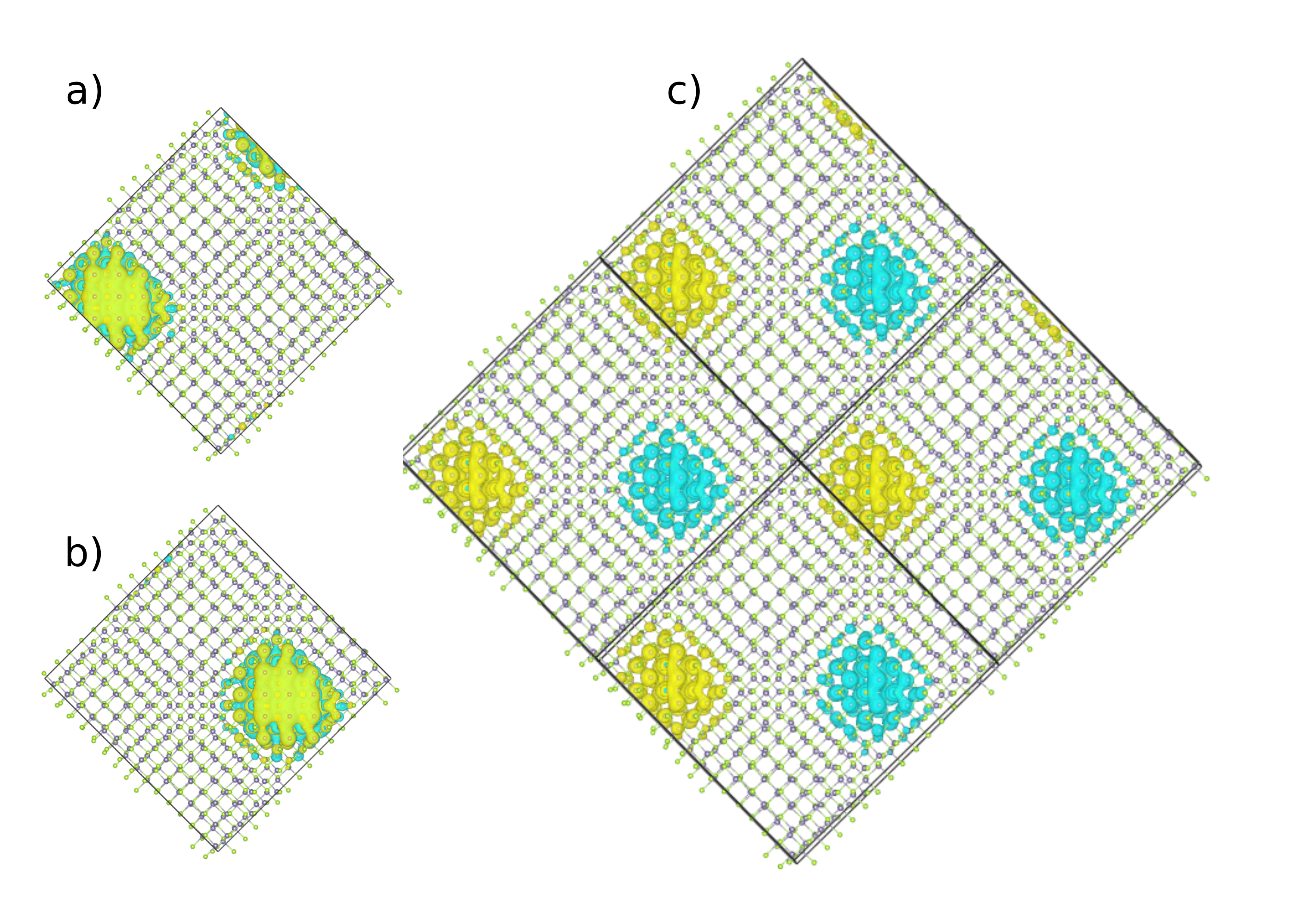}
    \caption{Wannier states a)–b)
corresponding to the two flat bands of a checkerboard $2\times1$ supercell of twisted bilayer GeSe. (c) The corresponding magnetization density obtained from DFT+U+V at half filling. Taken from Ref.~\cite{wcbz-lbr1}}\label{fig:dft_uv_mlwf}
\end{figure}

The capability of \octopus in combination with \texttt{Wannier90} is demonstrated in Fig.~\ref{fig:dft_uv_mlwf}. It illustrates a sequence of calculations performed with \octopus, in conjunction with \texttt{Wannier90}, to describe twisted bilayer physics. First, a LDA calculation is performed on the system, here twisted bilayer GeSe. The flat band of this system is then wannierized using the \texttt{oct-wannier90} utility and \texttt{Wannier90}. Finally, the Wannier states are constructed on the real-space grid and then used to build the localized subspace on which on-site Hubbard $U$, and intersite $V$ are added at the DFT+U+V level~\cite{PhysRevB.102.155117}. At half filling, this leads to a checker-board pattern for the magnetization in the   doubled unit cell, as predicted by the square 2D Hubbard model. The interaction parameters are computed ab initio and self-consistently, based on the screened Coulomb matrix elements evaluated in the Wannier basis. More details can be found in Ref.~\cite{wcbz-lbr1}.
Other interesting applications include for instance the modification of the Fermi surface caused by light-induced band renormalization in MoTe2, which some of us demonstrated using \octopus and \texttt{Wannier90}~\cite{beaulieu2021ultrafast}. These examples demonstrate how the \texttt{Wannier90} integration in \octopus can be used for describing systems as complex as twisted bilayer systems with hundreds of atoms, or light-induced non-equilibrium dynamics in correlated materials.


\subsection{Time-dependent Wannier functions}

\begin{figure}[h]
    \centering
    \begin{subfigure}[b]{0.48\textwidth}
        \includegraphics[width=\linewidth]{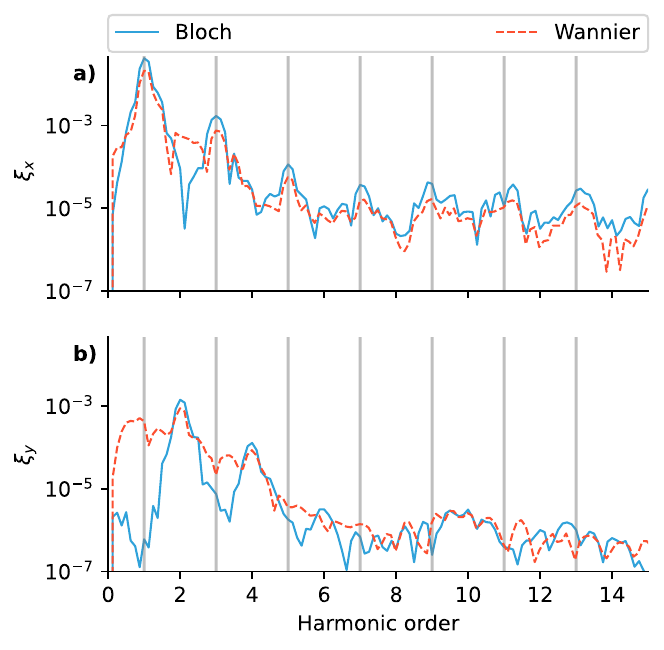}
        \caption{}\label{subfig:tdwannier_hBN_hhg}
    \end{subfigure}
    \hfill
    \begin{subfigure}[b]{0.48\textwidth}
        \includegraphics[width=\linewidth]{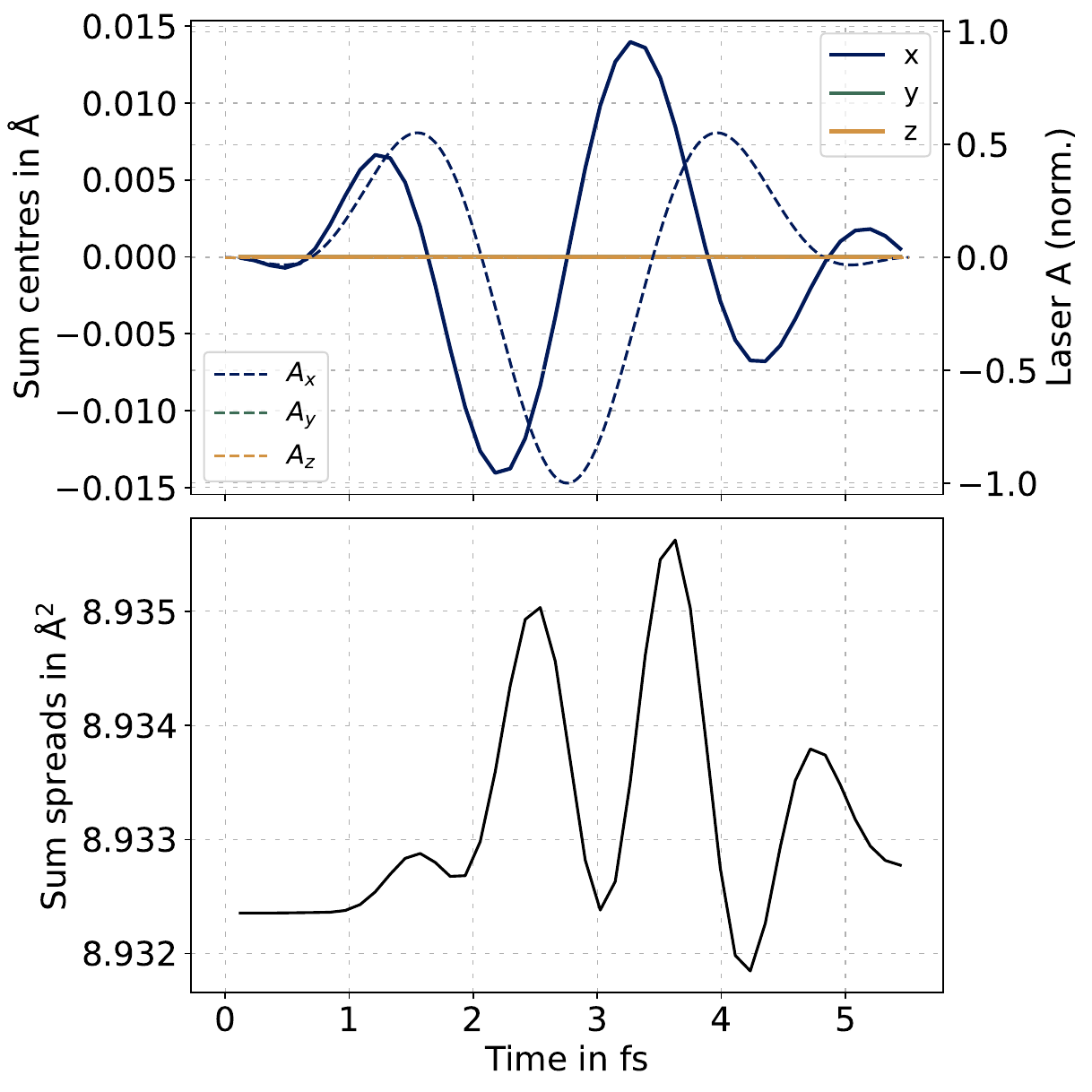}
        \caption{}\label{subfig:tdwannier_silicon}
    \end{subfigure}
    \caption{Application of time-dependent Wannier functions.\ a) High-harmonic spectra from time-dependent Wannier function approach (red) applied to monolayer hBN, compared with full TDDFT calculation (blue). The high-harmonic yield $\xi(t) = \frac{d\mathbf{J}(t)}{dt}$ is extracted from the WF centers following the modern theory of polarization~\cite{souzaDynamicsBerryphasePolarization2004}. Taken from Ref.~\cite{leEfficientTimedependentWannier2025}. b) Evolution of Wannier centres and spreads in bulk silicon under a weak laser pulse. Localization using \texttt{libwannier90} yields minimized spreads.}
    \label{fig:tdwannier}
\end{figure}

Here we discuss two approaches for time-dependent WFs implemented in \octopus. When localization is not required, \octopus supports the construction of time-dependent Wannier functions by applying the initial Wannier gauge, $U (t = 0)$, to the
propagated time-dependent Kohn-Sham states \cite{leEfficientTimedependentWannier2025}. The core idea is to build $U(t)$ from $U(t=0)$ by accumulating the phase of the time-dependent Kohn-Sham states:
\begin{equation}\label{eq:tdwannier_phase}
    U_{mn}^{(\mathbf{k})}(t) = U_{mn}^{(\mathbf{k})}(t=0) f_{m\mathbf{k}}^{\text{int}}(t) ,
\end{equation}
where the phase $f^{\text{int}}_{m\mathbf{k}}(t)$ is given by
\begin{equation}
    f_{m\mathbf{k}}^{\text{int}}(t) = e^{\frac{i}{\hbar}\int_0^t \epsilon_{m\mathbf{k}}(\tau)\, d\tau}
\end{equation}
and then transforming the time-dependent Kohn-Sham states with $U(t)$ to obtain the time-dependent WFs. This provides an efficient real-space representation for analyzing observables, such as polarization currents and high-harmonic spectra, over time scales during which the propagated Wannier functions
remain sufficiently localized as to retain a clear physical interpretation. Applied to monolayer hBN, this method reproduced high-harmonic spectra in close agreement with a full-fledged TDDFT calculation~\cite{leEfficientTimedependentWannier2025}, see Fig.~\ref{subfig:tdwannier_hBN_hhg}. \newline
Additionally, we include the option to maximally localize the time-dependent Kohn–Sham states at any point along a TD trajectory by interfacing \octopus with the \texttt{Wannier90 v4} library (here termed \texttt{libwannier90}~\cite{wannier90_github}). Prior to the availability of this library interface, such functionality would have required either
cumbersome file-based I/O with the standalone \texttt{Wannier90} program, or a full internal implementation of the Wannierization procedure. \newline
Following construction of a set of Wannier functions at t=0, each successive Wannierization is seeded by the U matrix of the preceding Wannierization, with Eq.~\eqref{eq:tdwannier_phase} applied to account for phase accumulation.
Since the localization procedure can become computationally expensive for dense time grids and large systems, snapshot Wannierization is performed only at selected time steps. \newline
In Fig.~\ref{subfig:tdwannier_silicon} we show the evolution of the sum of Wannier centres and spreads for bulk silicon under the influence of a weak laser pulse with $\lambda=826.6 \text{nm}$ and intensity $I=10^{9} \frac{\text{W}}{\text{cm}^2}$, using a $24\times24\times24$ k-point grid, spacing of $0.3 \text{a.u.}$ and $\Delta t=0.05\text{a.u.}$ as timestep. It shows the expected oscillation of the Wannier centres which is the origin of high-harmonic generation, following modern theory of polarization~\cite{souzaDynamicsBerryphasePolarization2004}. In addition, the spreads remain minimal due to the \texttt{libwannier90} localization.

\section{Improved Algorithms}

\subsection{Chebyshev-Filtered Subspace Eigensolver}

A major update in Octopus 17 is the change in our default eigensolver to the Cheybshev polynomial filtering method. Chebyshev polynomials of the first kind, $T_n(x)$, are recursively defined functions that grow exponentially outside the interval $[-1,1]$. This property makes them well suited to constructing spectral filters which, when applied iteratively to a trial subspace, amplify the desired part of the spectrum. 

The application of an affine mapping shifts and scales the Hamiltonian such that the unwanted spectral range lies within $[-1,1]$, while the targeted low-energy states lie outside this interval, and are therefore selectively enhanced. The filtered vectors then span an approximation to the invariant subspace associated with these states. To obtain the true eigenstates of the Hamiltonian, rather than linear combinations of them, a rotation of the states in the filtered subspace is performed, as usual for iterative eigensolvers\cite{zhou2011bounding, zhou2014chebyshev}.

\subsubsection{Estimating Spectral Bounds}

The initial spectral bounds are estimated using a $k$-Lanczos algorithm, which is equivalent to a Lanczos decomposition of the Hamiltonian into a $k \times k$ tridiagonal matrix, $T_k$, starting from a random vector. The upper and lower bounds are defined as:
\begin{equation}
b_{\mathrm{up}} = \lambda_{\max}(T_k) + |e_k^T z| \, \|f\|_2 ,
\end{equation}
and
\begin{equation}
e_{\min} = \lambda_{\min}(T_k) - |e_k^T z| \, \|f\|_2 ,
\end{equation}
respectively, where $\lambda_i(T_k)$ are the eigenvalues of $T_k$, $|e_k^T z|$ is the last component of the eigenvector associated with $\lambda(T_k)$, and $f$ is the residual. The limits are defined according to \cite{zhou2011bounding} rather than the conservative analogues given by \cite{zhou2014chebyshev}. Since $|e_k^T z| \leq 1$, this shifts the lower limit downward and the upper limit upward, making them safer but less tight.

By default, we define the $k$-Lanczos algorithm to run until the residual correction to the upper spectral limit is below $2\%$ of the current upper-bound estimate:
\begin{equation}
\frac{|e_k^T z_{\max}| \, \|f\|_2}{b_{\mathrm{up}}} < 0.02 ,
\end{equation}
or until a hard limit of 30 steps is reached. We find that evaluating this criterion with an empirically chosen tolerance gives a simple, robust heuristic for indicating when the Lanczos upper limit is stable enough for filtering. The bound estimation assumes a state-independent Hamiltonian, which is valid for conventional and generalized-KS hybrid functionals. However, we note that this is insufficient for orbital-dependent operators (e.g. RDMFT), and requires extending to per-state estimation, or some cutoff-based safe upper bound. 
During the self-consistent cycle, we update the upper bound of the Hamiltonian using this algorithm, and the search vector is recycled from the previous step.

\subsubsection{Refinement of Spectral Bounds and the Chebyshev Filtering Algorithm}

If an LCAO calculation was performed to obtain a guess of the eigenvalues and eigenvectors, the lower bound of the Hamiltonian and the energies of the highest desired states are used to define the affine transformation of the filtering. Otherwise, these quantities are estimated from the lower and higher bounds obtained with the $k$-Lanczos algorithm, following the method originally proposal in Ref.~\cite{zhou2014chebyshev}. In this approach, which can be used in all cases, Octopus diagonalizes the Hamiltonian in the initial-guess subspace to obtain the corresponding Ritz values. The lower bound of the filter is then set to the largest Ritz value, whilst the scaling reference is set to the smallest Ritz value. If the largest Ritz value exceeds the upper bound estimated from the Lanczos calculation, the upper bound is recomputed.

This defines the interval $[a,b]$ used by the Chebyshev filter, where $a$ is the current estimate of the cutoff between the wanted and unwanted parts of the spectrum, and $b=b_{\mathrm{up}}$ is the estimated spectral upper bound. The scaling reference $a_L$ is chosen from the wanted part of the spectrum, typically as the smallest Ritz value.

During this first SCF step, the subspace is then iteratively filtered and re-diagonalised until the residual norms, $\|H\psi_i - \lambda_i \psi_i\|$,  fall below the convergence tolerance. This produces approximate eigenvectors that are sufficiently accurate to initialise the subsequent SCF iterations. The procedure is described in Alg. \ref{alg:firstscf_iterative_chebyshev_filter}.

\begin{algorithm}
\caption{$[\Psi] = \text{firstscf\_iterative\_chebyshev\_filter}$}
\label{alg:firstscf_iterative_chebyshev_filter}
\begin{algorithmic}[1]

\State \textbf{Initialization:}
\Statex \quad \textbf{k-Lanczos:} Obtain initial $b_{\rm up}$.
\Statex \quad \textbf{Diagonalize Hamiltonian in the subspace}
\State \textbf{Filtering Loop} (for $i = 1$ to $N_{\rm max}$):
\Statex \quad \textbf{Bound Update:} If $\lambda_{\rm max} \ge b_{\rm up}$, compute a new estimate of $b_{\rm up}$. 
\Statex \quad \textbf{$a_L = \lambda_{min}$}; \textbf{$b_{\rm low} = \lambda_{\rm max}$} 
\Statex \quad \textbf{Chebyshev filter of $\Psi$}: Filter of order m, scaled with $a_L$.
\Statex \quad \textbf{Normalization of $\Psi$}
\Statex \quad \textbf{Subspace Update:}
$G = \Psi^\dagger (H \Psi)$, $S = \Psi^\dagger \Psi$. Solve $G\, Q = S\, Q\, D$. $\Psi = \Psi\, Q$ .
\Statex \quad \textbf{Convergence Check:} Exit if the norm of the residuals
  $\|H\psi_i - \lambda_i \psi_i\| <$ tolerance.
\end{algorithmic}
\end{algorithm}

Following this, and for all subsequent self-consistent steps, Octopus uses the scaled Chebyshev filtering algorithm described in Ref.~\cite{zhou2014chebyshev}. The scaled recurrence avoids explicitly forming large Chebyshev polynomial values while retaining the amplification of eigen-components lying below $a$.

Importantly, for efficiency, we modified the original algorithm: states are normalized at the end of the filtering step, but no orthogonalization is performed. Instead, for all steps, we replace the usual subspace diagonalization with a generalized eigenvalue problem. This combines two $O(N_p^3)$ operations, orthonormalization and subspace diagonalization, into a single operation, making the overall algorithm more efficient. For numerical stability, the states are normalized after filtering to preserve the conditioning of the generalized eigenvalue problem.

\subsubsection{Dynamic Determination of the Chebyshev Polynomial Degree}

Rather than applying Chebyshev polynomials of the same degree to all the states, and for all self-consistent iterations, \octopus determines it  dynamically, based on the ``Chebyshev oracle'' of Levitt and Torrent\cite{levitt2015parallel}. The oracle predicts the residual of a given state after filtering at degree $n$ using the residual of this state for the last self-consistent iteration. The degree $n$ is increased until one of the following criteria is met, or the maximum degree is reached (25 by default):
\begin{itemize}
    \item Conditioning guard: $|T_n(\epsilon_1)| > 0.01/\epsilon_{\text{machine}}$, where $\epsilon_1$ is the lowest eigenvalue. A polynomial order that amplifies the bottom of the spectrum by $\sim 1/\epsilon$ ill-conditions the overlap matrix $S$ in the subsequent generalized Rayleigh–Ritz step.
    \item Don't over-filter: The residual should not reduce by more than an order of magnitude per SCF step. This avoids over-converging eigenstates when the density is far from the convergence. 
    \item Target reached: the predicted residual falls below the requested tolerance. 
\end{itemize}
The oracle therefore yields an approximately optimal polynomial degree for each state, and a single value is assigned to each block of states, as the SCF algorithm performs state-batched operations. While this approach inherently creates load imbalance, it also significantly reduces the number of iterations required during the first steps of the self-consistent calculation. For spinor calculations, however, we found the oracle to be detrimental, as high-quality eigenvectors are required to stabilize clean spin textures.

\subsection{Kerker Preconditioning}

In the SCF procedure, preconditioning is typically used to suppress long-wavelength changes in the electron density, stabilising the calculation and improving convergence. In a plane-wave basis, this corresponds to applying a weighting function to the density or potential residual during the mixing procedure \cite{kresse1996efficient}. The density mixing, this typically looks like, for Broyden mixing,
\begin{equation} \label{eq.rho_kerker_in_G}
\rho_{in}^{m+1}(\mathbf{G}) = \sum_i^{m} \alpha_i \rho_{in}^{i}(\mathbf{G}) + P(\mathbf{G}) \sum_i^m \alpha_i R_\rho^i(\mathbf{G}) \,,
\end{equation}
where $R_\rho^i(\mathbf{G})$ is the density residual. Here, $\mathbf{G}$ is a reciprocal lattice vector, $\alpha_i$ weights the contributions from the SCF history, $m$ is the current SCF iteration, and $P(\mathbf{G})$ defines the preconditioning function. The well-known Kerker preconditioner is given by
\begin{equation}
P(\mathbf{G}) = a \frac{|\mathbf{G}|^2}{|\mathbf{G}|^2+\lambda^2}.  
\end{equation}
The preconditioner is defined such that $\lim_{|\mathbf{G}| \ll \lambda} P(\mathbf{G})= 0$, suppressing long-wavelength components of the density residual that could otherwise induce charge sloshing. The coefficients $a$ and $\lambda$ are  empirically chosen parameters, with $\lambda$ determining which values of $|G|$ are damped.

Octopus implements a real-space analogue of \eqref{eq.rho_kerker_in_G}. Following \cite{shiihara2008real}, we start from the modified Poisson equation used by Manninen et al.\cite{manninen1975electrons}, written in terms of the input and output potentials of the SCF procedure, and the potential residual, $R_v = v_{out} - v_{in}$. This leads to a real-space Broyden mixing scheme in terms of the density:
\begin{equation} 
\rho_{in}^{m+1}(\mathbf{r}) = \sum_i^{m} \alpha_i \rho_{in}^{i}(\mathbf{r}) + a (\mathbf{L} - \lambda^2 \mathbf{I})^{-1} \mathbf{L} \sum_i^m \alpha_i R_\rho^i(\mathbf{r}).  
\end{equation}
where  $\mathbf{L}$ is the discretized Laplacian. In practice, the inverse of $(\mathbf{L} - \lambda^2 \mathbf{I})$ is not formed explicitly. Instead, one expresses the preconditioned residual, $\mathbf{R}^\prime$ as the product of the preconditioner and the unscreened residual, and rearranges to give an Helmhotz equation:
\begin{equation} 
(\mathbf{L} - \lambda^2 \mathbf{I}) \mathbf{R}^\prime = \mathbf{L} \mathbf{R}.  
\end{equation}
The application of the real-space Kerker preconditioner is therefore reduced to solving a sparse linear system for the preconditioned residual, $\mathbf{R}^\prime$ and then to use it in the usual Broyden mixing expression, or other mixing schemes supported by \octopus.

\subsection{Linear and improved tetrahedron methods}

In periodic systems the crystal momentum $\bm{k}$ that is defined to be in the first Brillouin zone (BZ) is a good quantum number. Accordingly, the eigenstates $\phi_{\bm{k}n}^\sigma$ and eigenenergies $\epsilon_{\bm{k}n}^\sigma$ can be labelled by the crystal momentum $\bm{k}$ in addition to spin flavor $\sigma$ and band index $n$.  To obtain expectation values derived from those quantities often a sum over the $\bm{k}$ points has to be evaluated.  Among the various observables of interest are the total energy, the electron density, the density of states, and response functions.

In an infinite solid, the $\bm{k}$ points are dense and the summation is replaced by an integral with a distribution function, for electrons the Fermi-Dirac function.  However, on a computer the simplest way to evaluate this integral is by sampling points, e.g., according to a special points method, like the one introduced by Monkhorst and Pack~\cite{Monkhorst1976}, and summing their contributions.  Sometimes the $\bm{k}$ mesh required for good convergence needs to be extremely fine, which can become computationally prohibitive.  Another possibility is the smearing method~\cite{Fu1983, Methfessel1989} in which the Fermi-Dirac function is replaced by a smeared out variant that doesn't have a sharp step, but this introduces another parameter, the broadening width for which convergence needs to be evaluated.

Yet another possibility is the linear tetrahedron method~\cite{Jepsen1971, Lehmann1972, Rath1975, MacDonald1979} which is a special case of the more general linear simplex method in 3 dimensions.  In this method, the  first BZ mesh on which the function is sampled is divided into simplices (line segments in 1D, triangles in 2D, or tetrahedra in 3D).  Inside each simplex the real shape of the function is not known, because it is only sampled at the vertices, so it is assumed that linear interpolation is sufficient.  The linearly interpolated function can then be integrated analytically.  At zero temperature the Fermi-Dirac function is a step function and therefore there is a sharp cutoff.  With the simplex method, the location of the cutoff inside the tetrahedron can be estimated and the remaining volume calculated much more accurately than in smearing techniques.  Because of linear interpolation the resulting expressions for the integral over the simplex only depend on the values of the function at the vertices of the same tetrahedron.  Another advantage of the tetrahedron method is that it can be written as a weighted sum where the weights can be precomputed
\begin{equation}
  \frac{\Omega}{(2\pi)^d} \int \mathrm{d}^dk \; f(\epsilon_{\bm{k}}) \mathop{{}\Theta}(\epsilon_{\mathrm{F}} - \epsilon_{\bm{k}})
  = \frac{1}{N} \sum_{i} w_i f(\epsilon_{\bm{k}_i})\,,
\end{equation}
with the volume of the first BZ $\Omega$, the Fermi energy $\epsilon_{\mathrm{F}}$, the number of vertices $N$, and weights determined by the linear simplex method $w_i$.

One extension to the regular simplex method is to apply Bl\"ochl's correction~\cite{Bloechl1994, Lee2002} to correct for the neglect of curvature in the linear interpolation.  This correction corresponds to a surface integral over the Fermi surface within the tetrahedron which can be represented as a density of states contribution
\begin{equation}
  d w_i = \sum_T D_T(\epsilon_F) \frac{1}{2 (d + 1)(d + 2)} \sum_{j=1}^{d+1} (\epsilon_j - \epsilon_i)\,,
\end{equation}
with the density of states contribution $D_T(\epsilon_F)$ of the simplex $T$.

There is another extension to the regular simplex method to compensate the lack of curvature which takes the values from neighboring simplices into account, dubbed the improved tetrahedron method~\cite{Kawamura2014}.  In this technique a third-order polynomial is fitted to the simplex and its nearest neighbors such that the coefficients are uniquely determined.  Then a linear interpolation to that third-order polynomial is obtained through a least squares optimization.  The resulting linear interpolation is closer to the original function, but is no longer continuous.

The regular simplex method with Blöchl correction and the improved simplex method are implemented in Octopus for one, two, and three dimensional systems and are currently available in the electron density calculation and in the density of states calculation.

\subsection{New and improved propagation algorithms}

\octopus solves the time-dependent Kohn-Sham equations with propagators as described
in \citet{castro_propagators_2004} and \citet{gomez_pueyo_propagators_2018}. We have
implemented two new propagators, using explicit second-order and third-order Magnus expansions,
and we have implemented iterative versions of the AETRS (approximate enforced time-reversal symmetry)
and CFM4 (fourth-order commutator-free Magnus expansion) propagators, making them effectively
implicit. Moreover, the CFM4 propagator has been extended to make it compatible with
Ehrenfest dynamics and Gauge field propagation.

We call the first new propagator \texttt{exp\_mid\_predictor} and it follows
\citet[Eq. 20]{casas_explicit_2006}. In terms of the Kohn-Sham equations, it can be written
for propagating from $t$ to $t+\Delta t$ as
\begin{equation}
\begin{split}
  \psi_0 & = \psi(t) \\
  H_1 & = H(t, \psi_0) \\
  \psi_1 & = \exp(-i \frac{\Delta t}{2} H_1) \psi_0 \\
  H_2 & = H(t + \frac{\Delta t}{2}, \psi_1) \\
  \psi(t+\Delta t) & = \exp(-i \Delta t H_2) \psi_0.
\end{split}
\label{eq:exp_mid_predictor}
\end{equation}
First, the Kohn-Sham states are predicted by a Forward-Euler step at the half step,
then they are used to compute the Hamiltonian that is to be applied for the full step.
In comparison to this, the \texttt{exp\_mid} propagator uses an extrapolation of the
potential from previous steps to predict the Hamiltonian at the half-step. The
new propagator is second-order and symmetric in time and thus similar to the default
ETRS propagator, but cheaper, because only two applications of the exponential
operator are needed.

We call the second new propagator \texttt{explicit\_magnus3} and it follows
\citet[Eq. 22]{casas_explicit_2006}. In terms of the Kohn-Sham equations, it can be written
for propagating from $t$ to $t+\Delta t$ as
\begin{equation}
\begin{split}
  \psi_0 & = \psi(t) \\
  H_1 & = H(t, \psi_0) \\
  \psi_2 & = \exp\left(-i \frac{\Delta t}{2} H_1\right) \psi_0 \\
  H_2 & = H(t + \frac{\Delta t}{2}, \psi_2) \\
  \psi_3 & = \exp\left(-i \frac{\Delta t}{4} (H_1+H_2)\right) \psi_0 \\
  H_3 & = H(t + \frac{\Delta t}{2}, \psi_3) \\
  \psi_4 & = \exp\left(-i \Delta t H_4\right) \psi_0 \\
  H_4 & = H(t + \Delta t, \psi_4) \\
  \psi(t+\Delta t) & = \exp\left( -i \Delta t \left(
  \frac{1}{6} \left( H_1 + 4 H_3 + H_4 \right) \right.\right. \\
    & \quad \left.\left.
    + \frac{i \Delta t}{12} \left( [H_1+H_2, H_3] + [H_2, H_4] \right)
  \right)
  \right) \psi_0.
\end{split}
  \label{eq:explicit_magnus3}
\end{equation}
It is much more expensive than the second-order schemes, but it can be stable at
larger timesteps because it is third-order in $\Delta t$. In general, propagators based
on the Magnus expansion are geometric integrators that preserve qualitative
features and are well suited for highly-oscillatory
equations\cite{casas_explicit_2006}, such as the Kohn-Sham equations.

In addition to these new propagators, we have extended the existing AETRS and CFM4 propagators
to provide implicit variants by iterating to self-consistency in each timestep.
The second-order propagator AETRS is defined by
\begin{equation}
  \psi(t+\Delta t) = \exp\left(-i \frac{\Delta t}{2} H(t+\Delta t)\right) 
  \exp\left(-i \frac{\Delta t}{2} H(t) \right) \psi(t),
\label{eq:aetrs}
\end{equation}
where $H(t+\Delta t)$ is obtained by extrapolating the potential from previous timesteps with
Lagrangian interpolation \cite{castro_propagators_2004}.
The fourth-order propagator CFM4 is defined by
\begin{equation}
  \psi(t+\Delta t) = \exp\left(-i \Delta t (\alpha_1 H(t_1) +\alpha_2 H(t_2)) \right) 
  \exp\left(-i \Delta t (\alpha_2 H(t_1) +\alpha_1 H(t_2)) \right) \psi(t),
\label{eq:cfm4}
\end{equation}
where $\alpha_{1,2} = \frac{3\mp 2\sqrt{3}}{12}$, $t_i = t_0 + c_i \Delta t$,
and $c_{1,2} = \frac{1}{2} \mp \frac{\sqrt{3}}{6}$
\cite{gomez_pueyo_propagators_2018}. Similar to AETRS, the Hamiltonian at
$t_{1,2}$ is obtained by extrapolating the potential from previous timesteps
using Lagrangian interpolation.

To provide implicit variants of these two propagators, we have added
a self-consistent iteration on top: after the first computation of $\psi(t+\Delta t)$, i.e., the Kohn-Sham states
at the new time, we compute the Kohn-Sham potential and add it to the nodes of the
Lagrange interpolation. Then we repeat the propagation step and update the Kohn-Sham
potential until the potential is self-consistent. This behavior is controlled by
the input variable \texttt{TDStepsWithSelfConsistency}, which is an integer
specifying the number of steps from the beginning for which the self-consistency
is applied. When set to the special value \texttt{all\_steps}, it is applied
throughout the full run. Moreover, the input variable \texttt{TDSCFThreshold}
controls the threshold of the self-consistency loop and is set to $10^{-6}$ by default.
We have found this option useful to get more stable propagations for
complicated functionals, such as mGGA or hybrid functionals.

\subsection{Enabling large grids with more than 1 billion grid points}

Recently, we have seen an increased need for grids with a large
number of points in \octopus, e.g., to resolve the wave functions of core
electrons with a small spacing or to run simulations of twisted light beams
interacting with large structures.
\octopus  treats functions on the grid as 1D arrays and stores the
associated spatial information in a mapping between a 1D array and $n$D space
(most often, calculation are done in 3D). Historically, indexing of these arrays has been done in the code with
integers of default kind in Fortran which have a size of 32 bit. The largest
representable number for signed 32-bit integers is $2^{31}-1\approx2.1\times 10^9$.
Despite this, the maximum number of grid points feasible so far has been lower
by a factor of 3 (or $n$ for $n$D), so about 700 million, because in
some parts of the code the grid size was multiplied by the dimension.
However, recent applications of the code necessitate larger grids, 
with needed resolutions reaching $1000^3$ or even $2000^3$, clearly above the
current possibilities.

To enable these applications, 64-bit integers are needed for indexing and
mapping the grid, thus providing a sufficient range of numbers.\footnote{Up to about
$2^{63}-1\approx9.2\times10^{18}$.}
We have chosen to transform only the indexing and mapping of the grid and all accesses
to global arrays to 64-bit integers, leaving access to local arrays (i.e., parts
of the distributed array in domain parallelization) as it is. This means that
local arrays are still bound by about 700 million grid points, but has the
advantage that code changes are more localized.

First, the grid generation, indexing,
domain parallelization and all parts touching global arrays have been changed to
use 64-bit integers.
In the process, many low-level parts needed to be adapted, such as IO routines,
the interface to METIS and ParMETIS, qshep for interpolation, the C implementation
of the stencil application, the CUDA implementation (especially transfers and
kernel launches), handling of boundary points, writing and reading restart data, 
and more. Moreover, all MPI collective and send/receive
calls have been wrapped into a separate module and special wrappers for
collectives with 64-bit displacements have been
implemented because they do not exist in the MPI 3.1 standard. Although
this functionality is part of the MPI 4 standard, that standard is not
yet widely supported by the major MPI implementations, thus we have not used
it yet, but we can easily switch in the future.

The global index arrays that are still needed and allocated on each process,
take up large amounts of memory for large grid sizes. To reduce the
memory footprint of these arrays, they are now only allocated in one process on
each node and then shared between all processes on a node; the implementation
uses MPI-3 shared memory windows.

Simulations using large grids need a lot of memory because the code still needs
to allocate some global arrays, such as the index arrays and the mapping between
mesh and cube. Thus, it is very advisable to use the PFFT library for solving
the Poisson equation as otherwise, additional global mesh functions are
allocated on each core. To compute the distribution of large grids, the 64-bit
versions of METIS and ParMETIS are required. Moreover, using only the ParDomains
strategy for parallelization helps to reduce the memory footprint.

For large grid sizes, output takes very long, so it should be kept to a minimum.
The problem is not only the raw IO, but also the computation required by certain
quantities or output formats.  Moreover, the output size can
be quite large; for cubic Si with a resolution of $1500^3$, the restart data
alone takes more than 600 GB. 
We note that in order to use the new functionality, \octopus needs to be built against 64-bit
variants of METIS and ParMETIS.
We have successfully tested that the implementation works on grids of sizes up to $1500^3$ for
TDDFT simulations and up to $2000^3$ for Maxwell simulations.

\subsection{Improving the imaginary-time propagation}

Solving the Kohn-Sham equations in imaginary time is an alternative method of obtaining
the ground state compared to the self-consistent field method
\cite{aichinger_fast_2005,lehtovaara_solution_2007,bader_solving_2013,flamant_imaginary-time_2019}.
We have improved our implementation of imaginary-time evolution in \octopus:
inside the SCF framework, we do one imaginary timestep per SCF step, after which we
do a subspace diagonalization combined with an orthonormalization. Moreover, we have
extended the Lanczos and Chebyshev methods for computing the exponential to deal with
imaginary timesteps. This means that the corresponding operator is a diffusion operator
with eigenvalues on the real axis instead of on the imaginary axis as for the default
time propagation.

In addition to the default forward-Euler scheme,
\begin{equation}
  \psi(t_{n+1}) = \exp\left(-\tau H(t_n)\right) \psi(t_{n}),
  \label{eq:imag_forward_euler}
\end{equation}
with the imaginary time step $\tau$, we have implemented the exponential midpoint scheme
\begin{equation}
  \psi(t_{n+1}) = \exp\left(-\tau H(t_n+\frac{\tau}{2})\right) \psi(t_{n}),
  \label{eq:imag_exp_mid}
\end{equation}
where we compute the Hamiltonian at the midpoint by extrapolating the potential
from the previous steps.
Moreover, we implement variable timestepping as proposed by \citet{aichinger_fast_2005}, Section 2.5. The timestep is reduced if the normalized error in the normalization energies compared to the previous iteration is smaller than the normalized error between normalization and expectation energies. Moreover, the error in the normalization energy is not allowed to become too small compared to the residual to avoid reducing the timestep too much. This leads to a more resilient behavior.

Testing on a wide range of systems for different parameter values leads us to the following conclusions:
\begin{itemize}
  \item 
    Usually, the default forward-Euler scheme is more robust and converges more often than the exponential midpoint scheme.
  \item 
    For many systems, large initial timesteps combined with variable timestepping lead to faster convergence.
  \item 
    The convergence behavior with respect to the timestep depends on the system, i.e., some systems converge with certain timesteps while others diverge.
  \item 
    Variable timestepping leads to better convergence for many systems, while it does not work for other systems. For small initial timesteps, the timestep is often not reduced at all.
\end{itemize}

In summary, the imaginary-time propagation takes more computational time than
the default Chebyshev filtering, similar to the conclusion by
\citet{flamant_imaginary-time_2019}. When trying to reduce that time by
increasing the initial timestep, the procedure might not converge at all, so
overall it is also less robust than the default Chebyshev filtering eigensolver.
When used with small timesteps, it can be an alternative for systems where the default
eigensolver fail to converge.

\subsection{Improvements to the exponential operator}

In \octopus, we have three methods of computing the exponential of the Hamiltonian:
Taylor expansion, Lanczos method, and Chebyshev expansion. The method is controlled
by setting the input variable \texttt{TDExponentialMethod}. These methods have been improved
in various aspects.

First, the Lanczos and Chebyshev methods now iterate until a given
error threshold is reached as specified by the input variables \texttt{TDLanczosTol}
and \texttt{TDChebyshevTol}, respectively, without having to specify a maximum order for the Krylov subspace or the polynomial expansion. For the Lanczos method\cite{saad_analysis_1992}, all arrays needed
for the Krylov subspace are allocated on the fly such that the Krylov subspace is extended
until the estimated error falls below the threshold. For the Chebyshev expansion, we use
the error bounds from \citet{lubich_quantum_2008}, Theorem 2.1 to 2.4, to estimate the number of iterations needed.
Second, the Chebyshev expansion has been extended to also treat imaginary timesteps which
needs different coefficients in the expansion and also different error estimates\cite{hochbruck_exponential_2010,bergamaschi_efficient_2000}.
Third, all three methods have been extended to support evaluation of the functions $\phi_1(z) = \frac{\exp(z) - 1}{z}$
and $\phi_2(z) = \frac{\phi_1(z) - 1}{z}$, functions that are needed for time propagation schemes
of inhomogeneous equations, such as, e.g., Maxwell's equation in Riemann-Silberstein form.
The Lanczos and Chebyshev methods now even support the evaluation of arbitrary functions
of operators (with certain properties for $z\to\infty$). For Lanczos, this is implemented
by evaluating the corresponding function on the matrix that is the projection of the
Hamiltonian onto the Krylov subspace. For Chebyshev, the coefficients are computed
using a discrete cosine transform of the function values evaluated at $-i \Delta t (A\cos(j \pi/N) + B)$,
where $N$ is the order, $j$ goes from 0 to $N-1$, $A$ is the half span of the
spectrum of the Hamiltonian and $B$ the middle point. After the transform, the $j$-th coefficient
is divided by $j$ and the first and last coefficients in addition by 2. 
The error is estimated according to \citet{lubich_quantum_2008}, Theorem 2.1 to 2.4.

These changes make the implementation of our exponential operators easier to
use, more accurate, and more versatile, improving our default exponential integration methods.

\subsection{Improved support for Casida calculations}
\octopus featured for a long time Casida calculations. However, up to recently, this was restricted to the local density approximation level. 
Recently, Panta and Strubbe implemented the TDHF method, to complement the Casida implementation~\cite{panta2024implementation}. The details of this implementation can be found in their publication and are therefore not reported here.

Apart from the implementation of TDHF, we also improved the support of semilocal functionals, more precisely adding the contribution of the GGA gradient to the exchange-correlation kernel $f_{\rm xc}(\mathbf{r}, \mathbf{r'})$. We also added the support of the averaged-density self-interaction correction (ADSIC)~\cite{Legrand_2002, 10.1063/5.0179087}.

Finally, we implemented the Casida response function with fractional occupation numbers. Previously it was implemented for integer occupation where a transitions from fully occupied to fully unoccupied orbitals were taken into account. With the new implementation it can handle the fractional occupation due to smearing by considering both energy difference and occupation numbers difference windows. By default both the windows have a cut-off of $10^{-12}$ (for energy the unit would be a.u.). We defined two variables where the user can control these windows using keywords \texttt{CasidaThresholdEnergy} and \texttt{CasidaThresholdOccupation}. The recommended tolerance for these windows is $\sim 10^{-6}-10^{-7}$.

\subsection{Further improvements}

Let us now comment on some further developments that were added to the \octopus code, beyond the main developments discussed above.
We complemented our support of the symmorphic symmetries in solids by the support of non-symmorphic operations, in the cases where they are compatible with the real-space grid. These operations are always detected, using the Spglib library. However, the non-symmorphic symmetries are only used for the electrons when the grid does not break them. For symmetrizing forces and the stress tensor, both symmorphic and non-symmorphic are employed when the user requests to use symmetries, as this allows to properly preserve the space-group of crystals when performing geometry optimizations for instance.\\
We also implemented the computation of the local orbital magnetization, i.e., the integral of the orbital angular moment operator over atom-centered spheres, following the same philosophy as for the local spin magnetic moment. 

\section{Technical Improvements}

\subsection{CMake Build System}

A significant infrastructure improvement is the recent introduction of a CMake-based build system. This replaces the previous autotools build system with a more configurable framework, improving portability across compilers, platforms, and high-performance computing environments. The adoption of the Ninja build backend results in considerably faster configuration and build times compared to the former combination of autotools and make.

The CMake build system provides clearer handling of external libraries and build variants, making it easier to configure Octopus for serial, MPI-parallel, and accelerator-enabled installations. This also simplifies integration with continuous-integration workflows and downstream packaging tools. Configuration options used by our continuous-integration builds are fully formalised in CMake presets. We found that the use of a minimal inheritance hierarchy makes preset options transparent and extensible.

Required numerical and scientific libraries include GSL, BLAS/LAPACK or MKL, FFTW or the MKL FFT interfaces, Libxc, and Spglib. Libxc and Spglib can optionally be built using CMake’s FetchContent mechanism, while METIS is packaged with Octopus, making it possible to build Octopus with minimal external dependencies present, on most desktop and HPC systems.

Optional dependency discovery is deliberately permissive: CMake links to all detected optional libraries and fails only if a dependency is marked as required. Custom \texttt{Find*.cmake} modules normalise external packages into imported targets, trying native CMake packages first before falling back to \texttt{pkg-config}. The supported optional libraries now include: ADIOS2, BerkeleyGW, CGAL, CUDA, DFTB+, ELPA, Fortuno, HIP, libvdwxc, METIS/ParMETIS, netCDF-Fortran, NLopt, PFFT/PNFFT, PSolver, SPARSKIT, and Wannier90. ADIOS2, Fortuno, and Wannier90 can also be obtained through FetchContent.

Testing is integrated through CTest. Unit tests are optional and use Fortuno~\cite{fortuno2024} through a single Fortran test executable. The existing application regression suite is enabled by default and wraps existing \texttt{.test} files, extracting labels and processor requirements to populate CTest metadata, including MPI/OpenMP processor counts and GPU resource groups.

\subsection{Python post-processing and Workflow Integration}

Postopus provides a Python post-processing interface for Octopus simulations~\cite{postopusgit}. It automatically discovers the files produced by an Octopus run; maps systems, calculation modes, and output quantities onto a navigable Python object model; and exposes numerical results as standard \texttt{pandas} or \texttt{xarray} objects. This enables fast data inspection and visualisation of Octopus outputs in both Jupyter notebooks and interactive visualisation workflows, while also facilitating scientific Python analysis without the need to develop custom file parsers.

Postopus also provides input generation and output parsing functionality for the Octopus AiiDA plugin~\cite{octopusaiida}, which can be used to automate and formalise complex workflows, such as ground-state calculations followed by time-dependent propagation, multiple TD trajectories, and phonon band-structure calculations. The use of AiiDA~\cite{huber2020aiida} for reproducible, provenance-tracked workflows rather than ad hoc collections of files and scripts allows these multi-step calculations to be executed, inspected, and reused in a systematic way.

\subsection{Continuous integration and continuous deployment}

Octopus has a comprehensive continuous integration (CI) and continuous
deployment (CD) setup, which has undergone major restructuring. The system has
been ported from Buildbot to GitLab CI/CD. This has led to significant
simplifications, primarily because the CI/CD configuration is now version
controlled in the same repository as the code. Both code and CI/CD configuration
can now be adapted simultaneously to keep them compatible and each branch uses
its own compatible CI/CD configuration.

\begin{figure}
  \includegraphics*[width=\linewidth]{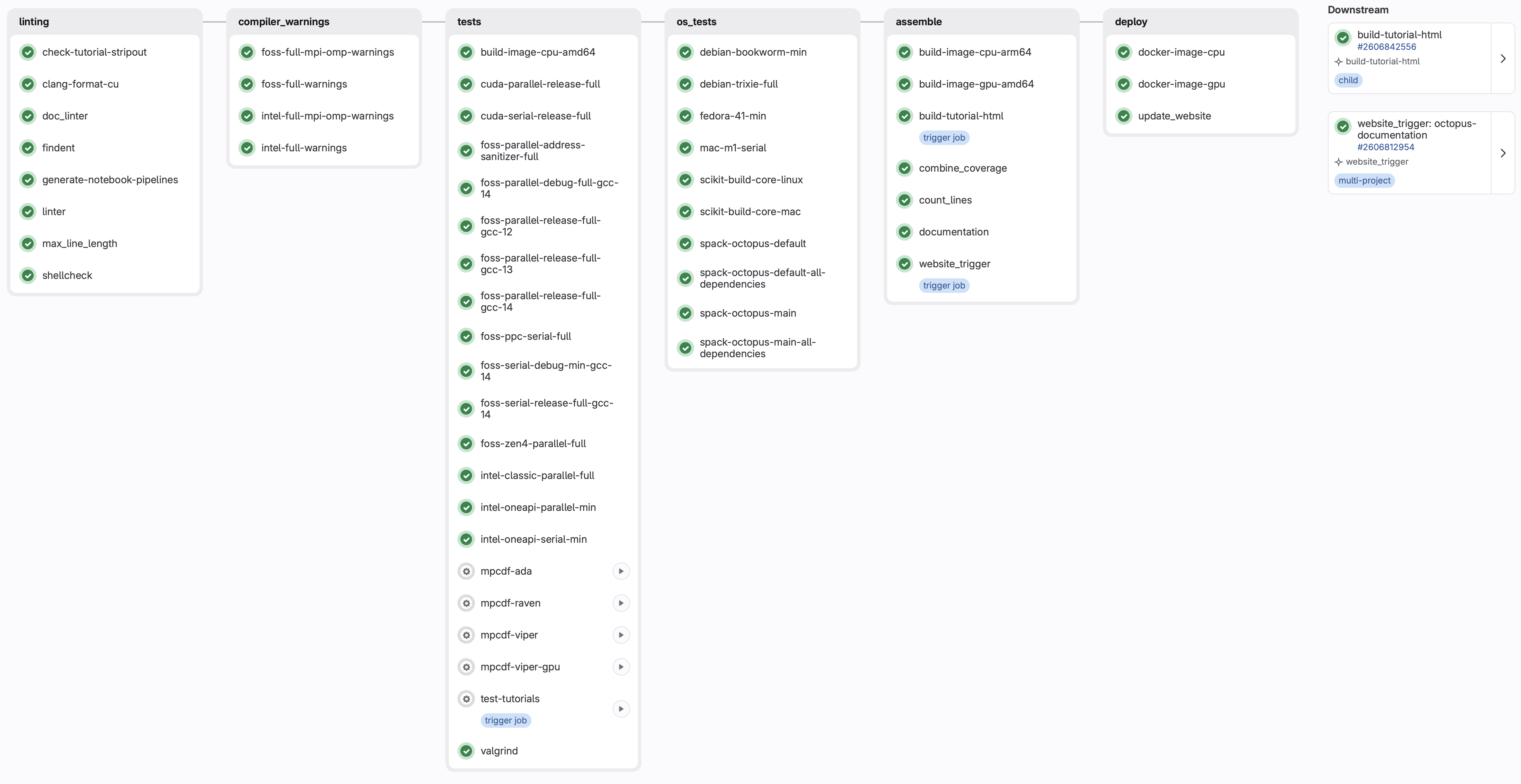}
  \caption{\label{fig:ci-pipeline}Screenshot of a scheduled pipeline executed for the main
    branch. Jobs are grouped into different stages, which are executed in order,
    trigger jobs run additional downstream pipelines. Execution of the whole
    pipeline takes about 2~hours.}
\end{figure}

For each commit a pipeline with about 50 different jobs is executed, the exact
number of jobs depends on the pipeline type (commit in a merge request, commit
on main, scheduled pipeline, etc.). Jobs are arranged in multiple stages, where
later stages only run once all previous jobs have finished successfully, an
overview is shown in Fig.~\ref{fig:ci-pipeline}. The following types of jobs are
run in order: (i)~Static analysis and style checks in the \emph{linting} stage.
(ii)~Compilation with many compiler warnings enabled in the
\emph{compiler\_warnings} stage; tests in this stage can only pass if no
warnings are emitted. (iii)~Execution of the Octopus test suite (approximately 250
regression and unit tests) in the \emph{tests} and \emph{os\_tests} stages
(which are executed in parallel) on different types of hardware and with
different software stacks, more details below. (iv)~Various “preparation” jobs
in the \emph{assemble} phase, such as building Docker images for different
architectures, building documentation, etc. (v)~Assembly and deployment of
artifacts in the \emph{deploy} phase, such as updating documentation on the web
server, and combining Docker images to multi-arch images and uploading them to
the GitLab container registry.

Octopus tests are executed on a variety of hardware and software environments to
increase the coverage breadth. The dedicated Octopus build farm provides Intel
and AMD CPU nodes, PowerPC nodes, NVIDIA GPU nodes, and Apple silicon nodes. All
machines run GitLab Runner, and groups of machines can be selected in the CI/CD
configuration via tags. Overall, more than 40 machines are used exclusively for
Octopus CI/CD. On the software side several different types of environments are
used. The majority of tests are run with dedicated software toolchains, that are
compiled with Spack~\cite{gamblin2015}, and provide GCC and Intel compilers in
different versions, all required and optional dependencies, and variants
without/with MPI, OpenMP, or CUDA. Besides regression testing a variety
of additional checks are performed with valgrind~\cite{Nethercote2003} and
address sanitizer~\cite{serebryany2012addresssanitizer}. On Apple silicon, the
software environment is installed with \texttt{brew} and Octopus is tested
against that software stack. To have stable test environments, despite the
rolling-release model of brew, we use Tart virtual machine images, which we
update monthly, and \texttt{gitlab-tart-executor} as executor for GitLab Runner,
which automatically starts an ephemeral virtual machine for each CI job. Further
tests are executed in Debian and Fedora Docker images with dependencies provided
from their respective package repositories. Finally, compilation of Octopus and
all dependencies in their most recent version directly with Spack is tested in
regular intervals.

To facilitate zero-install execution of Octopus, Docker images are provided for
the architectures amd64 (Intel/AMD), arm64 (Apple silicon) and GPU (NVIDIA
CUDA). Images are compiled as part of the CI/CD pipelines and made available in
the GitLab container registry for all Octopus releases as well as a nightly
image for the main branch~\cite{octopusdocker}.

\subsection{Tutorials on Jupyter notebooks}

Octopus comes with about 50 tutorials. These are currently being ported from static
markdown files to Jupyter notebooks~\cite{Granger2021}. The new format provides a
number of advantages~\cite{Fangohr2019,Beg2021}. Users can now run tutorials online on Binder without
having to install anything, each tutorial now is a fully self-contained file so
users only need to download a single notebook to run a complete tutorial,
notebooks combine setup of input files, execution of Octopus and data analysis
with \texttt{postopus}, and tutorials can now be tested in the CI. The CI setup
works as follows. First, a Docker image for the current commit is created, so
that Octopus only has to be compiled once. Second, each notebook is executed in
that image using either pytest~\cite{pytest} with the nbval
extension~\cite{fangohr2020testing} when working on tutorials in merge requests
or jupyter nbconvert to convert notebooks into static HTML for the Octopus
website. The jobs for executing notebooks are generated dynamically and executed
in a so-called downstream pipeline, which ensures that newly added tutorials are
automatically tested in CI. Compiling Octopus once for the Docker image and
subsequent parallel execution of the tutorials helps keeping the pipeline
walltime as short as possible and makes better use of the available CI/CD
hardware.

\subsection{AMD Support}

In 2024, the Max Planck Computing and Data Facility installed Viper, a new HPC system that includes several nodes equipped with AMD MI300A APUs. On this hardware the CPU and GPU reside on the same silicon and share a unified pool of RAM. As several other HPC centers have likewise adopted AMD GPUs, support for this vendor has become necessary and has been implemented in Octopus.

While NVIDIA provides \texttt{CUDA} as its GPU programming framework, AMD offers its own counterpart, \texttt{HIP}. Porting to AMD hardware would, in principle, require rewriting both the infrastructure and the kernels in this new language. However, \texttt{HIP} is designed as an abstraction layer whose API closely mirrors \texttt{CUDA}: depending on the target vendor, it dispatches either to \texttt{CUDA} or to the \texttt{ROCm} stack (AMD's GPU software platform).

Octopus exploits this similarity to introduce \texttt{HIP} support through simple macro definitions. Because most \texttt{HIP} functions share the same signatures and usage conventions as their \texttt{CUDA} counterparts, macro-based substitution in GPU kernels is both straightforward and effective, while keeping the required code changes to a minimum.

In Octopus the GPU kernels are compiled at run time with a JIT compiler (namely \texttt{nvrtc} for NVIDIA and \texttt{hiprtc} for AMD). Since automated tools such as \texttt{HIPIFY} do not fully support the conversion of these JIT functions and their associated functionality, we opted for a manual conversion instead.

In addition, several specific optimizations targeting the \texttt{AMD} vendor have been implemented. For instance, at the kernel level, \texttt{HIP} supports native FP64 in-warp reductions, while in \texttt{CUDA} these reductions should be manually implemented via in-warp programming (in-warp intrinsics).

\subsection{\texttt{Libxc} on GPUs}

\texttt{Libxc} is an open-source library that provides a unified and portable collection of exchange-correlation functionals for DFT~\cite{Libxc_paper1, Libxc_paper2}. It implements more than 700 functionals and exposes them, together with their derivatives, through a consistent interface. By separating the implementation of the functionals from that of the host electronic-structure code, \texttt{Libxc} has become a de facto standard, adopted by many widely used quantum-chemistry and condensed-matter packages. Octopus relies on it for the entire xc contribution to the Kohn-Sham problem.

Octopus is increasingly running on GPU-accelerated architectures, where evaluating the exchange-correlation functional on the host CPU introduces a bottleneck and costly host-device data transfers. Following recent developments in \texttt{Libxc} that allow to compute the functionals on the device, an initial GPU implementation has been added.

The limiting factor in the current CPU implementation is not the cost of the functional evaluation itself, but the host-device transfers before and after each \texttt{Libxc} call, which prevent scaling in multi-GPU simulations. A useful GPU implementation therefore requires the data passed to \texttt{Libxc} to reside permanently on the device throughout the simulation. As a first step, the current version improves the residency of the density on the GPU.

\subsection{Compiler support for LLVM Flang}

Even though the Fortran programming language was created in the 1950s, it is still the prevailing language in many high-performance computing applications today. Most of electronic structure code, such as Octopus, form the foundations of modern materials science, are written in Fortran, and it is not realistically feasible to rewrite these massive and complex computer programs in another language.  In light of today's advances in software development, Fortran might be viewed as outdated, but the language has a rich history and inspired many of the programming paradigms that we take for granted today.  In more recent iterations of Fortran's standardization process features were added to bring the language more on par with the current ways to develop software, such as object oriented programming.  The continued interest by the scientific community is the main driver behind new efforts to build modern compilers for Fortran.

The LLVM Project is a collection of modular and reusable compiler and toolchain technologies~\cite{Lattner2004}. LLVM also ships with a Fortran compiler named Flang, which is currently under active development, but is already being used in production as the basis for the AMD Optimizing C/C++ and Fortran Compilers (AOCC) and the next version of the NVIDIA HPC Fortran, C++ and C Compilers.  We have made sure that Octopus successfully compiles with the upstream Flang compiler, in order to be ready to support any compiler that is based on Flang in the future.

\subsection{Distribution of precompiled binaries via PyPI}

An important aspect for onboarding new users to effectively utilize Octopus is the installation process.  The installation itself is one of the main entry barrier and separates scientists from running their first simulation.  An easy installation procedure is also attractive for students and scientists with no or little experience in compiling and packaging, as it allows them to quickly deploy Octopus on the system of their choice with ease.  Thus, software distribution becomes a decisive factor in the success of a code and building a community around it.

On high-performance computing systems, Octopus is usually compiled from scratch, making use of hardware-aware compiler optimization and pre-installed high-performance libraries to exploit the maximum performance of the computational resources.  For new users, running Octopus on a laptop to complete tutorials or explore small systems, this barrier to entry is unreasonably high, having to potentially fight with local dependency resolution or missing dependencies, that altogether renders the installation annoying and inaccessible to inexperienced users.

In the Python ecosystem the distribution of precompiled binaries has prevailed, in particular with the growing popularity of machine learning frameworks, which tend to have similarly complicated build systems as Octopus.  The main limiting factor in publishing binaries is compatibility, but the Python Packaging Authority (PyPA) has set a standard for shipping of binaries as part of Python packages~\cite{pep513} to work on many systems.  While Octopus itself does not have a Python interface, we use the PyPA standard to package Octopus using scikit-build-core~\cite{Schreiner2024} and distribute it through the Python Package Index (PyPI), such that users with a working Python installation can easily install Octopus via a simple \texttt{pip install} command.  The set of dependencies provided with this installation is only minimal, but sufficient to run the tutorials and get started with simple calculations.

\subsection{A new restart file format}

\octopus has a long-standing issue when reading and writing restart data
for the wavefunctions: it creates one file per $\mathbf{k}$-point and state, which
can easily lead to several million files for simulations of periodic systems
with a highly resolved  $\mathbf{k}$-space. This is especially problematic on parallel
file systems, as used in HPC machines, because of the overhead from handling
metadata. We have now implemented a new restart file format based on
the ADIOS2 library\citep{adios2_2020} which is specifically tuned for
scientific data management on HPC machines. This solves the problem
of creating too many small files and improves the performance of the writing and reading of restart files in general.

To write out the wavefunctions, we create a big array of all states (corresponding to the both bands and  $\mathbf{k}$-points)
within ADIOS2 and create pointers with the API directly to the memory
of the batches which is the data structure in \octopus that holds the
wavefunction data. After setting everything up, reading or writing is
triggered by a single call, allowing ADIOS2 to aggregate the data
internally. We use the default BP5 file format with default parameters.

We have benchmarked the new implementation by running two examples
on the Raven supercomputer at the Max Planck Computing and Data Facility.
All runs have used 16 nodes with 8 MPI ranks and 9 OpenMP threads per node,
i.e., 128 MPI ranks in total. The first example is a large supercell
of a 2D-periodic layer of hBN with 2.8 million grid points, about 2700 states
and one $\mathbf{k}$-point. For the old restart format, this results in about 2700 files,
whereas it is only about 50 for the new implementation. For the ground
state with only domain parallelization, writing the restart files was
faster by a factor of 90 with the new implementation. For the time-dependent
simulation with a mixed parallelization (domains 16, states 8), reading the
restart files was faster by a factor of 7.5, writing by a factor of 8.8.
The second example is cubic Si with about 27000 grid points, 24 states and
64000  $\mathbf{k}$-points and a parallelization only over  $\mathbf{k}$-points. This results in about
1.5 million files for the old format and about 50 for the new format.
Writing the ground-state restart files was faster by a factor of 6.3, and
reading and writing the time-dependent restart files was faster by factor
of 4.2 and 4.9, respectively.

Thus, the new implementation is much faster across different examples
and it does not stress parallel file systems as much anymore by creating
fewer files. The ADIOS2-based format is now used by default if \octopus
is built with support for the ADIOS2 library.

\section{Use of artificial intelligence (AI)}
We use AI to proof-read the paper and find typographic errors as well as mistyped references.\\
Concerning the source code of the Octopus, some of the core developers started to use AI agents to enhance their contributions to the code. This has led to a more stringent review process, only performed by human reviewers, and a strong increase of the coverage of the testsuite and requirements in terms of code quality and documentation. These strict rules have been agreed by all core developers, and the guidelines are stated explicitly in the code documentation, in order to preserve the high-quality level of the Octopus code.

\section{Conclusion}

In conclusion, we have reported many recent developments in \octopus. 
These span widely different areas of electronic structure — lattice dynamics, excitons in materials, novel functionals, core-level spectroscopy, and light–matter coupling in and out of cavities, to name a few. Taken together, they show that \octopus remains, after 25 years, a lively and active ecosystem for exploring new ideas. The simplicity of the real-space real-time approach, combined with the flexibility of the implementation, allows the code to keep evolving rapidly, and new algorithms — from improved ground-state eigensolvers to novel time propagators — can be introduced with relative ease.\\
We have also described developments on the technical side: the port to AMD GPUs, a new restart format that improves file-system performance, and support for modern LLVM compilers. Equally important, the interfaces between \octopus and the wider ecosystem of electronic-structure libraries, such as Wannier90 and \phonopy, open new avenues for combining real-time TDDFT with established analysis tools.\\
The changes reviewed here are already available to all users. Further developments are planned in the \octopus roadmap and will be made available continuously in future releases.




\begin{acknowledgments}
The authors would like to thank all the people that have contributed to the development of \octopus over the last two decades. E.I.A. and J.R. acknowledge support from the International Max
Planck Research School. LK acknowledges funding from the EU's NextGenerationEU through the Recovery and Resilience Plan for Slovakia, under project No.~09I05-03-V02-00034.
S.V.B.J. acknowledges support from the Alexander von Humboldt Foundation. S.N. acknowledges funding from Deutsche Forschungsgemeinschaft (DFG) (Project ID 549456677). This work was supported by the European Research Council (ERC-2024-SyG-101167294; UnMySt), the Cluster of Excellence Advanced Imaging of Matter (AIM), Grupos Consolidados y Alto Rendimiento UPV/EHU, Gobierno Vasco (IT1453-22). We acknowledge support from the Max Planck-New York City Center for Non-Equilibrium Quantum Phenomena. The Flatiron Institute is a division of the Simons Foundation.
\end{acknowledgments}

\bibliography{biblio}

\end{document}